\documentclass[universe,review,accept,moreauthors,pdftex]{Definitions/mdpi} 
\firstpage{1} 
\pubvolume{1}
\issuenum{1}
\articlenumber{0}
\pubyear{2023}
\copyrightyear{2023}
\externaleditor{Academic Editors: Daniela Saadeh, Katarina Markovic}
\datereceived{} 
\dateaccepted{} 
\datepublished{} 
\hreflink{https://doi.org/} 

\usepackage{graphicx}
\usepackage{dcolumn}
\usepackage{bm}
\usepackage{aas_macros}
\usepackage[caption=true]{subfig}

\def\MP{M_{\rm Pl}}
\newcommand{\dd}{\mathrm{d}}

\DeclareMathOperator{\cd}{cd}

\newcommand{\customElder}{Elder, Vardanyan, et al. \cite{Elder:2019yyp} }

\Title{Modeling and testing screening mechanisms in the laboratory and in space}
\TitleCitation{Modeling and testing screening mechanisms in the laboratory and in space}

\Author{Valeri Vardanyan $^{1,2,\ddagger}$*\orcidA{} and Deaglan J. Bartlett $^{3,}$*\orcidB{}
\AuthorNames{Vardanyan, V.; Bartlett, D.J.}}
\address{%
$^{1}$ \quad Kavli Institute for the Physics and Mathematics of the Universe (WPI), UTIAS, The University of Tokyo, Chiba 277-8583, Japan\\
$^{2}$ \quad Center for Data-Driven Discovery, Kavli IPMU (WPI), UTIAS, The University of Tokyo, Kashiwa, Chiba 277-8583, Japan\\
$^{3}$ \quad CNRS \& Sorbonne Universit\'{e}, Institut d’Astrophysique de Paris (IAP), UMR 7095, 98 bis bd Arago, F-75014 Paris, France\\
}
\corres{Correspondence: valeri.vardanyan@ipmu.jp (V.V.); deaglan.bartlett@iap.fr (D.J.B.)}

\abstract{The non-linear dynamics of scalar fields coupled to matter and gravity can lead to remarkable density-dependent screening effects. In this short review we present the main classes of screening mechanisms, and discuss their tests in laboratory and astrophysical systems. We particularly focus on reviewing numerical and technical aspects involved in modeling the non-linear dynamics of screening. In this review, we focus on tests using laboratory experiments and astrophysical systems, such as stars, galaxies and dark matter halos.
}

\begin{document}
\tableofcontents

\section{Introduction}
\label{sec:introduction}

The accelerated expansion of the Universe, discovered over two decades ago \cite{SupernovaCosmologyProject:1998vns,SupernovaSearchTeam:1998fmf}, still lacks a satisfactory theoretical explanation. Scalar fields have played a significant role in modeling cosmic acceleration, and provide a relatively simple playground for establishing novel phenomenological features. Such scalar fields have been considered in two broad contexts, namely, Dark Energy (see e.g. Ref.~\cite{Copeland:2006wr}) and modifications of Einstein's General Relativity (GR) (see e.g. Refs.~\cite{Silvestri:2009hh,Clifton:2011jh,Joyce:2014kja,Koyama:2015vza}). The defining feature of modified gravity theories is the non-trivial coupling of scalar fields to gravity and matter. In modified gravity, therefore, one expects long-range ``fifth'' forces across a very broad range of scales, manifesting themselves from microscopically small scales to cosmologically large ones, characterized by
the present-day expansion rate $H_0$. 

In the cosmological regime, the phenomenology can be well-understood with linear perturbation theory methods. Particularly, the primary modifications are expressed as scale-and-time-dependent gravitational coupling and lensing potential (see e.g. Refs.~\cite{Ade:2015rim, Pogosian:2016pwr}). Ongoing and upcoming large scale structure surveys are the primary testing ground for cosmological phenomenology of modified gravity. However, in order to have consistent dynamics on all the scales of interest, fifth forces should be screened away in regions much denser than
the mean cosmological environments. Otherwise, such fifth forces would be ruled out on the grounds of local tests of gravity, which provide stringent constraints on additional interactions \cite{Will:1993ns}.

The inherently non-linear dynamics of non-trivially coupled scalar fields can lead to remarkably diverse effects depending on the environment where they operate. Particularly, it has been realized that non-minimally coupled scalar fields with or without higher-derivative terms can lead to the so-called screening effects in dense environments. This environmental dependence allows the field to induce large deviations from GR on relatively poorly constrained cosmological structure formation and gravitational lensing scales, while having smaller effects in regimes where GR has been tested more precisely. 

The most widely considered working scenario, known as \textit{chameleon mechanism,} has been proposed and investigated in Refs.~\cite{Khoury:2003rn,Khoury:2003aq,Brax:2004qh}. A related scenario is the so called \textit{symmetron screening,} introduced and studied in Refs.~\cite{Hinterbichler:2010es,Hinterbichler:2011ca}. Both of these classes operate due to non-linearities in the potential sector. Non-linearities in the derivative sector can also lead to density-dependent screening, for example, the \textit{Vainshtein mechanism} \cite{Vainshtein_1972,Nicolis_2009,Babichev:2013usa}.

Due to their non-linear nature, the consistent modeling of screened fifth forces is a challenging task. Analytical progress is possible in certain situations, although such approaches are always problem-specific, and are hard to generalize. Numerical techniques have played a significant role in understanding screening phenomenology both in cosmological and astrophysical scenarios, and in laboratory experiments. In this review we focus on summarizing the technical approaches, both analytical and numerical, to screened fifth force phenomenology. We refer the reader to already existing excellent reviews covering theoretical and phenomenological aspects of screening mechanisms in greater detail. Among these references \citet{Joyce:2014kja} provide a thorough review of modified gravity theories and their phenomenological aspects, \citet{Koyama:2015vza} focuses on cosmological phenomenology, \citet{Burrage:2016bwy,Burrage:2017qrf} provide detailed overview of chameleon and partly also symmetron tests, \citet{Sakstein:2018fwz} and \citet{Baker:2019gxo} focus on astrophysical tests, \citet{Brax:2021wcv} focus on effective field theory approach.

Instead of giving a detailed account of all the theoretical aspects of screening mechanisms, here we will focus on three representative classes. Namely, we will briefly introduce the chameleon, symmetron, and Vainshtein mechanisms in Section~\ref{sec:theories}. In Section~\ref{sec:numerical} we detail state of the art numerical methods used in studies of screened modified gravity. We particularly discuss Finite-Difference and and Finite-Element algorithms, as well as techniques for calculating the forces on extended objects. Section~\ref{sec:lab_tests} is dedicated to laboratory tests. We particularly discuss direct-force measurement experiments, such as Casimir and torsion balance tests, as well as indirect measurements, such as atomic interferometry and atomic spectroscopy experiments. In Section~\ref{sec:astrophysical_tests} we focus on a broad range of astrophysical tests. We particularly review the fifth force effects on stellar structure and evolution, galaxy morphology and dark matter halos structures. In this section we also provide a review on newly emerging field of research which aims at producing screening maps of large-scale structure using constrained dark matter simulations. We conclude in \Cref{sec:conclusions}.

\section{Summary of theories}
\label{sec:theories}

\subsection{Thin-shell scenarios}
\label{subsec:thin_shell}

The simplest way to couple matter to scalar fields is via the universal conformal coupling. Chameleon \cite{Khoury:2003rn,Khoury:2003aq,Brax:2004qh} and symmetron \cite{Hinterbichler:2010es,Hinterbichler:2011ca} mechanisms, two of the most studied scenarios in the literature, rely on such conformal couplings. More concretely, we consider a generic theory described by the action
\begin{equation}
	S_\varphi = \int \sqrt{-g} \;\mathrm{d}^4 x  \left[  \frac{M_\mathrm{Pl}^2}{2}R - \frac{1}{2} \nabla^\mu \varphi  \nabla_\mu \varphi - V(\varphi)\right] + S_\mathrm{M}\left(\widetilde{g}_{\mu \nu}, \Psi \right),
\end{equation}
where $M_\mathrm{Pl}$ is the Planck mass, $R$ is the Ricci scalar corresponding to the Einstein-frame metric $g_{\mu\nu}$, and $S_\mathrm{M}$ is the action for the matter fields. The real-valued scalar field $\varphi$ is assumed to be coupled to $g_{\mu\nu}$, while matter fields $\Psi$ couple to the Jordan frame metric $\widetilde{g}_{\mu \nu}$. The two metrics are related by the following conformal transformation
\begin{equation}
	\widetilde{g}_{\mu \nu} = A^2(\varphi) g_{\mu \nu}, 
\end{equation}
where $A(\varphi)$ is a generic function of the scalar field. The model is fully specified in terms of the coupling function $A(\varphi)$ and the potential $V(\varphi)$. 

The scalar-field equation of motion is given by
\begin{equation}
	\square \varphi  = V_{, \varphi} - \frac{\beta}{M_\mathrm{Pl}}A^4 (\varphi)\widetilde{T} \equiv V^\mathrm{eff}_{,\varphi}(\varphi),\;\text{with}\; \beta(\varphi) \equiv M_\mathrm{Pl}\frac{\mathrm{d}\log A}{\mathrm{d}\varphi},
\label{eq:eom_phi}
\end{equation}
where $\widetilde{T}$ is the trace of the matter stress-energy tensor in the Jordan frame and we have introduced the effective potential $V^\mathrm{eff}(\varphi)$ for convenience. We note that the Einstein-frame stress-energy tensor is not conserved, however, it is often useful to introduce a matter density variable which is conserved in Einstein frame. Particularly, $\rho \equiv A^3\widetilde{\rho}$ can be shown to be conserved with respect to the Einstein-frame metric. As a result, the effective potential takes the form
\begin{align}
    V^\mathrm{eff} = V(\varphi) + \rho A(\varphi).
\end{align}

Due to the non-trivial coupling, test particles feel an additional ``fifth'' force per unit mass given by
\begin{equation}\label{eq:force}
	\vec{F}_{\varphi} = -\vec{\nabla} \log A(\varphi). 
\end{equation}

\begin{figure*}[!t]
    \centering
    \includegraphics[width=0.45\textwidth]{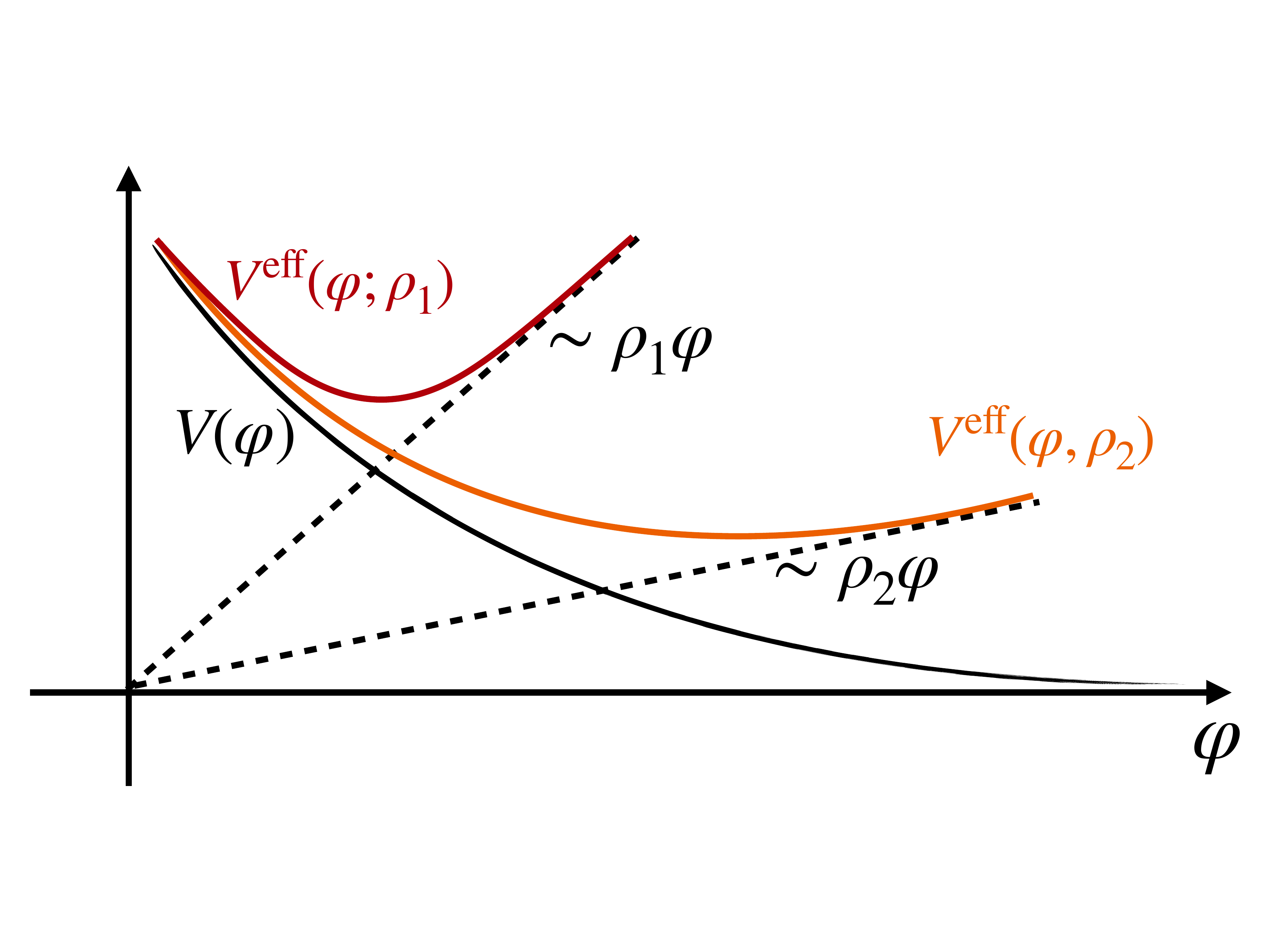}
    \includegraphics[width=0.45\textwidth]{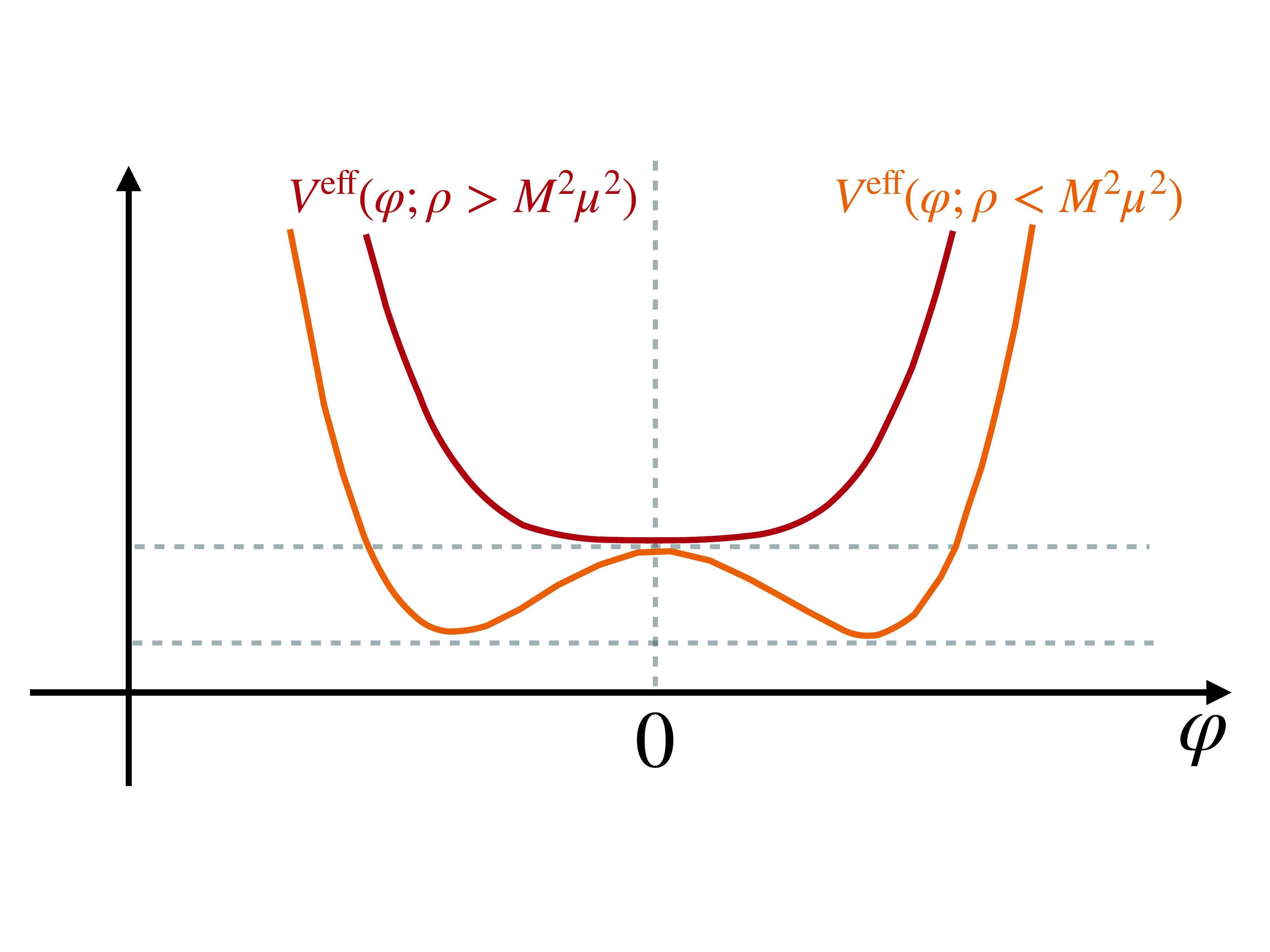}
    \caption{Schematic illustration of chameleon (left panel) and symmetron (right panel) mechanisms. In high-density regions chameleons become more massive, while the symmetrons operate due to driving the field value to zero in such environments.}
    \label{fig:cham_symm}
\end{figure*}

The qualitative realization of screening varies depending on the form of the coupling function $A(\varphi)$. Particularly, the sum of runaway potentials and coupling functions in chameleon mechanism leads to environmentally-dependent effective mass for the scalar field. The larger mass in higher-density regions leads to shortened range of interactions. In contrast, the cymmetron mechanism and its modifications rely on a suppression of the field expectation value in denser regions. The two mechanisms are schematically illustrated in Figure~\ref{fig:cham_symm}.

Prototypical examples of chameleons and symmetrons are given by 

\begin{enumerate}
    \item $V(\varphi) = \Lambda^4\left(1 + \frac{\Lambda^n}{\varphi^n}\right),\quad A(\varphi) = e^{\beta \varphi / M_\mathrm{Pl}}$, with const $\Lambda, \beta$ and $n$ (chameleons),
    \item $V(\varphi) = -\frac{1}{2} \mu^2 \varphi^2 + \frac{1}{4} \lambda \varphi^4,\quad A(\varphi) = 1+ \frac{1}{2}\frac{\varphi^2}{M^2}$, with const $\mu, \lambda$ and $M$   (symmetrons).
\end{enumerate}

The models described above and their variations have highly non-linear equations of motion. While we will focus on methods for solving such equations in non-trivial scenarios, it is useful to recall that the most essential properties of solutions can already be demonstrated by a spherically symmetric, static top-hat dust configuration with radius $R$. The density profile is taken to be $\rho(r) = \rho\Theta(R - r)$, with $\Theta$ denoting a top-hat function. For concreteness, we will focus on symmetrons below. The exposition mainly follows Ref.~\cite{Elder:2019yyp}. In this one-dimensional configuration, the Symmetron equation of motion is
\begin{align}
    \varphi^{\prime\prime} + \frac{2}{r}\varphi^\prime = V_{,\varphi} + A_{,\varphi}\rho, 
\end{align}
where primes denote derivative with respect to the radial coordinate $r$. We are interested in finding the field profile satisfying the following boundary conditions
\begin{align}
    \varphi^\prime(r = 0) = 0,\;\varphi(r\rightarrow\infty) = \varphi_0,
\end{align}
where the vacuum expectation value in a $\rho = 0$ environment is given by $\varphi_0 = \mu / \sqrt{\lambda}$. 

First, consider a sphere with density higher than the spontaneous symmetry breaking density $\rho_\mathrm{SSB} \equiv \mu^2 M^2$. The field profile can be obtained analytically by approximating the system as a free scalar field with a potential centered at $\varphi = 0$. This leads to
\begin{align}
\varphi_\mathrm{in}(r) = \mathcal{A} \frac{R}{r} \sinh \left( \mu r \zeta \right)~,
\end{align}
where $\zeta \equiv \sqrt{ \frac{\rho}{\mu^2 M^2} - 1}$, and $\mathcal{A}$ is an undetermined constant.

Outside the object, $r>R$, the field behaves as a free field with a quadratic potential centered at $\varphi = \varphi_0$, and the field profile is given by
\begin{align}
\varphi_\mathrm{out}(r) = v + \mathcal{B} \frac{R}{r} e^{- \sqrt 2 \mu r}~,
\end{align}
with $\mathcal{B}$ being an undetermined constant. Matching the two solutions and their first derivatives at $r = R$ fixes the constants $\mathcal{A}$ and $\mathcal{B}$ as follows
\begin{align}
\mathcal{B} &= - v e^{\sqrt 2 \mu R} \frac{\mu R\zeta \coth \left(\mu R\zeta  \right) - 1}{\mu R\zeta \coth \left(\mu R\zeta  \right) + \sqrt 2 \mu R}~,\\
\mathcal{A} &= \frac{1}{\sinh{\mu R \zeta}}\left(v + \mathcal{B}e^{-\sqrt{2}\mu R}\right).
\end{align}

A limit of interest is the very strongly perturbing source with $\rho \gg \mu^2 M^2$. In this limit for the coefficient $\mathcal{B}$ we have
\begin{align}
B = - v \left( 1 - \frac{\tanh \sqrt \alpha}{ \sqrt \alpha} \right)~,
\end{align}
where the dimensionless parameter $\alpha \equiv \rho R^2/M^2$ controls the characteristic outer shell thickness within which the field profile is varied significantly, $\Delta R/R = 1/\alpha$.

It is instructive to consider large and small $\alpha$ limits. The coefficient $\mathcal{B}$ takes the form
\begin{align}
    B(\alpha \ll 1) = - v \frac{\alpha}{3},\quad B(\alpha \gg 1) = - v \left( 1 - \frac{1}{\sqrt \alpha} \right)~.
\end{align}

The force between the sphere of mass $m_\mathrm{sphere}$ and a test particle with mass $m_\mathrm{test}$ is then given by
\begin{align}
F &\approx - \frac{v^2}{4 \pi M^4} \frac{m_\mathrm{test} m_\mathrm{sphere}}{r^2},\; \alpha \ll 1,
\label{eq:force-unscreened}\\
F &\approx - \frac{v^2}{4 \pi M^4} \frac{m_\mathrm{test} (\lambda_\mathrm{sphere} m_\mathrm{sphere})}{r^2},\;\alpha \gg 1,
\label{eq:force-screened}
\end{align}
where in the last result we have introduced the screening factor $\lambda_\mathrm{sphere} \equiv 3/\alpha$. The essential assumption behind these results is that the test object does not alter the field profile of the sphere significantly. The important take-away is that for large and dense objects, $\alpha \gg 1$, the fifth force on test particles is substantially suppressed due to the small screening factor $\lambda_\mathrm{sphere} \ll 1$.

It is also instructive to compare the scalar force with the Newtonian gravitational interaction:
\begin{align}
\frac{F_\phi}{F_\text{N}} = 2 \left( \frac{v M_\mathrm{Pl}}{M^2} \right)^2,\; \alpha \ll 1,\;\;\;
\frac{F_\phi}{F_\text{N}} = 2\lambda_\mathrm{sphere} \left( \frac{v M_\mathrm{Pl}}{M^2} \right)^2,\;\alpha \gg 1.
\label{eq:force-screened-newt}
\end{align}
The screening factor representation will be useful when discussing experimental configurations involving atoms and other test objects.

\subsection{Galileons}
\label{subsec:galileons}

Thus far, we have focused exclusively on \textit{thin shell} screened modified gravity theories, where either the mass of the scalar field or the coupling of a test particle to the field is dependent on the environment, such that the fifth force is suppressed in regions with higher density or gravitational potential. We now consider the second main class of modified gravity theories: \textit{kinetically} screened theories. This includes models exploiting, for example, the Vainshtein \citep{Vainshtein_1972} or K-mouflage \citep{Babichev_2009} mechanisms. Extensively studied models in this class are the Dvali-Gabadadze-Porrati (DGP) models \cite{DGP_2000}. In these cases, screening relies on higher order operators in the action, which may be beyond the validity of the relevant effective field theory. We do not dwell on this potential inconsistency in this section, and instead discuss the phenomenology which such models exhibit since future, consistent models may display similar behaviour, and thus the results described here can be transferred to such models.

The canonical example of a Vainshtein-screened theory is the galileon \citep{Nicolis_2009}: scalar-tensor theories where the scalar, $\varphi$, obeys the galileon symmetry $\varphi \to \varphi + a + b_\mu x^\mu$, where $a$ and $b_\mu$ are constants. The cubic galileon has the action
\begin{equation}
    \label{eq:galileon action}
    S = \int \dd^4 x \sqrt{-g} \left[ \frac{R}{16 \pi G} 
    - \frac{1}{2} c_2 \nabla_\mu \varphi \nabla^\mu \varphi 
    - \frac{c_3}{\mathcal{M}^3} \square \varphi \nabla_\mu \varphi \nabla^\mu \varphi - \mathcal{L}_{\rm m} \right],
\end{equation}
for constants $c_2$ and $c_3$ and where $\mathcal{M}^3 = M_{\rm Pl}H_0^2$. The resulting equations of motion are second-order, and thus this is a special case of Horndeski theory \citep{Horndeski_1974}. In the mostly minus signature, the case $c_2 < 0$ -- dubbed the self-accelerating branch -- has a non-canonical kinetic term and can thus self-accelerate, removing the necessity for a cosmological constant \citep{Babichev_2013}, whereas this cannot be achieved by the normal branch ($c_2 > 0$). Despite the negative $c_2$, the non-linear kinetic terms in  Eq.~(\ref{eq:galileon action}) admit ghost-free self-accelerating de Sitter solutions \citep{Nicolis_2009}, and conditions on model parameters exist which prevent such instabilities arising in scalar and tensor perturbations \citep{deFelice_2010}. Despite providing theoretically viable self-accelerating models, observations of the Integrated Sachs-Wolfe effect \citep{Renk_2017} make such a scenario unlikely for the cubic model, and measurements of the speed of gravitational waves with GW170817 \cite{GW170817,Ezquiaga_2017} significantly constrain quartic and quintic galileons. Even if galileons cannot drive the Universe's acceleration, the fifth-force due to galileons has interesting phenomenology, which we describe below.

In astrophysical and laboratory tests considered in this review, one can neglect terms suppressed by the Newtonian potential and its spatial derivatives, and work in the quasi-static approximation to obtain the equation of motion \citep{Barreria_2013}
\begin{equation}
    \label{eq:cubic galileon EOM}
        \nabla^2 \varphi + \frac{1}{3 \beta_1 a^2 \mathcal{M}^3} \left[ \left( \nabla^2 \varphi \right)^2 - \nabla_i \nabla_j \varphi \nabla^i \nabla^j \varphi \right]
        = \frac{M_{\rm Pl}}{3 \beta_2} 8 \pi G_{\rm N} a^2 \delta\rho_\mathrm{m},
\end{equation}
where $i \in \{1,2,3\}$ and
\begin{equation}
    \label{eq:beta definitions}
    \beta_1 = \frac{1}{6c_3} \left[ - c_2 - \frac{4 c_3}{\mathcal{M}^3} \left( \ddot{\bar{\varphi}} + 2 H \dot{\bar{\varphi}}\right)  + \frac{16 \pi G c_3^2}{\mathcal{M}^6} \dot{\bar{\varphi}}^4 \right],
    \quad
    \beta_2 = 2 \frac{\mathcal{M}^3M_{\rm Pl}}{\dot{\bar{\varphi}}^2} \beta_1.
\end{equation}
Here we are considering perturbations, $\varphi$, about a background, $\bar{\varphi}$. It is common to rewrite Eq.~(\ref{eq:cubic galileon EOM}) in terms of the cross-over scale, $r_{\rm C} \equiv \left( \beta_1 a^2 \mathcal{M}^3 \right)^{-1}$ and coupling $\alpha \equiv \MP{} / (3 \beta_2)$, instead of $\beta_1$ and $\beta_2$. Note that these parameters are function of time, but for low-redshift astrophysical tests the temporal evolution is typically neglected.

The acceleration experienced by a test object due to the fifth force is
\begin{equation}
    \bm{a}_5 = - \alpha Q \nabla \varphi,
\end{equation}
for
\begin{equation}
    Q = \int T^{\mu}{}_\mu \dd^3 x,
\end{equation}
where $T^{\mu\nu}$ is the stress-energy tensor of the test object \citep{Hui_2009}. For a non-relativistic object of mass $m$, $Q\approx m$ and thus the coupling to the field is the same as for gravity. In general, however, $Q<m$ since $T^{\mu}{}_\mu$ excludes the gravitational binding energy. The extreme case is for black holes, where $Q=0$, which is a manifestation of the no hair theorem for galileons \citep{Hui_2012,Hui_2013}. This is often phrased as a violation of the Strong Equivalence Principle (SEP), since the coupling is a function of an object's composition. However, we note that the presence of an additional field means that this interpretation can be misleading; one would not say that the SEP is violated when an electrically charged and a neutral object take different trajectories in the presence of an electromagnetic field.

Despite the presence of non-standard kinetic terms in the equation of motion, for spherically symmetric systems, one can express the left hand side of Eq.~(\ref{eq:cubic galileon EOM}) as a total derivative, and thus derive the radial variation of $\varphi$ to be \citep{Schmidt_2010}
\begin{equation}
    \label{eq:galileon_spherical}
    \frac{\dd \varphi}{\dd r} = - \frac{GM\alpha}{r^2} \left( \left( \frac{r}{r_{\rm v}} \right)^3 - \sqrt{\left( \frac{r}{r_{\rm v}} \right)^6 + 4 \left( \frac{r}{r_{\rm v}} \right)^3} \right),
\end{equation}
where $M(r)$ is the  mass (perturbation) enclosed at radius $r$, and the Vainshtein radius is defined to be 
\begin{equation}
    \label{eq:Vainshtein radius}
    r_{\rm v} = \frac{4}{3}GM(r)\alpha r_{\rm C}^2,
\end{equation}
which is itself a function of $r$. 
In this case, we see that well within the Vainshtein radius ($r \ll r_{\rm v}$), the fifth force is suppressed by a factor $(r/r_{\rm v})^{3/2}$ relative to gravity, whereas far from a source the effect of the galileon is equivalent to increasing the gravitational constant by a factor $\Delta G / G_{\rm N} = 2 \alpha^2$. Similar behaviour is observed for other variants of the galileon, but with different powers of $r/r_{\rm v}$. A solar mass object has a Vainshtein radius of $r_{\rm v} \sim \mathcal{O}\left(100 {\rm \, pc}\right)$ \citep{Sakstein_2017}, allowing this model to evade solar system tests, as we are well within the regime where the fifth-force is suppressed.

Given the large Vainshtein radius of the Sun, one may suppose that the galileon field is screened in all regions of the Universe and thus any attempt to constrain such a model is futile, as there would not be any observational consequences with this level of suppression. There are two main reasons why this is incorrect.

First, we note that Eq.~(\ref{eq:Vainshtein radius}) depends on the enclosed mass at a given radius, and is thus similar to Gauss' law. Therefore, within an extended mass distribution, the Vainshtein radius is smaller than for the corresponding point mass, allowing the galileon field to remain partially unscreened. This can lead to observational effects at distances $d \gtrsim 0.1 R_{200}$, where $R_{200}$ is the typical size of a dark matter halo \citep{Schmidt_2010}. Outside the mass distribution, the two fields are clearly equal. 

Second, even if the field sourced by a given object is screened, it is important to remember this object lives in an external environment. The galileon symmetry means that, if we have a solution for the galileon field $\varphi$, we can obtain another solution with $\partial_\mu \varphi \to \partial_\mu \varphi + b_\mu$, for constant $b_\mu$. This allows the galileon field sourced by large scale structure to remain unscreened, since this is approximately linear across the region of interest. This has been confirmed by cosmological simulations \cite{Falck_2014}, which demonstrate that $\varphi$ has linear dynamics on $\gtrsim 10 {\rm \, Mpc}$ if $r_{\rm C} \simeq 6 {\rm \, Gpc}$ \citep{Cardoso_2008,Schmidt_2009,Khoury_2009,Chan_2009}.

\section{Numerical methods and technical considerations}\label{sec:numerical}

\subsection{Relaxation method}\label{subsec:relaxation}

For many problems of interest the density configurations can be assumed to be static, and the scalar field equations of motion in e.g. Eq.~(\ref{eq:eom_phi}) or (\ref{eq:cubic galileon EOM}) can be considered as non-linear elliptical boundary value problems. Such an approximation is very well-motivated in the context of laboratory experiments as the experimental density configurations are kept static. The quasi-static assumption has also been widely considered in the astrophysical and cosmological contexts (see e.g. \cite{Davis:2011pj,Clampitt:2011mx,Brax:2012nk}) and has been verified in $N$-body simulations \cite{Llinares:2013jua,Noller:2013wca}. Qualitatively, the validity of this assumption can be understood by noticing that the typical timescale of the dynamics scales as $\sim \lambda_0/c$, with $\lambda_0$ denoting the vacuum Compton wavelength of the field, whereas the typical time-variation of matter in the cosmological context is given by $c/H_0$, which is several orders of magnitude larger than $\lambda_0/c$ in many relevant applications.

In order to introduce the relaxation algorithms which have been used in a number of analyses of screened modified gravity let us for concreteness focus on a 1-dimensional symmetron theory. The exposition below partly follows Ref.~\cite{Vardanyan:2019gjt}. In numerical evaluation, it is useful to introduce dimensionless variables. We particularly introduce the normalized field variable $\chi \equiv \varphi/\varphi_0$ and consider the following equation:
\begin{align}\label{eq:symmetron}
	\frac{d^2\chi}{dr^2}+ \frac{2}{r} \frac{d\chi}{dr} = \frac{1}{2\lambda_0^2} \left[ \left( \frac{\rho(r, t)}{\rho_{ssb}} -1 \right)\chi + \chi^3 \right]\,,
\end{align}
where the vacuum Compton wavelength for symmetrons is given by
\begin{align}\label{eq:lambda_0}
	\lambda_0 = \frac{1}{\sqrt{2}\mu}.
\end{align}
We impose the standard boundary conditions $\partial \chi/\partial r = 0$ at  both $r = 0$ and $r \rightarrow \infty$.

We consider a discretization on a regular grid of size $h$ and employ a second-order discretization scheme. Higher order discretization schemes have been implemented and tested in e.g. \cite{Contigiani:2018hbn}, without encountering significant performance differences in most of the problems of interest. The discretized equation can be written as
\begin{align}
    \mathcal{L} [\chi_{i+1}, \chi_{i-1};\chi_{i}] = 0,
\end{align}
with
\begin{align}
    \mathcal{L} [\chi_{i+1}, \chi_{i-1};\chi_{i}] \equiv \mathcal{D}_\mathrm{K}[\chi_{i+1}, \chi_{i-1};\chi_{i}] -\mathcal{D}_\mathrm{P}[\chi_{i}, \rho_i].
\end{align}
Here we have introduced the following dicretizations of the Laplace operator and the potential term
\begin{align}
    \mathcal{D}_\mathrm{K} \equiv \frac{ \chi_{i+1}+ \chi_{i-1} -2\chi_{i} }{h^2}
    +\frac{2}{ r_i }\frac{\chi_{i+1}-\chi_{i-1}}{2h},\;\mathcal{D}_\mathrm{P} = \frac{1}{\lambda_0^2}\left(\left(\frac{\rho_i}{\rho_{\text{ssb}}}-1\right)\chi_i+\chi_i^3 \right).
\end{align}

In order to obtain a solution to the non-linear discretized problem an iterative method should be employed. The particular structure of the operator $\mathcal{L}$ suggests the use of the  Newton-Gauss-Seidel relaxation method. This has been a standard method used for obtaining the scalar field solutions in $N$-body simulations with additional fifth forces (see later). 

The starting point of the iterative relaxation is the arbitrarily chosen initial ``guess'' for $\chi$. At each iteration $n$ we use the current estimate $\chi_n(i)$ in order to obtain an improved estimate $ \chi^{\mathrm{new}}(i)$ according to
\begin{align}\label{eq:discretization_2}
    \chi^{\text{new}}(i)=\chi_n(i) - 
    \left.
        \frac{\mathcal{L}(\chi(i))}{\partial\mathcal{L}(\chi(i))/\partial\chi(i)}
    \right\vert_{\chi(i) = \chi_n(i)}.
\end{align}

In most of the practical problems, this iteration itself does not converge. Instead, one should typically use a fraction of $ \chi^{\text{new}}(i)$ as a current estimate:
\begin{align}\label{eq:relax_upgrade}
	\chi_{n+1}(i) = \omega \chi^{\text{new}} + (1-\omega)\chi_{n},
\end{align} 
where $0<\omega\leqslant 1$ is a constant hyper-parameter which should be tuned for each specific model separately. The optimal value of this parameter is problem-specific and is often chosen by empirically leveraging between being able to converge and the rate of change for $\chi_n$ between consecutive iterations.     

Relaxation iterations are repeated multiple times until convergence is reached according to a pre-determined criterion. Two examples of convergence criteria are often employed in the literature. These are the global residual defined as
\begin{align}
    \mathcal{R}_1 \equiv \sqrt{\sum_{i}\mathcal{L}[\chi(i+1), \chi(i-1);\chi(i)]^2},
\end{align}
and the mesh-averaged change of the field profile between consecutive iterations 
\begin{align}
\mathcal{R}_2 \equiv \sqrt{\sum_{i}(\chi^{\text{new}}(i)-\chi^{\text{old}}(i))^2}.
\end{align}
One could, for example, choose $\mathcal{R}_2$ to reach a multiple of the numerical precision for the given implementation. 

\begin{figure*}
    \centering
    \includegraphics[width=1\textwidth]{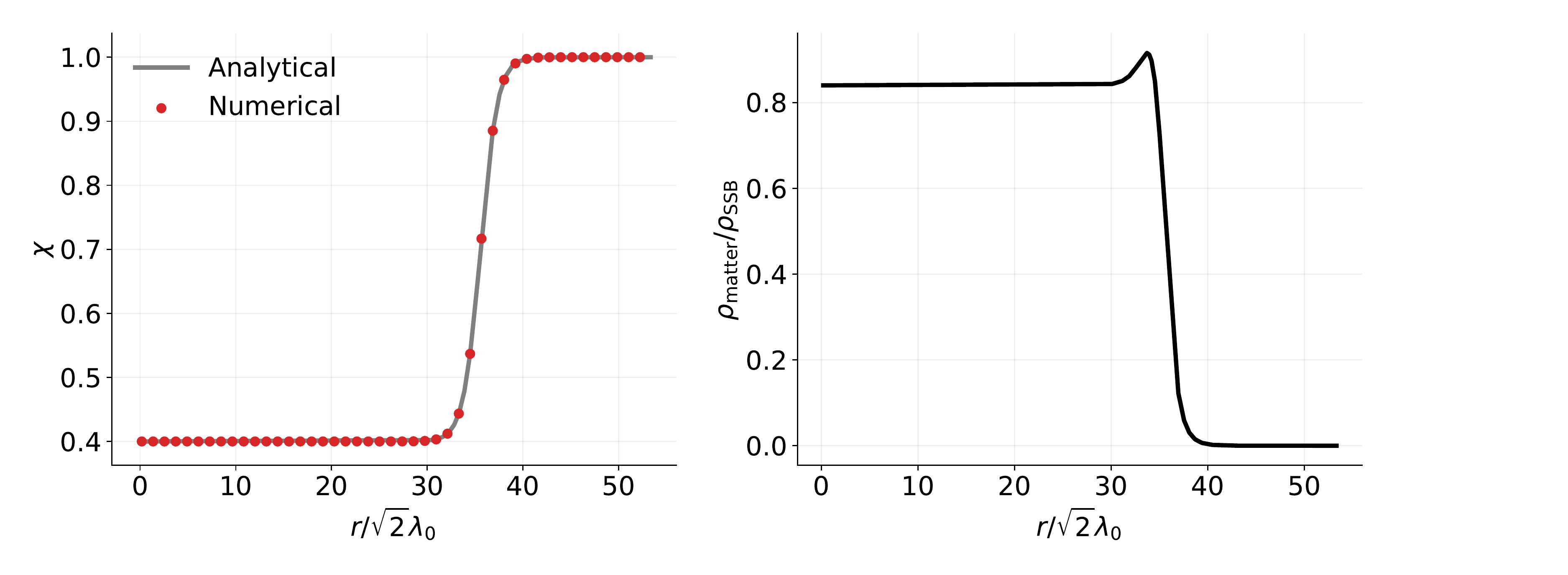}
    \caption{Validation of the numerical solver. The gray line in the left panel represents an arbitrary non-zero function. The right panel depicts the reconstructed density profile obtained by inserting the function from the left panel into Eq.~(\ref{eq:symmetron}). The red dots in the left panel show the numerical result for a problem with a density profile from the right panel. Figure adapted from \cite{Vardanyan:2019gjt}.}
    \label{fig:tanh_profile}
\end{figure*}

In order to validate the performance of the solver, and asses the choice of the convergence criterion one can use analytically solvable configurations. Let us consider two such examples. In order to construct the first analytical solution note that we can always plug a non-zero field profile in Eq.~(\ref{eq:symmetron}) and reconstruct a unique density profile that serves as a source for the mentioned field profile. As an example, we can choose $\chi \sim \tanh(r)$ and solve for the density configuration using Eq.~(\ref{eq:symmetron}). The gray line in the left panel of Fig.~\ref{fig:tanh_profile} shows the chosen field profile. The right panel of the same figure demonstrates the corresponding reconstruction of the density profile. The red dots in the left panel are the result of the numerical integration using the density profile from the right panel as an input source in our solver. As one can see, the numerical integration successfully matches the expected analytical field configuration.  
 
As our second example, we consider the configuration of two parallel plates with infinitely high density, separated by a vacuum gap \cite{Upadhye:2012rc}. Let the coordinate perpendicular to the plates be $z$ with the gap width being $\Delta z$ and the plate surfaces being placed at $-\Delta z/2$ and $+\Delta z/2$. The field equation of motion in this setup is given by 
\begin{equation}\label{eq:symmetron_casimir}
	\frac{d^2\chi}{d\hat{z}^2} = \left[ \left( \tilde{\rho}(\hat{z}) -1 \right)\chi + \chi^3 \right]\,,
\end{equation}
where we have additionally defined $\hat{z} \equiv z/\sqrt{2} \lambda_0$ and $\tilde{\rho}(\hat{z}) \equiv \rho(\hat{z})/M^2\mu^2$ is assumed to be infinitely large inside the plates and zero in the gap. We can integrate this equation once in a $z$-interval where the density is constant. Choosing two subsequent intervals being $(0, \Delta \hat{z}/2)$ and $(\Delta \hat{z}/2, \infty)$ we can show that the value of the field on the plate surface $\chi_\mathrm{s}$ is zero up to negligible corrections of the order of the ratio of the vacuum matter density to the plate density. Then, choosing an interval $(0, \hat{z})$ with $\hat{z} < \Delta \hat{z}/2$ we obtain
\begin{align}\label{eq:elliptic}
    \hat{z} = \frac{1}{\sqrt{1 - \frac{\chi^2_\mathrm{g}}{2}}}\left[ \mathcal{F}\left(\pi/2, \sqrt{\frac{\chi_\mathrm{g}^2}{2 - \chi_\mathrm{g}^2}}\right) - \mathcal{F}\left(\sin^{-1}\frac{\chi}{\chi_\mathrm{g}}, \sqrt{\frac{\chi_\mathrm{g}^2}{2 - \chi_\mathrm{g}^2}}\right)\right],
\end{align}  
where $\mathcal{F}$ is the elliptic integral of the first kind, and $\chi_\mathrm{g}$ is the field value in the middle of the gap. 

\begin{figure}
    \centering
    \includegraphics[width=0.8\columnwidth]{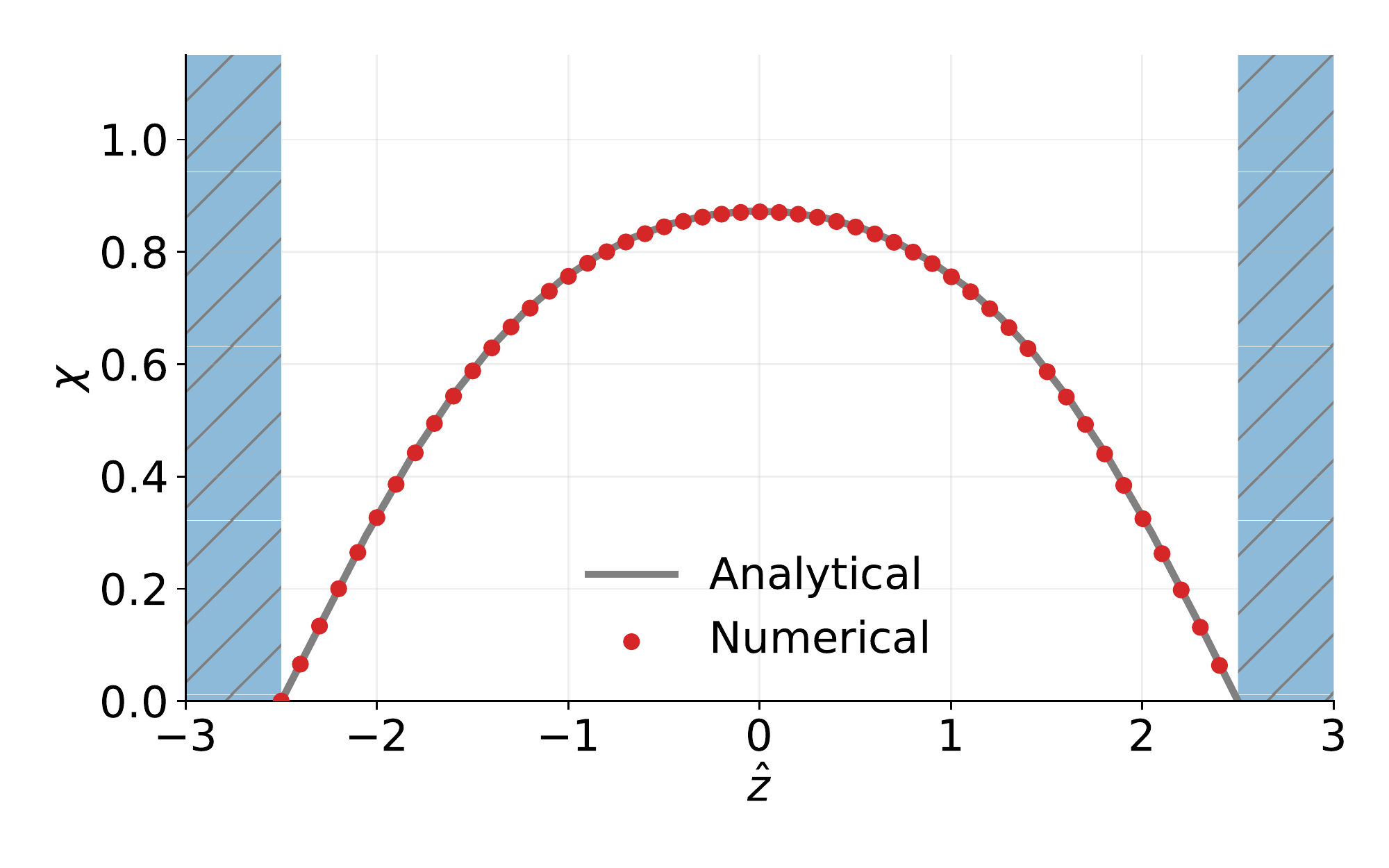}
    \caption{Analytical (gray line) and numerical (red) solutions of the symmetron field profile inside a gap between two parallel plates. Figure adapted from \cite{Vardanyan:2019gjt}.}
    \label{fig:casimir}
\end{figure}

Fixing $\hat{z}$ to $\Delta \hat{z}/2$ and setting $\chi = 0$ as one of our boundary conditions we can numerically solve for $\chi_\mathrm{g}$. Having the latter we will then have $\chi$ as a function of $\hat{z}$ in the gap expressed in terms of the Jacobi elliptic function. The numerical relaxation solution of Eq.~(\ref{eq:symmetron_casimir}) in the gap subject to boundary conditions $\chi(-\Delta \hat{z}/2) = 0 = \chi(+\Delta \hat{z}/2)$ is depicted in Fig.~\ref{fig:casimir} (red dots). This is in sub-percent-level agreement with the exact analytical solution obtained from Eq.~(\ref{eq:elliptic}).

Variations of the relaxation method have been implemented in N-Body simulation codes. The state of the art codes are \texttt{ECOSMOG} \cite{Li_2012}, \texttt{ISIS} \cite{Llinares:2013jza} and \texttt{MG-GADGET} \cite{Puchwein:2013lza}. See also Ref.~\cite{Winther:2015wla} for a code-comparison study. In highly demanding setups, such as N-body simulations, the scalar solvers can be further accelerated due to multigrid techniques. This relies on the observation that relaxation iterations efficiently reduce small-wavelength errors. Moving the problem to a coarser grid helps in efficiently reducing the longer-wavelength error modes in the original problem. Such multigrid techniques are implemented in the state-of-the-art Modified Gravity N-body codes. These codes additionally support adaptive mesh refinement.

\subsection{Finite element codes}\label{subsec:FEM}

The finite element method (FEM) offers a very rich toolkit for solving partial differential equations. FEM-based methods are often employed in engineering tasks, while their use is astrophysics and cosmology has been limited. In this section we offer an introduction to the main FEM ideas and their applications to screening. 

Let us consider the following general problem on a domain $\Omega$ with boundary $\partial \Omega$
\begin{align}
    \nabla^2 u(x) = -f(x, u(x)),\;x\in \Omega\\
    u(x) = u_\mathrm{B},\;x\in \partial\Omega,
\end{align}
where $f(x,u)$ is a source which can generally depend non-linearly on both $x$ and $u$, and $u_\mathrm{B}$ is the value of the solution at the boundary.

The first step in FEM is to reformulate the boundary-value problem as a variational problem. In order to achieve this, we multiply the equation by a well-defined ``test'' function $v(x)$:
\begin{align}
    \int_\Omega v(x)\nabla^2 u(x) = -\int_\Omega f v, 
\end{align}
where $\int_\Omega$ is an integral over the domain $\Omega$. The test function $v(x)$ should satisfy certain properties, and we will particularly assume it vanishes on the entire boundary $\partial \Omega$. Next, by employing the Green's theorem, the derivative order can be reduced, resulting in the following ``weak-form'' equation:
\begin{align}\label{eq:weak_form}
    \int_\Omega \nabla u(x) \nabla v(x) = \int_\Omega f v, 
\end{align}
which should be satisfied for all $v(x)$ in the relevant function space.

In practice, we seek a solution to a discretized problem, and instead of the functions $u(x)$ and $v(x)$ we consider the functions $u_h(x)$ and $v_h(x)$ belonging to a subspace of the original function space. We consider a basis expansion of both $u_h(x)$ and $v_h(x)$ as follows
\begin{align}\label{eq:FEM_decomposition}
    u_h(x) = \sum_i u_ie_i(x),\;v_h(x) = \sum_i v_i e_i(x),
\end{align}
where $u_i$ and $v_i$ are the values at vertices $\{P_i\}$ of the discretely triangulated grid of interest, and the basis functions can be chosen to satisfy $e_i(P_j) = \delta_{ij}$.

Inserting basis expansions into Eq.~(\ref{eq:weak_form}) we arrive at
\begin{align}\label{eq:weak_form_basis}
    \sum_i \left(\int_\Omega \nabla e_i(x) \nabla v_j(x)\right)u_i = \int_\Omega f\left(x, \sum_iu_ie_i(x)\right) v_j. 
\end{align}

This system can be conveniently rewritten in a matrix form as

\begin{align}\label{eq:FEM_matrix}
    \mathbf{M} \mathbf{u} = \mathbf{b},
\end{align}
where
\begin{align}
    \mathbf{M}_{ij} = \int_\Omega \nabla e_i(x) \nabla v_j(x),\;\mathbf{b}_{j} = \int_\Omega f v_j.
\end{align}

Since $f$ is in general a non-linear function of unknown $u$ we need to rely on iterative non-linear solvers. Two general approaches are described below.

\textbf{Picard iteration:} The basic idea behind Picard iterations is to linearize the right-hand-side of Eq.~(\ref{eq:FEM_matrix}) around an initial guess $u_0$, and solve the resulting linear system of equations. Given an intermediate estimate $u_k$, the new estimate for $u$ is computed as follows. First, the non-linear function $f(x, u_\mathrm{new})$ is replaced by its linear expansion around $u_k$:
\begin{align}\label{eq:FEM_linear_u}
    f(x, u_\mathrm{new}) \rightarrow f(x, u_{k}) - u_k\frac{\partial}{\partial u_k}f(x, u_{k}) + u_\mathrm{new}\frac{\partial}{\partial u_k}f(x, u_{k}).
\end{align}
Substituting this in Eq.~(\ref{eq:FEM_matrix}) and using Eq.~(\ref{eq:FEM_decomposition}) again we obtain
\begin{align}\label{eq:FEM_linear_picard}
    \left[\mathbf{M} + \mathbf{B}_k\right]\mathbf{u}_\mathrm{new} = \mathbf{C}_k,
\end{align}
where the matrix $\textbf{B}_k$ and vector $\mathbf{C}_k$ are defined as
\begin{align}
   \textbf{B}_k &=  -\int_\Omega \frac{\partial f(x, u_{k})}{\partial u_k} e_i(x) v_j(x),\\
   \textbf{C}_k &=  \int_\Omega\left(f(x, u_{k}) - u_k\frac{\partial}{\partial u_k}f(x, u_{k})\right) v_i(x).
\end{align}
Having solved for $\mathbf{u}_\mathrm{new}$, the new estimate for the field profile is obtained as in Eq.~(\ref{eq:relax_upgrade}).

As a concrete example let us derive the corresponding linearized system for the power-law chameleon model described earlier. The equation of interest in dimensionless variables reads as
\begin{align}
    \alpha \nabla^2 \chi = -\chi^{-(n+1)} + \tilde{\rho},
\end{align}
where $\alpha$ here is a dimensionless quantity determined in terms of the chameleon parameters, $\chi$ is a field variable normalized by the value minimizing the effective potential in the ambient space, and $\tilde{\rho}$ is the matter density normalized by the density of the ambient space; see Ref.~\cite{Briddon_2021} for detailed definitions. The weak-form equation of motion is of the form
\begin{align}\label{eq:chameleon_weak}
    \alpha\int_\Omega \nabla \chi(x) \nabla v_j(x)u_i = \int_\Omega \chi^{-(n+1)} v_j - \int_\Omega \tilde\rho v_j, 
\end{align}
Following Eq.~(\ref{eq:FEM_linear_u}), we expand the non-linear term around the $k$-th estimate of the solution, yielding
\begin{align}
\chi^{-(n+1)}_\mathrm{new} \rightarrow (n+2)\chi^{-(n+1)}_k - (n+1)\chi^{-(n+2)}_{k}\chi_\mathrm{new}.
\end{align}
In this case, the matrix $\textbf{B}_k$ and vector $\mathbf{C}_k$ in Eq.~(\ref{eq:FEM_linear_picard}) take the form
\begin{align}
   \textbf{B}_k &=  \frac{n+1}{\alpha}\int_\Omega \chi_k^{-(n+2)} e_i(x) v_j(x),\label{eq:FEM_B_cham}\\
   \textbf{C}_k &=  \frac{n+2}{\alpha}\int_\Omega \chi_k^{-(n+1)} v_i(x) - \frac{1}{\alpha}\int_\Omega \tilde{\rho}(x) v_i(x).\label{eq:FEM_C_cham}
\end{align}
Note that our notation here slightly differs from Ref.~\cite{Briddon_2021}. The resulting linear system can then be iterated over until convergence is achieved.

\textbf{Newton iteration:} In the Newton method, our goal is again to start from an initial guess and iteratively converge to the solution. Particularly, given the weak-form equation $\mathcal{L}\left(u, v_j\right) = 0$, we linearize the left-hand side around the current estimate for the solution, but now solve for the increment $\delta u \equiv u_\mathrm{new} - u_k$. Eq.~(\ref{eq:FEM_linear_picard}) now changes into
\begin{align}\label{eq:FEM_linear_newton}
    \left[\mathbf{M} + \mathbf{B}_k\right]\delta\mathbf{u} = -\mathbf{M}\mathbf{u}_k - \mathbf{D}_k,
\end{align}
where $\mathbf{M}$ and $\mathbf{B}$ are defined as in the Picard method, while $\mathbf{D}$ reads as
\begin{align}
   \textbf{D}_k =  \int_\Omega f(x, u_{k}) v_i(x).
\end{align}
For the chameleon example discussed earlier, the vector $\textbf{B}$ is given by Eq.~(\ref{eq:FEM_B_cham}), and the quantity $\textbf{D}$ coincides with $\textbf{C}$ in Eq.~(\ref{eq:FEM_C_cham}) but without the factor of $n + 2$ in the first term. Once the solution for $\delta\mathbf{u}$ is found, the field estimate for the next iteration is obtained as
\begin{align}
    u_{k+1} = u_k + \omega\delta\mathbf{u}.
\end{align}
Here $0<\omega\leqslant 1$ is a hyperparameter analogous to the one in Eq.~(\ref{eq:relax_upgrade}), and controls the rate of iterative convergence.

In both Eqs.~(\ref{eq:FEM_linear_picard}) and (\ref{eq:FEM_linear_newton}) the matrix $\mathbf{M}$ should only be computed once before the iterative evaluation starts. The matrix $\mathbf{M} +  \mathbf{B}_k$, however, can in general be very large, especially for higher-dimensional problems. Inverting the linear equations with standard exact methods, therefore, is not always practical. In practice, additional iterative methods can be employed at each level of the Picard or Newton iteration in order to approximate the solutions to the linear systems of equations. 

A class of powerful iterative methods for approximating the solutions of large linear systems of the form $\mathbf{A}\mathbf{x} = \mathbf{b}$ can be derived by treating the problem as a minimization problem for the function $f(\mathbf{x}) = \frac{1}{2}\mathbf{x}^\mathrm{T}\mathbf{A}\mathbf{x} - \mathbf{x}^\mathrm{T}\mathbf{b}$. Such a problem can be efficiently solved with a conjugate gradient algorithm, which relies on starting with an initial guess $\mathbf{x}_0$, and updating the solution by minimizing the function $f$ along mutually \textit{conjugate} directions $\{\mathbf{p}_i\}$ with $\mathbf{p}_i^\mathrm{T}\mathbf{A}\mathbf{p}_j = 0$ if $i \neq j$. The latter construction ensures that minimizing solution along a direction $\mathbf{p}_k$ also minimizes the function along all the directions constructed in the previous iterations.

Currently, there are three publicly available, well-documented, and user-friendly FEM implementations for screening mechanisms in \texttt{Python}. The computational backbones of these are derived from the open-source FEM frameworks \texttt{FEniCS} \cite{AlnaesEtal2015,LoggEtal2012} (implemented both in \texttt{C++} and \texttt{Python}) and \texttt{SfePy} \cite{Cimrman2018} (implemented in \texttt{Python}). The \texttt{$\varphi$-enics} code \citep{Braden_2021}, which is built on top of the \texttt{FEniCS} library, offers a 1-dimensional implementation of Vainshtein models. A more recent \texttt{FEniCS}-based code, \texttt{SELCIE}, has been presented in Ref.~\citep{Briddon_2021} and offers an implementation of chameleon theories in $2$ and $3$ spatial dimensions. The package has been employed for computing the chameleon field profiles in cosmic voids \cite{Tamosiunas_2022_voids} and around halos \cite{Tamosiunas_2022_halos}.

The outer boundary conditions are set by the asymptotic behaviour of the scalar field infinitely far away from matter sources. Since numerical implementations are limited to finite simulation boxes, the boundary conditions are typically imposed at distances \textit{sufficiently} large compared to the Compton wavelength of the field. This approach is practical for many laboratory setups and isolated astrophysical configurations but can be computationally prohibitive if simulation domain is required to be very large. Adaptive mesh refinement techniques are often employed to reduce the computational cost in such situations. Alternatively, Ref.~\cite{Levy:2022xni} presents a \texttt{SfePy}-based FEM package \texttt{femtoscope} which offers a rigorous support for accurately implementing boundary conditions at infinity. The approach exploits a coordinate mapping from the infinite computational domain to a finite one, hence making the problem with an unbounded domain computationally tractable \cite{Grosch1977}.

Earlier examples of using FEM solvers in the context of screening mechanisms can be found in Ref.~\cite{Burrage:2017shh} for chameleon theories and in Ref.~\cite{Elder:2019yyp} for symmetrons. Particularly, using \texttt{FEniCS}-based FEM solvers, \citet{Burrage:2017shh} simulated non-trivial density configurations in order to optimize for fifth force effects in laboratory experiments. \customElder employed the commercially available FEM implementations in \texttt{MATLAB} alongside the custom-developed relaxation-based solvers in order to model symmetron fifth forces in Casimir experiments.

\subsection{Forces on extended objects}

Once the field profiles are computed, they can be used to calculate the fifth forces. So far we have only addressed the calculation of the fifth force on test particles which do not alter the ambient scalar field profile significantly. In realistic experiments, however, the relevant forces are often between extended, massive, and dense object. For example, in the Casimir experiments to be described later, the relevant force is either between two parallel plates or a plate and a large sphere. Calculation of the scalar force on extended objects is nuanced and we dedicate this subsection on introducing a method for systematically calculating the force. The discussion follows Ref.~\cite{Hui_2009}; see also \cite{Elder:2019yyp} for examples in the context of non-trivial laboratory configuration.

The starting point is the separation of the energy-momentum tensor $T_{\mu \nu}$ into the matter contribution $T_{\mu \nu}^\mathrm{matter}$ and scalar field contribution $T_{\mu \nu}^\varphi$. The quantity of interest is the $3$-momentum in a spatial volume $\mathcal{V}$ which is barely enclosing the surface of the extended object of interest and is therefore primarily dominated by the matter energy-momentum tensor:
\begin{align}
P_i = \int_\mathcal{V} \mathrm{d}^3 x T_i^{~0}~.
\end{align}

The force on the object can be computed from the time derivative of the $3$-momentum. The latter can be written as: 
\begin{align}
\dot P_i = - \int_\mathcal{B} \mathrm{d}^2 \sigma_j T_i^{~j}~,
\label{eq:force-general}
\end{align}
where we have made use of the energy-momentum conservation, and the integral is over the surface $\mathcal{B}$ of the volume $\mathcal{V}$ with area element $\mathrm{d}^2 \sigma_i$. 

\begin{figure}[!th]
    \centering
    \includegraphics[width=0.45\textwidth]{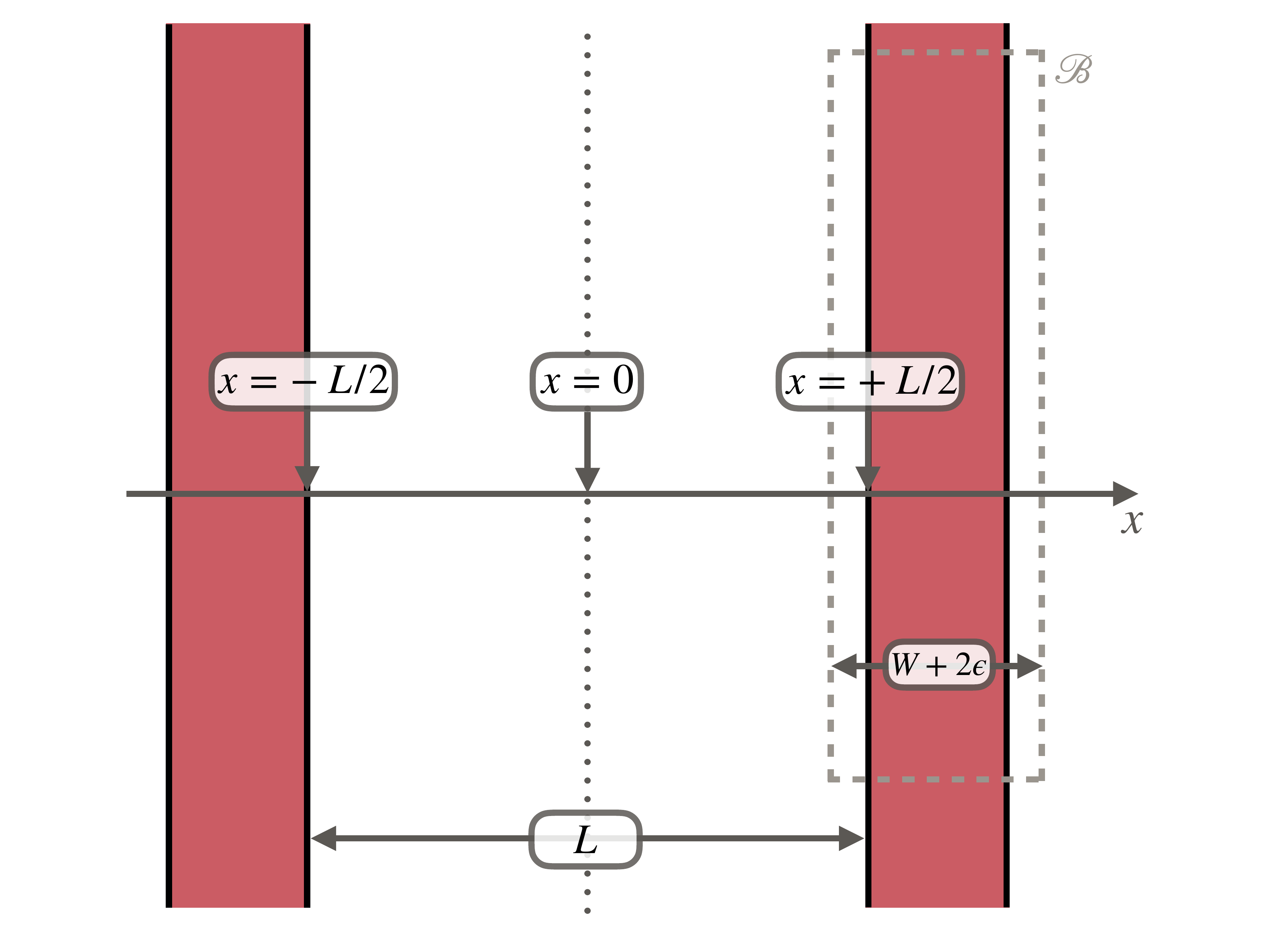}
    \includegraphics[width=0.45\textwidth]{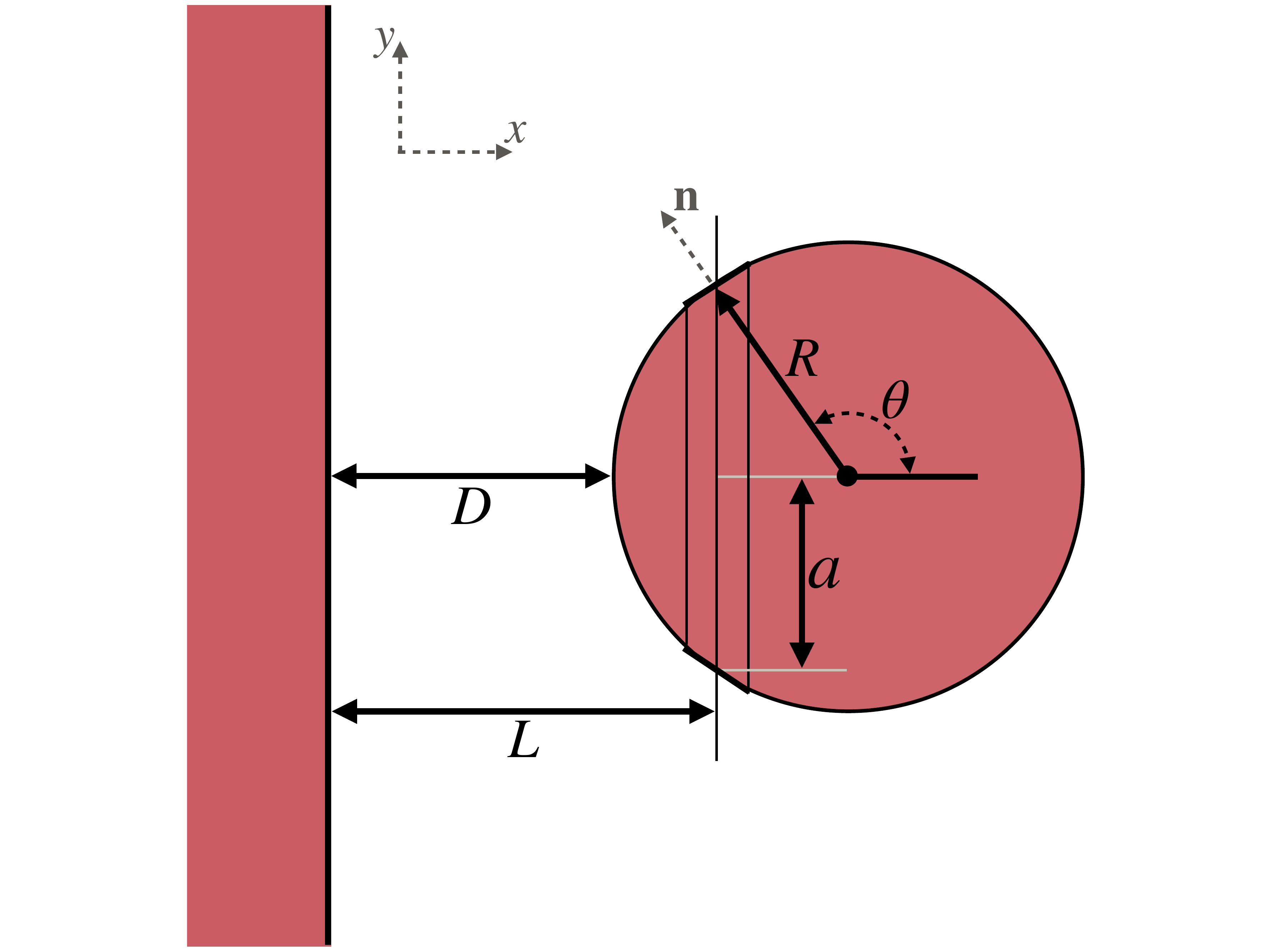}
    \caption{Density configurations relevant to Casimir force experiments. The left configuration demonstrates the choice of the integration box used for estimating the scalar force on the plate. The right panel demonstrates the proximity-force calculations discussed in Section~\ref{sec:lab_tests}. Figure adapted from \cite{Elder:2019yyp}.}
    \label{fig:plates}
\end{figure}

As a specific example, let us consider a prototypical Casimir experiment with two infinitely-dense, parallel plates, separated by a vacuum gap. The left panel of Fig.~\ref{fig:plates} depicts the configuration and the surface $\mathcal{B}$ of a rectangular box. Our primary interest in this example is the calculation of the Symmetron pressure on one of the plates, and we can make use of the analytical field profile derived using Eq.~(\ref{eq:elliptic}).

Given the specific choice of the enclosing surface, the matter of the plate does not contribute to the integral in Eq.~(\ref{eq:force-general}). For the Symmetron field, we have
\begin{align}\label{eq:EMT_tensor}
T^{\varphi}_{\mu \nu} = \partial_\mu \varphi \partial_\nu \varphi + \eta_{\mu \nu} \left( - \frac{1}{2} (\partial \varphi)^2 - V(\varphi) \right)~,
\end{align}

From the symmetry of the problem, the only component of the force is in $x$-direction, and it reads as
\begin{align}
\dot P_x = \frac{1}{2} \left[\varphi_\mathrm{in}^\prime\left(\frac{L}{2}\right)\right]^2 -  \frac{1}{2} \left[\varphi_\mathrm{out}^\prime\left(\frac{L}{2} + W\right)\right]^2~,
\end{align}
where ``in'' and ``out'' denote the solutions inside and outside of the gap. The latter can be obtained by using the general form for the solution inside and assuming an infinitely wide gap. Primes denote derivatives with respect to the $x$-coordinate.

The fifth force per area $A$ on the plate the follows: 
\begin{align}
\frac{F}{A} = \frac{\mu^4}{4 \lambda} \left( \varphi_0^2 (2 - \varphi_0^2) - 1 \right)~,
\label{eq:exact-pressure}
\end{align}
where a useful identity for Jacoby elliptic functions has been employed
\begin{align}
    \left( \frac{d }{dx} \cd(x, m) \right)^2 = (1 - \cd(x, m)^2)(1 - m^2 \cd(x, m)^2)~.
\end{align}

An alternative approach for calculating the pressure would have been to calculate the energy density stored in the scalar field configuration, and take the derivative with respect to the plate separation $L$. Direct computation shows agreement with the result presented in Eq.~(\ref{eq:exact-pressure}).

\section{Laboratory tests}
\label{sec:lab_tests}

Laboratory experiments that are sensitive to the effects of weak forces are capable of providing useful constraints on the screening mechanisms. We will schematically divide such experiments into two broad categories. The experiments in the first category constrain the screening mechanisms due to the fifth force effects on test objects, such as atoms or neutrons. Such experiments are referred to as ``indirect measurements'' below, and include atomic interferometry \cite{Burrage:2014oza,Burrage:2015lya,Hamilton:2015zga,Burrage:2016rkv,Jaffe:2016fsh,Sabulsky:2018jma}, atomic spectroscopy \cite{Brax:2010gp,Brax:2022olf} and bouncing neutron \cite{Cronenberg:2018qxf} experiments. On the other hand, the experiments that try to directly measure the fifth forces between extended objects are referred to as ``direct force measurements'', which include Casimir \cite{Decca:2005qz,Decca:2003zz,Almasi:2015zpa,Sedmik:2018kqt,Brax:2010xx} and torsion balance experiments \cite{Adelberger:2002ic}.

\begin{figure}[!t]
\centering
\includegraphics[width = \textwidth]{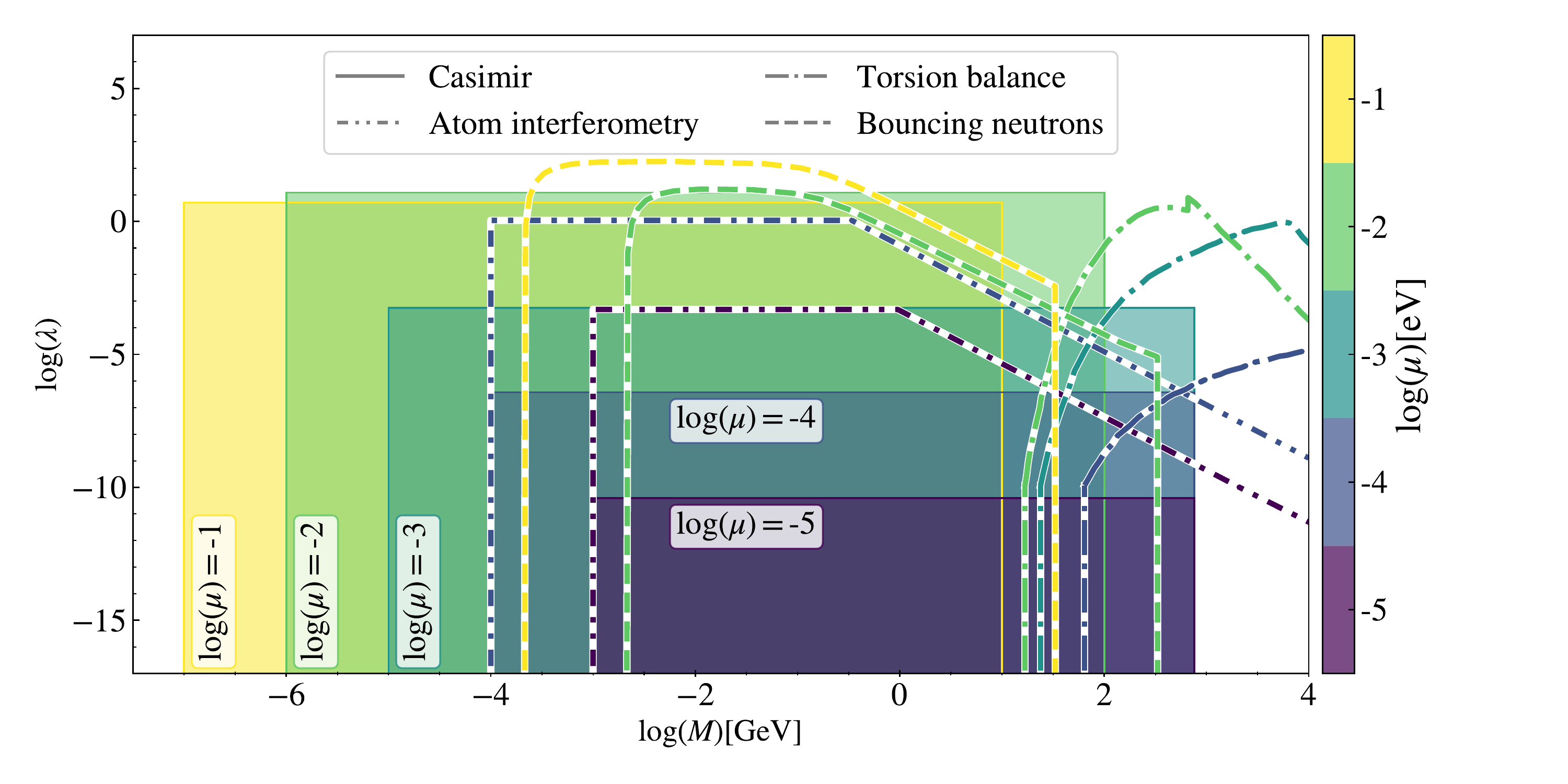}
\caption{\small Summary of laboratory constraints on symmetron parameters. Shaded regions and regions under the curves are ruled out by the respective experiment. Atomic interferometry, torsion balance and ultracold bouncing neutron results are reproduced from Refs.~\cite{Jaffe:2016fsh}, \cite{Upadhye:2012rc} and \cite{Cronenberg:2018qxf}, respectively. The Casimir constraints should be interpreted as realistic forecast results. Figure adapted from \cite{Elder:2019yyp}.}
\label{fig:constraints}
\end{figure}

We make this distinction primarily because of the unique technical challenges associated with making robust and reliable theoretical predictions for those two classes of experiments. While in the first class of experiments, the analytical approximations have proven to be very useful and accurate enough, the direct force measurements often involve very complex density geometries, and their analysis requires more careful numerical modeling. Despite differences in modeling, it is important to combine the constraints derived in different experiments since they often cover complementary regions of the parameter space. Fig.~\ref{fig:constraints} represents a summary of constraints on symmetron models from laboratory experiments. A similar summary for $n = 1$ chameleon models can be found in Ref.~\cite{Brax:2022uiv}.

\subsection{Direct force measurements}\label{subsec:direct_force}

\textbf{Casimir Experiments:} The primary purpose of Casimir-force experiments is the measurement of the force due to modified vacuum energy density in presence of conducting extended bodies (see \cite{Sparnaay1958,Blokland1978,Lamoreaux:1996wh} for historical realizations of the experiment, and \cite{Decca:2005yk,Decca:2003zz} for modern and more precise experiments). Such experiments, however, can also be used for constraining additional (classical) forces \cite{Decca:2005qz,Decca:2003zz}. It is therefore natural to consider the chameleon and symmetron fifth forces in the context of such experiments. Modern Casimir experiments probe distance scales of $\approx 10 \mu$m and are therefore sensitive to large effective scalar-field masses.

The main theoretical complication lies in the difficulty of obtaining accurate estimates of scalar field profiles and fifth forces in complex configurations of realistic experiments. While the earlier Casimir force measurements were performed using two parallel plates, see e.g. Ref~\cite{Almasi:2015zpa}, many modern setups can be approximated as a sphere placed in front of the source plate \cite{Mohideen:1998iz,Harris:2000zz,Lamoreaux:1996wh,Decca:2005yk,Chen:2014oda} (see, however, \cite{Bressi:2002fr,Zou2013} for modern experimental configurations with planar geometry). The schematic setup of interest is depicted in the right panel of Fig.~\ref{fig:plates}. 

A careful treatment of systematic effects is essential to produce reliable constraints on fifth-force models. For example, lateral forces and roughness are not captured by the proximity force calculation \citep{Neto_2005,Rodrigues_2006} and the surfaces used have a finite conductivity \citep{Lambrecht_2000}, both of which must be accounted for. Since the focus of this review is on the reconstruction of the scalar fifth force profiles, we do not discuss these effects further here, but see e.g. \citet{Reynaud_2017} for a review.

Analytical progress can be made in the limiting cases of very small and large spheres. The exposition below focuses on symmetrons, and largely follows \cite{Elder:2019yyp}. In the significantly simpler small-sphere limit, $R \ll \mu^{-1}$, we can treat the sphere as a test mass that does not significantly alter the ambient field profile sourced by the plate. The force on the sphere with mass $m_\mathrm{sphere}$, in this case, can then be computed as
\begin{align}
    \vec F = - \frac{ \lambda_\mathrm{sphere} m_\mathrm{sphere} }{M^2} \varphi \vec \nabla \varphi~,
\end{align}
where $\lambda_\mathrm{sphere}$ is the screening factor introduced in Subsection \ref{subsec:thin_shell}, and $\varphi$ is the analytically solvable field profile of the isolated plate (see the end of Subsection \ref{subsec:relaxation}). As a result we can obtain an accurate analytical force estimate when $R \ll \mu^{-1}$.

The situation is more subtle in the opposite limit of a very large sphere, $R \gg \mu^{-1}$. In this limit, the configuration can be approximated by a collection of parallel plates of varying gap openings by dividing the sphere into co-centric rings facing the plate. The exact (although not explicit) parallel-plate solutions can then be used to obtain the total force on the sphere. This approach is essentially the extension of the proximity force approximation widely employed for calculating the quantum Casimir forces (see e.g. \cite{Dalvit:2009gw,Krause:2007zz}), and it has been used for studying Chameleon models in Casimir experiments \cite{Brax:2007vm}.

Given the exact parallel-plate pressures $P$ for each of the co-centric rings, the total force can be expressed as
\begin{align}
F = - 2 \pi R^2 \int_{\pi / 2}^\pi \mathrm{d} \theta ~ \sin \theta \cos \theta P(L)~,
\end{align}
where $L = D + R + R \cos \theta$ is the $x$-distance between the plate and a ring (see Fig.\ref{fig:plates} for the notation).

We expect both the approximate approaches to fail in the intermediate regime in between of $R \gg \mu^{-1}$ and $R \ll \mu^{-1}$. For such intermediate radii we should rely on exact numerical integration of the field profiles, which can be performed using any of the methods explained in Section~\ref{sec:numerical} with sufficient care toward choosing an appropriately large box size so that the boundary effects do not alter the physical conclusions. Ref.~\cite{Elder:2019yyp} presents a detailed analysis of the plate-sphere configuration in symmetron models and finds that the proximity approximation provides an acceptable accuracy for large-enough spheres. The analysis also validates the results of the large and small sphere approximations. A similar analysis has been performed for chameleon models \cite{Brax:2007vm,Brax:2022uiv}. Interestingly, it has been demonstrated that, unlike the case of symmetrons, the proximity force approximation in the case of chameleons does not reproduce the numerical results even in the limit of large spheres. For chameleons, the approach based on the screening factor calculation for the sphere leads to more reliable analytical estimates \cite{Brax:2022uiv}. This stresses once more the strong model-dependence of screening mechanisms.

\textbf{Torsion Balance:} One version of torsion-balance experiments \cite{Adelberger:2002ic} consists of two parallel cylinders. The bottom disc is turned uniformly while the upper one is a torsion pendulum and can rotate freely. The discs have regularly spaced holes with missing mass. In presence of a fifth force the torsion pendulum would experience a torque when its holes move across the holes of the lower rotating disc. The scalar field configuration is challenging to obtain in such a complex geometry. Particularly, the parallel-plane results can be used to obtain qualitatively robust results \cite{Brax:2014zta}. However, edge effects require a refined analytical modeling and numerical treatment.

Initial results have been obtained by numerically solving the field profile using the conjugate gradient minimization of the Hamiltonian of the system \cite{Upadhye:2006vi}. This direct numerical approach, however, can become prohibitive in certain parts of parameter space which require very fine grid sizes compared to the geometry of the problem. Refs.~\cite{Upadhye:2012qu,Upadhye:2012rc} offered an approximate treatment of the system by introducing the so-called 1-dimensional plane-parallel approximation. The essential idea is to approximate the field on the pendulum cylinder surface by the corresponding surface field in a parallel-plate setup with the gap size corresponding to the distance of the point of interest and the nearest point on the bottom cylinder. Such an approach is similar to the proximity force approximation discussed earlier, and gives qualitatively correct results on the torque, although it quantitatively underestimates the numerical results.

\subsection{Indirect measurements}

\textbf{Atomic Interferometry:} Atomic interferometry is a sensitive probe of gravitational free-fall acceleration. Such experiments are realized inside a vacuum chamber with a spherical source near its center. Atomic matter waves are beam-split and travel in free fall until their interference. The primary observable of interest is the accumulated phase difference between the split matter waves, which is linearly proportional to the acceleration of atomic clouds.

In the context of screening mechanisms, the acceleration is due to both gravity and the fifth force. The entire experiment is carried out inside a vacuum chamber with thick and dense walls. As a result the field profile inside the chamber is independent of the ambient field profile outside. In order to induce fifth forces, a macroscopic source with e.g. spherical shape is placed near the center of the chamber. The total acceleration can be determined using the screening factors $\lambda_\mathrm{atom}$ and $\lambda_\mathrm{source}$ of the atom and the source, and for Chameleon theories can be written as  
\begin{equation}
a_\varphi + a_\mathrm{grav} =  \frac{G_\mathrm{N}m_\mathrm{source}}{r^2}\left(1 + \lambda_\mathrm{atom}\lambda_\mathrm{source} \left(\frac{M_\mathrm{Pl}}{M}\right)^2\right).
\end{equation} 
This expression includes both the gravitational acceleration $a_\mathrm{grav}$ and the acceleration $a_\varphi$ due to fifth forces. A central feature of such configurations is that, while the source could be screened and have $\lambda_\mathrm{source} \ll 1$, the atoms are typically not, $\lambda_\mathrm{atom} = \mathcal{O}(1)$, and the overall fifth-force acceleration is not doubly-suppressed.

Atomic interferometry experiments in the context of screened fifth force searches have been carried out by two independent groups \cite{Hamilton:2015zga, Jaffe:2016fsh, Sabulsky:2018jma}. Analytical modeling approximations have been extensively validated using exact numerical integration of the scalar field equations in the context of such experiments \cite{Elder:2016yxm}.

\textbf{Bouncing Neutrons and other methods:} Experiments including neutrons have been designed in order to measure the energy levels of quantum-mechanical particles in the gravitational field. In presence of an additional fifth force, the energy levels are perturbed and the parameters of the screening mechanisms can be strongly constrained. Such bouncing neutron experiments have been analyzed in the context of symmetrons in \cite{Cronenberg:2018qxf}.

A logically similar method is to compute the shifts in atomic energy levels due to the fifth force on electrons \cite{Brax:2010gp,Brax:2022olf}. The effect can be analytically described using the screening factor of the nucleus. In both the bouncing neutron and atomic spectroscopy methods the effect of the fifth force constitutes in perturbing the Hamiltonian of either the neutrons or electrons. Once the perturbed Hamiltonian is computed, the shifts in energy levels can be estimated by perturbation theory.

\section{Astrophysical scales}
\label{sec:astrophysical_tests}

\subsection{Stars}

\subsubsection{Hydrostatic equilibrium - stellar structure and evolution}

In the presence of a fifth force, the structure and evolution of stars will be altered compared to GR, allowing stellar objects to be utilised as probes of such modifications. The gravitational collapse of a star is halted by its internal pressure such that, if gravity is stronger, then the required outward pressure must also increase, so stars are forced to burn their fuel at a faster rate. This shortens the lifetime of the star and will make them appear brighter. The opposite would occur if the modification to gravity weakened the attractive force within the star.

To quantitatively study the impact of modifications to gravity, the only equation one needs to alter is the criterion for hydrostatic equilibrium \citep{Koyama_2015,Saito_2015}
\begin{equation}
    \frac{1}{\rho} \frac{\dd P}{\dd r} = - \frac{G M}{r^2} - \frac{\Upsilon}{4} G \frac{\dd^2 M}{\dd r^2},
\end{equation}
for pressure $P$, density $\rho$ and where $M$ is the mass enclosed at radius $r$. In standard gravity $\Upsilon = 0$, and is more generally given in terms of the $\alpha_B$, $\alpha_T$ and $\alpha_H$ parameters \citep{Gleyze_2015} as
\begin{equation}
    \Upsilon = \frac{4 \alpha_H^2}{\alpha_H - \alpha_T - \alpha_B \left( 1 + \alpha_T \right)}.
\end{equation}
The mere existence of stars already requires $\Upsilon > -2/3$, since below this value gravity becomes repulsive near the center of the star \citep{Saito_2015}.

In the simplest models, one supposes that a star has a polytropic equation of state, $P = K \rho^{\frac{n+1}{n}}$, for positive constants $K$ and $n$. One can then define dimensionless variables in terms of the central density $\rho_{\rm c}$ and pressure $P_{\rm c}$ to be
\begin{equation}
    r = r_{\rm c} \xi, \quad 
    \rho = \rho_{\rm c} \theta (\xi)^n, \quad
    P = P_{\rm c} \theta (\xi)^{n+1}, \quad
    r_{\rm c}^2 = \frac{(n+1) P_{\rm c}}{4 \pi G \rho_{\rm c}^2},
\end{equation}
and thus arrive at the modified Lane-Emden equation \citep{Koyama_2015}
\begin{equation}
    \frac{1}{\xi^2} \frac{\dd}{\dd \xi} \left[ \left( 1 + \frac{n}{4} \Upsilon \xi^2 \theta^{n-1} \right) \xi^2 \frac{\dd \theta}{\dd \xi} + \frac{\Upsilon}{2} \xi^3 \theta^n \right] = - \theta^n.
\end{equation}
which has boundary conditions $\theta(0) = 1$, $\theta^\prime(0) = 0$.

To make progress, one now needs to couple this equation with other equations describing stellar evolution. By using analytic approximations for the temperature profile and assuming a polytropic index $n$, \citet{Koyama_2015} demonstrated that increasing $\Upsilon$ resulted in a reduction of the stellar luminosity at fixed mass, with stronger modifications for low mass stars since these are supported by gas pressure, rather than radiation pressure. This constrains $\Upsilon$ to be $\lesssim \mathcal{O}(1)$.
Similarly, \citet{Sakstein_2015,Sakstein_2015_dwarf} showed that the weakening of gravity within astrophysical bodies in scalar-tensor theories will increase the minimum mass a star must have to sustain hydrogen burning which, given the observed masses of Red Dwarf stars (which are greater than $0.08 M_{\odot}$ \citep{Segransan_2000}), lead to similar constraints, ruling out $\Upsilon \gtrsim 1.6$. Modifications of gravity can also change the radius of brown dwarfs from their GR prediction ($0.1 R_{\odot}$) to arbitrarily large values if gravity is weaker, or down to $0.078R_{\odot}$ if it is stronger.

To go beyond these semi-analytic models, one must solve all the relevant stellar structure equations simultaneously. This is commonly achieved by modifying the gravity model in the publicly available Modules for Experiments in Stellar Astrophysics (\texttt{MESA}) code \citep{MESA_2011,MESA_2013,MESA_2015,MESA_2018,MESA_2019,MESA_2022}.

After making such modifications, one can investigate the effect on different aspects of stellar evolution. For example, by considering the RGB phase, \citet{Chang_2011} saw that, in Chameleon models, red giants can be significantly smaller (by tens of percent) and hotter (by hundred of Kelvins) than their GR counterparts. This effect is particularly pronounced in red giants since main sequence stars are much denser and thus have a suppressed scalar charge, whereas the outer envelope of red giants remains unscreened.  If one considers $G^3$ galileons \citep{Gleyze_2015} exhibiting Vainstein screening, one finds that it is the main-sequence stars which are most strongly affected \citep{Koyama_2015}, and have a smaller effective temperature than GR stars at the same point in the Hertzsprung–Russell diagram. While such an effect could be degenerate with metallicity, modified gravity theories leave distinct evolutionary tracks in the Hertzsprung–Russell diagram, allowing these objects to constrain such models. In the context of general modified gravity theories \citet{Saltas:2019ius} and \citet{Saltas:2022ybg} have investigated the changes in the solar sound speed using analytical approximations and \texttt{MESA} simulations. These studies have concluded that precise helioseismic observations are able to improve the constraints on the fifth force coupling strength $\Upsilon$, and have derived an approximate constraint of $-10^{-3} < \Upsilon < 5\times 10^{-4}$ at the $2\sigma$ confidence level.

\subsubsection{Out of equilibrium - stellar oscillations}

Until now we have only considered stars in equilibrium. Moving beyond this, suppose one can consider small radius perturbations $\delta r$ in the star which obey the equation
\begin{equation}
    \frac{\partial^2 \delta r}{\partial t^2} = - \frac{1}{\rho} \frac{\partial P}{\partial r} - \frac{\dd \Phi}{\dd r}.
\end{equation}
In the absence of modified gravity, the period of small oscillations is found to be
\begin{equation}
    \Pi = \frac{2 \pi}{\sqrt{4/3 \pi G \rho \left( 3 \gamma - 4 \right)}}.
\end{equation}
Hence any enhancement in the strength of gravity ($G \to G + \Delta G$ with $\Delta G > 0$) would reduce the period \citep{Jain_2013}. The change in the period can be as large as 30\% for $\mathcal{O}(1)$ couplings in chameleon screened theories \citep{Sakstein_2013}. This would be important for Cepheids -- massive stars which pulsate with periods ranging from days to weeks when they pass through the instability strip -- since the relation between their luminosity and period is used to determine their distance. This allows one to measure the strength of gravity in e.g. the Large Magellanic Cloud \citep{Desmond_2021} by modeling the internal properties of Cepheids with \texttt{MESA}.

Moreover, since Cepheids are used in the construction of the distance ladder for local measurements of $H_0$, an incorrect distance estimation can bias the inferred value of $H_0$. Indeed, \citet{Desmond_2019} found that a fifth force with 5-30\% the strength of gravity can reduce the Hubble tension. A similar effect can be obtained by instead calibrating to the TRGB \citep{Desmond_2020_TRGB}; if the hydrogen burning shell become unscreened, then the luminosity at the TRGB would be reduced \citep{Sakstein_2019} and thus distances based on this feature would be overestimated. Fifth-forces can therefore reduce the locally inferred value of $H_0$ thus reducing or alleviating the Hubble tension. \citet{Hogas:2023vim} have studied the period-luminosity relation in the context of the symmetron theory. While this theory has been shown to worsen the Hubble tension, the distance ladder measurements impose strong constrains on the symmetron Compton wavelength.

Using the same notation as above, for Vainshtein screened theories, one can write $\delta r = r \chi(r) \exp(i\omega t)$ and linearise the pressure and potential to obtain a modified Linear Adiabatic Wave Equation \citep{Sakstein_2017_osc}
\begin{equation}
    \label{eq:LAWE}
    \frac{\dd}{\dd r} \left[ r^4 \left( \Gamma P_0 + \Upsilon \pi G r^2 \rho_0^2\right) \frac{\dd \chi}{\dd r} \right] + 
    r^3 \chi \frac{\dd}{\dd r} \left[ \left(3 \Gamma - 4 \right) P_0 \right] + 
    r^4 \rho_0 \omega^2 W \left( r, \Upsilon \right) \chi = 0,
\end{equation}
where
\begin{equation}
    \Gamma = \left( \frac{\partial \log P}{\partial \log \rho} \right)_{\rm adiabatic}, \quad
    W \left(r , \Upsilon \right) = 1 - \frac{\Upsilon \pi r^3 \rho_0 M}{2 \left( M + \pi r^3 \rho_0 \right)^2},
\end{equation}
and a subscript `0' denotes an unperturbed quantity (see \citet{Sakstein_2013} for the chameleon-screened version of this equation). For stars in GR, this results in unstable stars if $\Gamma < 4/3$, whereas in beyond Horndeski theory there is a second instability for sufficiently large $\Upsilon$, since this will result in $W < 0$ \citep{Sakstein_2017_osc}. Using \texttt{MESA}, \citet{Sakstein_2017_osc} showed that the modification to the period-luminosity relation for Cepheids also holds for these theories, and thus the distances inferred using this method can be compared to that from the TRGB to constrain $\Upsilon$.

\subsection{Screening maps}
\label{subsec:screening maps}

When determining whether a given object is screened or unscreened, one often computes properties of the local gravitational field: the potential, $\Phi$, for chameleons, the acceleration, $\bm{a}$, for kinetically screened theories, or the local curvature (e.g. the Kretschmann scalar, $K$) in the presence of Vainshtein screening. These all have contributions from the object itself, but also from an object's local environment. Particularly for astrophysical tests, it is therefore imperative to produce maps of these quantities to identify the screened and unscreened regions of the local Universe.

The Poisson equation for the potential, $\phi$, in the presence of a density perturbation, $\delta \rho$, is modified in $f(R)$ gravity to be 
\begin{equation}
    \nabla^2 \phi = 4 \pi G a^2 \delta \rho + \left( \frac{4 \pi G a^2 \delta \rho}{3} + \frac{a^2}{6} \delta R \right) \equiv 4 \pi G a^2 \delta \rho_{\rm eff},
\end{equation}
where $\delta R$ is perturbation to Ricci scalar and $a$ is the scale factor. By integrating $a^2 \delta \rho_{\rm eff}$ to obtain a dynamical mass, $M_{\rm D}$, and comparing to true ``lensing'' mass, $M_{\rm L}$, one can calculate a proxy for the degree of screening, $\Delta_m \equiv M_{\rm D} / M_{\rm L} - 1$ \citep{Zhao_2011,Schmidt_2010_dyn}. In this case, $\Delta_{\rm m} \in [0, 1/3]$, where $\Delta_{\rm m} = 0$ corresponds to a perfectly screened object, and $1/3$ is completely unscreened. Although this offers a useful metric for quantifying the degree of screening, we often do not have access to both the lensing and dynamical masses of objects, so must rely on alternative methods.

\citet{Cabre_2012} were the first to produce maps of the local gravitational potential for use in quantifying the degree of screening. For a set of galaxies of mass $M_i$ and virial radii $r_i$, they compute the external contribution to $\Phi$ to be
\begin{equation}
    \Phi_{\rm ext} = \sum_{d_i < r_i + \lambda_{\rm c}} \frac{G M_i}{r_i},
\end{equation}
where $\lambda_{\rm c}$ is the Compton wavelength of the field, and $d_i$ is the distance to the $i$\textsuperscript{th} galaxy. One must introduce a cut-off based on the Compton wavelength, otherwise the above sum would diverge. This screening map covered $z < 0.05$ and the Sloan Digital Sky Survey (SDSS) footprint, since it was produced predominantly from a compilation of the SDSS group catalog \citep{Yang_2007}, alongside other catalogs \citep{Abell_1989,Ebeling_1996,Karachentsev_2004,Lavaux_2011}. The approximations used were tested and calibrated against $N$-body simulations \citep{Zhao_2011_nbody}.

A more sophisticated analysis was performed by \citet{Desmond_2018_maps}, who extended these methods to produce full-sky maps up to $\sim 200 {\rm \, Mpc}$, as well as maps of $\bm{a}$ and $K$. Galaxies from the 2M++ galaxy catalog \citep{Lavaux_2011} were matched to 
\textsc{rockstar} \citep{Behroozi_2013} halos from the DARKSKY-400 simulation \citep{Skillman_2014} using the abundance matching method of \citet{Lehman_2017}. Assuming these halos have Navarro–Frenk–White (NFW) density profiles \citep{NFW_1997}, $\Phi$ and $\bm{a}$ were computed using all objects within $\lambda_{\rm c}$ of a given point using standard formulae \citep{Cole_1996}. To compute the curvature, $K$ was approximated to be the sum of point mass contributions; this neglects the extended nature of halos and the non-linearity of this quantity, so should be treated as an order of magnitude approximation. To account for galaxies in halos which are too faint to be observed, correction factors 
for each quantity were applied by comparing to results from $N$-body simulations. To account for mass which does not reside in halos, a single constrained density field inferred using the Bayesian Origin Reconstruction from Galaxies (\textsc{borg}) algorithm \citep{BORG_1,BORG_2,BORG_3,BORG_4,Lavaux_2016} applied to the 2M++ galaxy catalog was used.

Instead of computing screening proxies, one could directly compute the profiles of the dynamical fields which result in a modification to gravity, although one must do this separately for each theory one wishes to test. As part of the LOCal Universe Screening Test Suite (\textsc{locusts}) project, \citet{Shao_2019} solved for the scalar field for $n=1$ Hu-Sawicki $f(R)$ gravity applying the \textsc{ecosmog} \citep{Li_2012,Li_2013_VS,Li_2013_quartic} code to a constrained $N$-body simulation from the \textsc{elucid} project \citep{Wang_2014,Wang_2016} (inferred using the SDSS DR7 galaxy catalog), using 20 $f_{\rm R0}$ values in the range $10^{-7}-10^{-6}$.

Both the \textsc{borg} and \textsc{elucid} density fields used to produce screening maps are constrained assuming no modification to gravity. Hence, these maps implicitly assume that the low redshift Universe has matter fields which are similar for all reasonable modified gravity and $\Lambda$CDM models on the scales of the reconstruction. The first constrained density fields which include the effects of modified gravity have now been produced \citep{Naidoo_2022} from CosmicFlows peculiar velocity data \citep{Kourkchi_2020}, through an extension of the \textsc{ICeCoRe} package \citep{Doumler_2013_III}. By applying a Wiener filter \citep{Zaroubi_1995} to obtain the linear density and velocity fields, applying the reverse Zeldovich approximation \citep{Doumler_2013_I}, then adding fluctuations in poorly constrained regions, they obtain estimates for the initial conditions at $z=49$. They then run simulations using the \textsc{mg-picola} \citep{Winther_2017} implementation of the COmoving Lagrangian Acceleration (COLA) method \citep{Tassev_2013} for nDGP and $f(R)$ models, which solves for particle trajectories in the frame of the reference given by Lagrangian perturbation theory. The current implementation neglects the scale-dependence of the growth function in $f(R)$ when generating constrained realisations, hence they find that it is better to generate constrained $f(R)$ simulations from $\Lambda$CDM initial conditions than the ones they infer. Future work should be dedicated to producing fully-consistent constrained simulations and moving away from the linear approximation required in this reconstruction \cite{Naidoo_2022}. Phenomenological approaches to incorporating screening effects in quasi-linear regimes of structure formation could offer additional insights in understanding the large-scale environmental effects on screening dynamics \cite{Fasiello:2017bot}.

\subsection{Galaxy morphology}

\subsubsection{Thin-shell screened theories}

In thin-shell screened theories where the degree of screening is determined by the Newtonian potential, one can define a cut-off, $\Phi_{\rm c}$, such that objects with potentials below this value are unscreened and experience a fifth-force, and all other objects are screened. If $\Phi_{\rm c}\sim 10^{-6}$, then main sequence stars will always be screened, since this is the potential at their surfaces, irrespective of their environment. However, if the local gravitational potential is below this value, the surrounding gas and dark matter will be unscreened, and hence feel an additional force. If there exists a gradient in the field responsible for the fifth force, then the gas and dark matter will be accelerated whereas the stars will not, leading to an offset between the centre of mass of the stars and the remaining mass of the galaxy. This offset will be stabilised by the induced gravitational potential, and this will also induce a warp in the stellar disk. Thus, two morphological signatures are to be expected in such a theory: (1) an offset parallel to the local gravitational field between the stars and gas of galaxies in regions of low gravitational potential  and (2) a warping of the stellar disk into a characteristic `U' shape \citep{Jain_2011}.

Initial studies \citep{Vikram_2013} of these phenomena utilising the screening maps of \citet{Cabre_2012} (see Subsection~\ref{subsec:screening maps}) and, combining data from the ALFALFA \cite{Giovanelli:2005ee} and SDSS \cite{SDSS:2008tqn} surveys, found no evidence for screened fifth forces, but the constraints were not competitive with other astrophysical probes. More recently, galactic morphology has played an increasingly important role in constraining such theories \citep{Desmond_2018_sep,Desmond_2018_warp,Desmond_2018_web}, with the screening maps of \citet{Desmond_2018_maps} being used to determine which objects are unscreened and to predict the magnitude and direction of the offsets and warps. The most recent constraints find that a fifth force with $\lambda_{\rm c} \geq 0.3 {\rm \, Mpc}$ must have a strength $\Delta G / G_{\rm N}$ at least as small as 0.8 at $1 \sigma$ confidence \citep{Desmond_2020}. $n=1$ Hu-Sawicki $f(R)$ with $f_{\rm R0} > 1.4 \times 10^{-8}$ is ruled out, resulting in practically all astrophysical objects being screened in the surviving region of parameter space, and hence all astrophysically relevant models are disfavoured. The modeling of other astrophysical effects for these constraints has been shown to be robust when calibrated against hydrodynamical $\Lambda$CDM simulations \citep{Bartlett_2021_cal}.

\subsubsection{Vainshtein screened theories}

One of the main astrophysical tests of the Vainshtein mechanism relies on the difference in coupling for non-relativistic matter ($Q=m$) and black holes ($Q=0$) and was proposed by \citet{Hui_2012}. This extreme difference means that, if the non-relativistic matter in a galaxy (stars, gas, dark matter) is falling down a gradient in the galileon field, then the supermassive black hole at its centre lags behind as it does not experience the fifth force. This offset is stabilised by the induced gravitational potential, and thus one would expect to see an $\mathcal{O}({\rm kpc})$ offset between the black hole near a galaxy's centre and its centroid, as measured by the stars.

\begin{figure}
    \centering
    \includegraphics[width=\columnwidth]{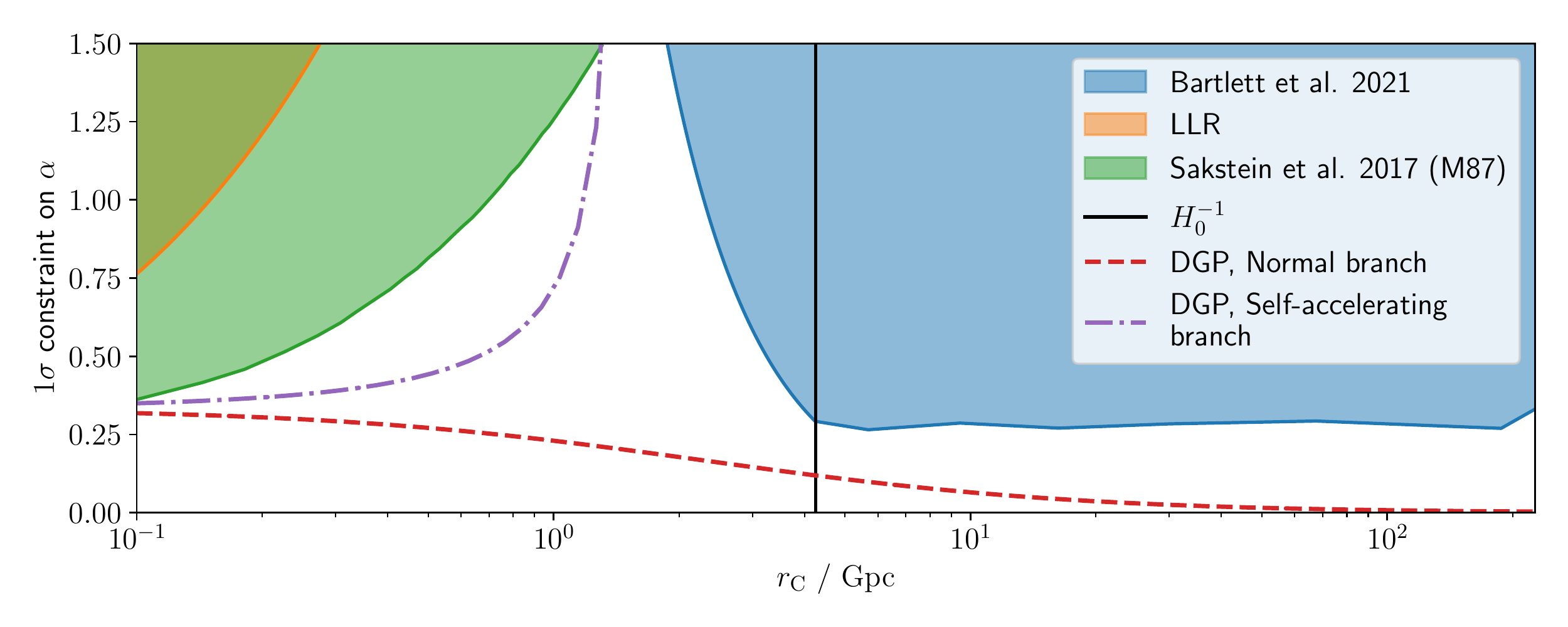}
    \caption{Constraints on the galileon coupling parameter, $\alpha$, as a function of cross-over scale, $r_{\rm C}$, from astrophysical tests. The shaded regions are excluded at $1\sigma$ confidence. For small $r_{\rm C}$, the constraints are driven by Lunar Laser Ranging (LLR; \citep{Khoury_2013,Murphy_2012}) and by considering the offset of the black hole in M87 \citep{Sakstein_2017}. For larger $r_{\rm C}$, the constraints are derived from the offsets in a larger sample of galaxies, where the galileon field is sources by large scale structure \citep{Bartlett_2021_VS}. Figure adapted from \citep{Bartlett_2021_VS}.}
    \label{fig:vs_constraints}
\end{figure}

This test was first performed by \citet{Sakstein_2017} by considering the black hole--galaxy offset in M87, utilising the analysis of \citet{Asvathaman_2017}. Here the galileon profile was determined analytically assuming spherical symmetry (Eq.~(\ref{eq:galileon_spherical})), which was shown to be an excellent approximation for cosmological simulations (see e.g. Ref.~\cite{Schmidt_2009}). As shown in Fig.~(\ref{fig:vs_constraints}), $\alpha\sim\mathcal{O}(1)$ was found to be excluded for cross-over scales $r_{\rm C} \lesssim 1 {\rm \, Gpc}$.

\citet{Bartlett_2021_VS} extended these results by comparing the galaxy centroids to the X-ray and radio positions of active galactic nuclei (AGN) in a sample of 1916 galaxies compiled in \citep{Bartlett_2021_HAGN}. They considered the galileon field sourced by large scale structure, utilising a suite of constrained $N$-body simulations to produce maps of dark matter in the nearby Universe to predict both the magnitude and direction of the offset. By comparing to observations, the constraints on $\alpha$ were found to be tighter than those obtained from M87, with $\Delta G / G_{\rm N} > 0.16$ ruled out at $1\sigma$ confidence (see Fig.~(\ref{fig:vs_constraints})), although these constraints are applicable to larger $r_{\rm C}$. We note that the constrained simulations were constructed assuming $\Lambda$CDM, and thus these constraints are only applicable for models which give rise to similar large scale structure as $\Lambda$CDM, and only affect objects on smaller scales. Furthermore, the Vainshtein mechanism was approximated as a hard cut-off, such that on large scales the galileon field traces the gravitational potential, but above some wavenumber the galileon field is set to zero. This is likely to be a less accurate approximation than that used by \citet{Sakstein_2017}, although the advantage of using a statistical sample is that the non-galileon effects which were ignored in \citet{Sakstein_2017} were empirically modelled and marginalized over.

\subsection{Halo properties}

In GR, gravity determines the formation and growth of dark matter halos. Their abundance and internal structure can be modified in presence of fifth forces, which might be partially unscreened especially for objects with smaller masses and on the outskirts of halos. Since halos are strongly non-linear structures, N-body simulations have been the primary framework for deriving a quantitative understanding of these effects.

Modeling halo structure in modified gravity theories has been the focus of extensive past work, such as Refs.~\cite{Schmidt:2008tn,Li:2011uw,Pourhasan:2011sm,Lombriser:2012nn,Lombriser:2013wta,Lombriser:2013eza,Shi:2015aya,Mitchell:2019qke} for chameleon theories, Refs.~\cite{Schmidt_2010,Barreira:2014zza,Falck_2014,Mitchell:2021aex,Pizzuti:2021brr} for Vainshtein mechanism, and Ref.~\cite{Clampitt:2011mx} for symmetrons. Additionally, environmental dependences for both chameleon and Vainshtein mechanisms have been studied in Refs.~\cite{Schmidt_2010_dyn,Falck:2015rsa}.

Screening efficiency depends on the halo mass, and in the case of chameleon theories less massive halos tend to be unscreened. For Vainshtein, on the other hand, screening is less sensitive to the mass, and practically all halos are screened \cite{Schmidt_2010_dyn,Falck:2015rsa}. As explained in Subsection~\ref{subsec:galileons}, even though the inner regions of NFW halos in Vainstein-screened theories are expected to be fully screened, one expects non-vanishing fifth force outside the virial radius. Simulations agree well with this theoretical prediction, with certain variation near the outskirts of the halos which is likely due to the breakdown of the NFW approximation \citep{Schmidt_2009,Khoury_2009,Schmidt_2010}.

As far as population-level mass distribution is concerned, it has been established that partially screened fifth forces induce significant modifications in the differential halo mass function~$\mathrm{d}n/\mathrm{d}log M_\mathrm{halo}$, see e.g. \cite{Schmidt:2008tn,Lombriser:2013wta,Shi:2015aya}. While the exact results depend strongly on the model parameters, a general trend is that the distribution of very heavy halos is not affected. The main difference can be seen below a certain mass threshold, where halos are more abundant than in General Relativity. At even lower masses, halos are more abundant in GR, because such smaller halos are more efficiently absorbed by larger ones.   

In $\Lambda$CDM, halos have universal density profiles described by the NFW parametrization:
\begin{align}
    \rho(r)_\mathrm{dm} = \frac{\rho_\mathrm{s}}{\frac{r}{r_\mathrm{s}}\left(1 + \frac{r}{r_\mathrm{s}}\right)^2},
\end{align}
where $\rho_\mathrm{s}$ is a characteristic density, and $r_\mathrm{s}$ is a characteristic scale for a halo. The compactness (concentration) of the profile is determined as the ratio of the virial radius $r_{200}$ and $r_\mathrm{s}$; $c \equiv r_{200}/r_\mathrm{s}$. The concentration and the characteristic density $\rho_\mathrm{s}$ are connected via
\begin{align}
    \rho_\mathrm{s} = \frac{200}{3}\rho_\mathrm{c}\frac{c^3}{g(c)},\quad
    4\pi r_\mathrm{s}^3\rho_\mathrm{s} = \frac{M_{200}}{g(c)}
\end{align}
where $\rho_\mathrm{c}$ is the critical density of the Universe, and $g(c)\equiv \log(1+c) - c/(1+c)$. 

Earlier studies, such as Refs.~\cite{Shi:2015aya,Lombriser:2013eza} have examined the halo profiles in $f(R)$ N-body simulations and found that NFW is still an appropriate description, especially for larger halos. The profiles however can be more compact compared to GR. Detailed investigation of such an effect can be found in e.g. Refs.~\cite{Shi:2015aya,Mitchell:2019qke} which found that halos with masses below a certain threshold are typically more concentrated. This, however, also depends on redshift, with the effect diminishing at relatively higher redshifts.

\subsection{Splashback}

In addition to modifying the halo concentrations and their mass function, fifth forces can also imprint non-trivial and potentially observable effects on the halo profiles. For larger, cluster-size halos this primarily concerns the outskirts of the density profiles, where fifth forces might be unscreened, hence significantly altering the dynamics of collapsing matter shells. In recent years, the \emph{splashback} feature has gained importance as a physically motivated boundary of halos. This feature corresponds to the dividing boundary of single-streaming and multi-streaming regions and manifests itself as a steepening at outer regions \cite{Diemer:2014xya}. The splashback location can in principle be inferred by measuring the radial distribution of subhalos, and assuming matter follows the same distribution \cite{More:2016vgs, Baxter:2017csy, Chang:2017hjt, Shin:2018pic}.
Additionally, dedicated lensing measurements can also be used \cite{Umetsu:2016cun, Contigiani:2018qxn}. 

The position of the splashback feature is sensitive to any additional interactions and has been investigated in the context of $f(R)$ gravity and nDGP by \citet{Adhikari:2018izo}. Here the authors have used both semi-analytical and N-body approaches to predict the changes in splashback location. For symmetrons, splashback has been investigated in \citet*{Contigiani:2018hbn} by extending the self-similar collapse approach \cite{Fillmore:1984, Bertschinger:1985pd}. The latter approach, while not quantitatively very accurate, allows gaining important insights into halo growth in presence of fifth forces. We will use this approach to demonstrate the changes in splashback location.          

We will focus on the collapse in Einstein de-Sitter (EdS) universe, and enforce the mass profile to satisfy a self-similar form:
\begin{equation}
 M(r, t) = M(R, t) \mathcal{M}(r/R),
\end{equation}
where $R(t)$ is the turn-around radius at time $t$, and $M(r, t)$ is the mass within the radius $r$. The mass withing turn-around radius grows with scale factor as $M(R, t) \sim a(t)^s$, with accretion rate $s$, and in EdS is a power-law in cosmic time $t$.

In absence of fifth force the shell variables can be rescaled such that all the shells satisfy the same equation of motion
\begin{align}
	\frac{d^2\lambda}{d\tau^2} &= - \frac{\pi^2}{8}\frac{\tau^{2s/3}}{\lambda^2}\mathcal{M}\left(\frac{\lambda}{\Lambda(\tau)}\right)\label{eq:lambda},\\
	\mathcal{M}(y) &= \frac{2s}{3} \int_{1}^{\infty} \frac{d\tau}{\tau^{1+2s/3}} \Theta\left( y - \frac{\lambda(\tau)}{\Lambda(\tau)} \right)\label{eq:M},
\end{align}
where we have labeled each shell by its turn-around time $t_\ast$ and radius $r_\ast$, and have introduced $ \lambda \equiv r/r_\ast, \tau \equiv t/t_\ast$, and $\Lambda = R(t)/r_\ast$. This system, complemented by initial conditions $\lambda(\tau)$ are $\lambda(\tau = 1) = 1$, $d\lambda/d\tau(\tau = 1)=0$, can be iteratively solved until a converged mass-profile is obtained. The time-dependence of the density profile is determined in terms of the critical density $\rho_\mathrm{c}(t)$ and $R(t)$. These equations can also be formulated for other $1$-dimensional configurations \cite{Fillmore:1984}, as well as tri-axial collapse \cite{Lithwick:2010ej}. The framework can also be extended to full $\Lambda$CDM backgrounds \cite{Shi:2016lwp}.

\begin{figure*}
\includegraphics[width=1\textwidth]{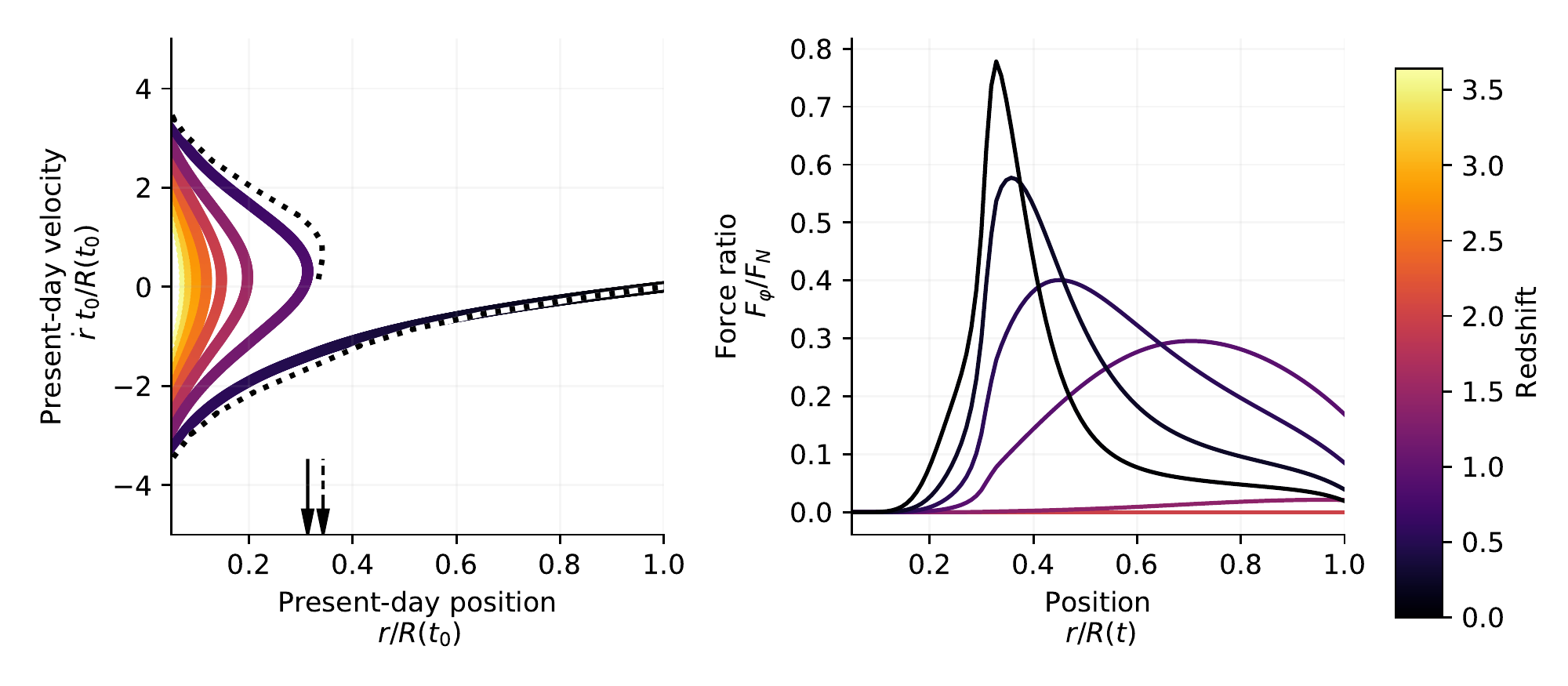}
\caption{Splashback location in presence of symmetron fifth forces. The left panel depicts the phase-space distribution of shells in a halo, with colors denoting their turn-around redshifts. The dotted line depicts the distribution in the symmetron model. Splashback locations with and without symmetron force are marked by dotted and solid arrows, respectively. The right panel shows the symmetron force profile relative to the Newtonian force evaluated at different redshifts. Figure adapted from \cite{Contigiani:2018hbn}.}
\label{fig:splashback}
\end{figure*}

Once the self-similar halo profile is obtained as function of redshift, the symmetron field profiles can be computed with one of the methods introduced in Section~\ref{sec:numerical}. The extra force on outer shells can be added in Eq.~(\ref{eq:lambda}) in order to study the dynamics of the outer-most collapsing shells. It should be noted that in the cosmological context presented here, the symmetron is fully screened at higher redshift. The unscreening redshift, $z_\mathrm{ssb}$, is determined by the condition $\rho_\mathrm{c}(z_\mathrm{ssb}) = \rho_\mathrm{ssb} = M^2\mu^2$. Fig.~\ref{fig:splashback} demonstrates the fifth-force profiles as a function of redshift, and the resulting effect on splashback location. From the right panel of the same figure it is clear that the innermost regions of the overdensity always stay screened. Substantial fifth forces start appearing in the outer regions when the $z \ll z_\mathrm{ssb}$ condition is satisfied. At some point, a thin shell forms and the force gets concentrated in a narrow radial region. Due to this behaviour there exists an optimal value of $z_\mathrm{ssb}$ which maximizes the change in splashback location due to fifth forces.

\section{Conclusions}
\label{sec:conclusions}

The discovery of late-Universe cosmic acceleration motivated the extensive studies of alternatives to Einstein's theory of General Relativity. Such modified gravity theories generically involve additional scalar degrees of freedom. When such fields couple non-trivially to gravity and matter fields, they exert additional, ``fifth'' forces on matter particles. The presence of fifth forces leads to remarkable variety of phenomenological effects, ranging from cosmologically large scales down to microscopic effects. 

At large and linear scales the novel effects have been relatively well-understood in terms of modified growth of structure, which in modified gravity can be scale- and time-dependent, and modified gravitational lensing. Significant progress has been made, first of all, by employing perturbation theory techniques in order to derive generic and largely model-independent observables, see e.g. \cite{Pogosian:2016pwr, Amendola:2016saw}. Moreover, Effective Field Theory approaches have been very successful for systematically deriving unified prescriptions of linear perturbations \cite{Gubitosi:2012hu,Gleyzes:2013ooa,Creminelli:2013nua,Bellini:2014fua}. Such unified approaches have been efficiently implemented in Einstein-Boltzmann numerical solvers \cite{Hu:2013twa,Raveri:2014cka,Zumalacarregui:2016pph,Bellini:2019syt}.

On smaller scales the dynamics of scalar fields exhibit highly non-linear behaviour, and lead to remarkable density-dependent screening effects. In this short review we have summarized the main classes of screening mechanisms and have provided a broad summary of numerical and analytical methods developed in the context of non-linear dynamics. Unlike the linear regime, in screening regimes scalar fields behave in a very model-dependent way. Even though at the theoretical level screening mechanisms can be categorized into distinct classes, such as chameleons, symmetrons, and Vainshtein mechanisms, their precise phenomenological descriptions are given at a model-by-model basis. Even very similar mechanisms, such as chameleons and symmetrons can often require different analysis approaches. An interesting and non-trivial example of this is the study of Casimir fifth forces in sphere-plane configurations, which closely resemble modern experimental setups. For example, the proximity-force approximation, explained in Subsection~\ref{subsec:direct_force} can provide accurate description for the system for symmetrons, while fail for chameleons.

Numerical approaches are typically required in order to asses the validity of analytical approximations for a specific model. Providing an introductory summary of numerical algorithms is among the central topics of this review. While current numerical solvers have been designed with specific models in mind, most of the numerical approaches are applicable for a broad range of theories. This motivates for developing unified solver packages. The Finite-Elements-Based solvers introduced in Subsection~\ref{subsec:FEM}, particularly, posses very favourable dimensional scaling properties. A unified FEM-based code can be applied not only to complex laboratory configurations, but also to astrophysical and cosmological scenarios.

Identifying novel regimes where modified gravity can be tested is another promising avenue. Particularly, as outlined in Subsection~\ref{subsec:screening maps}, screening maps of the large-scale structure are useful for systematically exploring regions with discovery potential. To this end, constrained simulations are important, in order to correctly account for environmental effects. In some of the existing studies GR-based constrained simulations have been employed. Self-consistent modified gravity constrained simulations, such as the one presented in \cite{Naidoo_2022}, will play an important role in producing precise screening maps.

Testing modified gravity theories in non-linear regimes will provide invaluable insights about the nature of gravitational interactions. At the same time, robust modeling of non-linear phenomena is often very challenging and extensive cross-domain effort is required. As the precision of both the laboratory and astrophysical tests is drastically improving, approximate methods need to be replaced with more accurate approaches.


\funding{V.V. is supported by the WPI Research Center Initiative, MEXT, Japan and partly by JSPS KAKENHI grant 20K22348. D.J.B. acknowledges support by the Simons Collaboration on ``Learning the Universe''.
}

\dataavailability{\Cref{fig:cham_symm,fig:tanh_profile,fig:casimir} are adapted from \citet*{Vardanyan:2019gjt}, \Cref{fig:plates,fig:constraints} are adapted from \customElder, \Cref{fig:vs_constraints} is adapted from \citet*{Bartlett_2021_VS}. \Cref{fig:splashback} is adapted from \citet*{Contigiani:2018hbn}.
}

\conflictsofinterest{The authors declare no conflict of interest.}

{}
\reftitle{References}
\bibliography{bibliography}

\begin{thebibliography}{999}

\bibitem[Perlmutter \em{et~al.}(1999)Perlmutter
  et~al.]{SupernovaCosmologyProject:1998vns}
Perlmutter, S.;  et~al.
\newblock {Measurements of $\Omega$ and $\Lambda$ from 42 high redshift
  supernovae}.
\newblock {\em Astrophys. J.} {\bf 1999}, {\em 517},~565--586,
  \href{http://xxx.lanl.gov/abs/astro-ph/9812133}{{\normalfont
  [astro-ph/9812133]}}.
\newblock {\url{https://doi.org/10.1086/307221}}.

\bibitem[Riess \em{et~al.}(1998)Riess et~al.]{SupernovaSearchTeam:1998fmf}
Riess, A.G.;  et~al.
\newblock {Observational evidence from supernovae for an accelerating universe
  and a cosmological constant}.
\newblock {\em Astron. J.} {\bf 1998}, {\em 116},~1009--1038,
  \href{http://xxx.lanl.gov/abs/astro-ph/9805201}{{\normalfont
  [astro-ph/9805201]}}.
\newblock {\url{https://doi.org/10.1086/300499}}.

\bibitem[Copeland \em{et~al.}(2006)Copeland, Sami, and
  Tsujikawa]{Copeland:2006wr}
Copeland, E.J.; Sami, M.; Tsujikawa, S.
\newblock {Dynamics of dark energy}.
\newblock {\em Int. J. Mod. Phys. D} {\bf 2006}, {\em 15},~1753--1936,
  \href{http://xxx.lanl.gov/abs/hep-th/0603057}{{\normalfont
  [hep-th/0603057]}}.
\newblock {\url{https://doi.org/10.1142/S021827180600942X}}.

\bibitem[Silvestri and Trodden(2009)]{Silvestri:2009hh}
Silvestri, A.; Trodden, M.
\newblock {Approaches to Understanding Cosmic Acceleration}.
\newblock {\em Rept. Prog. Phys.} {\bf 2009}, {\em 72},~096901,
  \href{http://xxx.lanl.gov/abs/0904.0024}{{\normalfont
  [arXiv:astro-ph.CO/0904.0024]}}.
\newblock {\url{https://doi.org/10.1088/0034-4885/72/9/096901}}.

\bibitem[Clifton \em{et~al.}(2012)Clifton, Ferreira, Padilla, and
  Skordis]{Clifton:2011jh}
Clifton, T.; Ferreira, P.G.; Padilla, A.; Skordis, C.
\newblock {Modified Gravity and Cosmology}.
\newblock {\em Phys. Rept.} {\bf 2012}, {\em 513},~1--189,
  \href{http://xxx.lanl.gov/abs/1106.2476}{{\normalfont
  [arXiv:astro-ph.CO/1106.2476]}}.
\newblock {\url{https://doi.org/10.1016/j.physrep.2012.01.001}}.

\bibitem[Joyce \em{et~al.}(2015)Joyce, Jain, Khoury, and
  Trodden]{Joyce:2014kja}
Joyce, A.; Jain, B.; Khoury, J.; Trodden, M.
\newblock {Beyond the Cosmological Standard Model}.
\newblock {\em Phys. Rept.} {\bf 2015}, {\em 568},~1--98,
  \href{http://xxx.lanl.gov/abs/1407.0059}{{\normalfont
  [arXiv:astro-ph.CO/1407.0059]}}.
\newblock {\url{https://doi.org/10.1016/j.physrep.2014.12.002}}.

\bibitem[Koyama(2016)]{Koyama:2015vza}
Koyama, K.
\newblock {Cosmological Tests of Modified Gravity}.
\newblock {\em Rept. Prog. Phys.} {\bf 2016}, {\em 79},~046902,
  \href{http://xxx.lanl.gov/abs/1504.04623}{{\normalfont
  [arXiv:astro-ph.CO/1504.04623]}}.
\newblock {\url{https://doi.org/10.1088/0034-4885/79/4/046902}}.

\bibitem[Ade \em{et~al.}(2016)Ade et~al.]{Ade:2015rim}
Ade, P.A.R.;  et~al.
\newblock {Planck 2015 results. XIV. Dark energy and modified gravity}.
\newblock {\em Astron. Astrophys.} {\bf 2016}, {\em 594},~A14,
  \href{http://xxx.lanl.gov/abs/1502.01590}{{\normalfont
  [arXiv:astro-ph.CO/1502.01590]}}.
\newblock {\url{https://doi.org/10.1051/0004-6361/201525814}}.

\bibitem[Pogosian and Silvestri(2016)]{Pogosian:2016pwr}
Pogosian, L.; Silvestri, A.
\newblock {What can cosmology tell us about gravity? Constraining Horndeski
  gravity with $\Sigma$ and $\mu$}.
\newblock {\em Phys. Rev.} {\bf 2016}, {\em D94},~104014,
  \href{http://xxx.lanl.gov/abs/1606.05339}{{\normalfont
  [arXiv:astro-ph.CO/1606.05339]}}.
\newblock {\url{https://doi.org/10.1103/PhysRevD.94.104014}}.

\bibitem[Will(1993)]{Will:1993ns}
Will, C.M.
\newblock {\em {Theory and experiment in gravitational physics}};  1993.

\bibitem[Khoury and Weltman(2004{\natexlab{a}})]{Khoury:2003rn}
Khoury, J.; Weltman, A.
\newblock {Chameleon cosmology}.
\newblock {\em Phys. Rev. D} {\bf 2004}, {\em 69},~044026,
  \href{http://xxx.lanl.gov/abs/astro-ph/0309411}{{\normalfont
  [astro-ph/0309411]}}.
\newblock {\url{https://doi.org/10.1103/PhysRevD.69.044026}}.

\bibitem[Khoury and Weltman(2004{\natexlab{b}})]{Khoury:2003aq}
Khoury, J.; Weltman, A.
\newblock {Chameleon fields: Awaiting surprises for tests of gravity in space}.
\newblock {\em Phys. Rev. Lett.} {\bf 2004}, {\em 93},~171104,
  \href{http://xxx.lanl.gov/abs/astro-ph/0309300}{{\normalfont
  [astro-ph/0309300]}}.
\newblock {\url{https://doi.org/10.1103/PhysRevLett.93.171104}}.

\bibitem[Brax \em{et~al.}(2004)Brax, van~de Bruck, Davis, Khoury, and
  Weltman]{Brax:2004qh}
Brax, P.; van~de Bruck, C.; Davis, A.C.; Khoury, J.; Weltman, A.
\newblock {Detecting dark energy in orbit: The cosmological chameleon}.
\newblock {\em Phys. Rev. D} {\bf 2004}, {\em 70},~123518,
  \href{http://xxx.lanl.gov/abs/astro-ph/0408415}{{\normalfont
  [astro-ph/0408415]}}.
\newblock {\url{https://doi.org/10.1103/PhysRevD.70.123518}}.

\bibitem[Hinterbichler and Khoury(2010)]{Hinterbichler:2010es}
Hinterbichler, K.; Khoury, J.
\newblock {Symmetron Fields: Screening Long-Range Forces Through Local Symmetry
  Restoration}.
\newblock {\em Phys. Rev. Lett.} {\bf 2010}, {\em 104},~231301,
  \href{http://xxx.lanl.gov/abs/1001.4525}{{\normalfont
  [arXiv:hep-th/1001.4525]}}.
\newblock {\url{https://doi.org/10.1103/PhysRevLett.104.231301}}.

\bibitem[Hinterbichler \em{et~al.}(2011)Hinterbichler, Khoury, Levy, and
  Matas]{Hinterbichler:2011ca}
Hinterbichler, K.; Khoury, J.; Levy, A.; Matas, A.
\newblock {Symmetron Cosmology}.
\newblock {\em Phys. Rev. D} {\bf 2011}, {\em 84},~103521,
  \href{http://xxx.lanl.gov/abs/1107.2112}{{\normalfont
  [arXiv:astro-ph.CO/1107.2112]}}.
\newblock {\url{https://doi.org/10.1103/PhysRevD.84.103521}}.

\bibitem[{Vainshtein}(1972)]{Vainshtein_1972}
{Vainshtein}, A.I.
\newblock {To the problem of nonvanishing gravitation mass}.
\newblock {\em Physics Letters B} {\bf 1972}, {\em 39},~393--394.
\newblock {\url{https://doi.org/10.1016/0370-2693(72)90147-5}}.

\bibitem[{Nicolis} \em{et~al.}(2009){Nicolis}, {Rattazzi}, and
  {Trincherini}]{Nicolis_2009}
{Nicolis}, A.; {Rattazzi}, R.; {Trincherini}, E.
\newblock {Galileon as a local modification of gravity}.
\newblock {\em \prd} {\bf 2009}, {\em 79},~064036.
\newblock {\url{https://doi.org/10.1103/PhysRevD.79.064036}}.

\bibitem[Babichev and Deffayet(2013)]{Babichev:2013usa}
Babichev, E.; Deffayet, C.
\newblock {An introduction to the Vainshtein mechanism}.
\newblock {\em Class. Quant. Grav.} {\bf 2013}, {\em 30},~184001,
  \href{http://xxx.lanl.gov/abs/1304.7240}{{\normalfont
  [arXiv:gr-qc/1304.7240]}}.
\newblock {\url{https://doi.org/10.1088/0264-9381/30/18/184001}}.

\bibitem[Burrage and Sakstein(2016)]{Burrage:2016bwy}
Burrage, C.; Sakstein, J.
\newblock {A Compendium of Chameleon Constraints}.
\newblock {\em JCAP} {\bf 2016}, {\em 11},~045,
  \href{http://xxx.lanl.gov/abs/1609.01192}{{\normalfont
  [arXiv:astro-ph.CO/1609.01192]}}.
\newblock {\url{https://doi.org/10.1088/1475-7516/2016/11/045}}.

\bibitem[Burrage and Sakstein(2018)]{Burrage:2017qrf}
Burrage, C.; Sakstein, J.
\newblock {Tests of Chameleon Gravity}.
\newblock {\em Living Rev. Rel.} {\bf 2018}, {\em 21},~1,
  \href{http://xxx.lanl.gov/abs/1709.09071}{{\normalfont
  [arXiv:astro-ph.CO/1709.09071]}}.
\newblock {\url{https://doi.org/10.1007/s41114-018-0011-x}}.

\bibitem[Sakstein(2018)]{Sakstein:2018fwz}
Sakstein, J.
\newblock {Astrophysical tests of screened modified gravity}.
\newblock {\em Int. J. Mod. Phys. D} {\bf 2018}, {\em 27},~1848008,
  \href{http://xxx.lanl.gov/abs/2002.04194}{{\normalfont
  [arXiv:astro-ph.CO/2002.04194]}}.
\newblock {\url{https://doi.org/10.1142/S0218271818480085}}.

\bibitem[Baker \em{et~al.}(2021)Baker et~al.]{Baker:2019gxo}
Baker, T.;  et~al.
\newblock {Novel Probes Project: Tests of gravity on astrophysical scales}.
\newblock {\em Rev. Mod. Phys.} {\bf 2021}, {\em 93},~015003,
  \href{http://xxx.lanl.gov/abs/1908.03430}{{\normalfont
  [arXiv:astro-ph.CO/1908.03430]}}.
\newblock {\url{https://doi.org/10.1103/RevModPhys.93.015003}}.

\bibitem[Brax \em{et~al.}(2021)Brax, Casas, Desmond, and Elder]{Brax:2021wcv}
Brax, P.; Casas, S.; Desmond, H.; Elder, B.
\newblock {Testing Screened Modified Gravity}.
\newblock {\em Universe} {\bf 2021}, {\em 8},~11,
  \href{http://xxx.lanl.gov/abs/2201.10817}{{\normalfont
  [arXiv:gr-qc/2201.10817]}}.
\newblock {\url{https://doi.org/10.3390/universe8010011}}.

\bibitem[Elder \em{et~al.}(2020)Elder, Vardanyan, Akrami, Brax, Davis, and
  Decca]{Elder:2019yyp}
Elder, B.; Vardanyan, V.; Akrami, Y.; Brax, P.; Davis, A.C.; Decca, R.S.
\newblock {Classical symmetron force in Casimir experiments}.
\newblock {\em Phys. Rev. D} {\bf 2020}, {\em 101},~064065,
  \href{http://xxx.lanl.gov/abs/1912.10015}{{\normalfont
  [arXiv:gr-qc/1912.10015]}}.
\newblock {\url{https://doi.org/10.1103/PhysRevD.101.064065}}.

\bibitem[{Babichev} \em{et~al.}(2009){Babichev}, {Deffayet}, and
  {Ziour}]{Babichev_2009}
{Babichev}, E.; {Deffayet}, C.; {Ziour}, R.
\newblock {k-MOUFLAGE Gravity}.
\newblock {\em International Journal of Modern Physics D} {\bf 2009}, {\em
  18},~2147--2154.
\newblock {\url{https://doi.org/10.1142/S0218271809016107}}.

\bibitem[{Dvali} \em{et~al.}(2000){Dvali}, {Gabadadze}, and
  {Porrati}]{DGP_2000}
{Dvali}, G.; {Gabadadze}, G.; {Porrati}, M.
\newblock {4D gravity on a brane in 5D Minkowski space}.
\newblock {\em Physics Letters B} {\bf 2000}, {\em 485},~208--214.
\newblock {\url{https://doi.org/10.1016/S0370-2693(00)00669-9}}.

\bibitem[{Horndeski}(1974)]{Horndeski_1974}
{Horndeski}, G.W.
\newblock {Second-Order Scalar-Tensor Field Equations in a Four-Dimensional
  Space}.
\newblock {\em International Journal of Theoretical Physics} {\bf 1974}, {\em
  10},~363--384.
\newblock {\url{https://doi.org/10.1007/BF01807638}}.

\bibitem[{Babichev} and {Esposito-Far{\`e}se}(2013)]{Babichev_2013}
{Babichev}, E.; {Esposito-Far{\`e}se}, G.
\newblock {Time-dependent spherically symmetric covariant Galileons}.
\newblock {\em \prd} {\bf 2013}, {\em 87},~044032.
\newblock {\url{https://doi.org/10.1103/PhysRevD.87.044032}}.

\bibitem[{de Felice} and {Tsujikawa}(2010)]{deFelice_2010}
{de Felice}, A.; {Tsujikawa}, S.
\newblock {Cosmology of a Covariant Galileon Field}.
\newblock {\em \prl} {\bf 2010}, {\em 105},~111301.
\newblock {\url{https://doi.org/10.1103/PhysRevLett.105.111301}}.

\bibitem[{Renk} \em{et~al.}(2017){Renk}, {Zumalac{\'a}rregui}, {Montanari}, and
  {Barreira}]{Renk_2017}
{Renk}, J.; {Zumalac{\'a}rregui}, M.; {Montanari}, F.; {Barreira}, A.
\newblock {Galileon gravity in light of ISW, CMB, BAO and H$_{0}$ data}.
\newblock {\em \jcap} {\bf 2017}, {\em 2017},~020.
\newblock {\url{https://doi.org/10.1088/1475-7516/2017/10/020}}.

\bibitem[{LIGO Scientific Collaboration} and {Virgo
  Collaboration}(2017)]{GW170817}
{LIGO Scientific Collaboration}.; {Virgo Collaboration}.
\newblock {GW170817: Observation of Gravitational Waves from a Binary Neutron
  Star Inspiral}.
\newblock {\em \prl} {\bf 2017}, {\em 119},~161101.
\newblock {\url{https://doi.org/10.1103/PhysRevLett.119.161101}}.

\bibitem[{Ezquiaga} and {Zumalac{\'a}rregui}(2017)]{Ezquiaga_2017}
{Ezquiaga}, J.M.; {Zumalac{\'a}rregui}, M.
\newblock {Dark Energy After GW170817: Dead Ends and the Road Ahead}.
\newblock {\em \prl} {\bf 2017}, {\em 119},~251304.
\newblock {\url{https://doi.org/10.1103/PhysRevLett.119.251304}}.

\bibitem[{Barreira} \em{et~al.}(2013){Barreira}, {Li}, {Hellwing}, {Baugh}, and
  {Pascoli}]{Barreria_2013}
{Barreira}, A.; {Li}, B.; {Hellwing}, W.A.; {Baugh}, C.M.; {Pascoli}, S.
\newblock {Nonlinear structure formation in the cubic Galileon gravity model}.
\newblock {\em \jcap} {\bf 2013}, {\em 2013},~027.
\newblock {\url{https://doi.org/10.1088/1475-7516/2013/10/027}}.

\bibitem[{Hui} \em{et~al.}(2009){Hui}, {Nicolis}, and {Stubbs}]{Hui_2009}
{Hui}, L.; {Nicolis}, A.; {Stubbs}, C.W.
\newblock {Equivalence principle implications of modified gravity models}.
\newblock {\em \prd} {\bf 2009}, {\em 80},~104002,
  \href{http://xxx.lanl.gov/abs/0905.2966}{{\normalfont
  [arXiv:astro-ph.CO/0905.2966]}}.
\newblock {\url{https://doi.org/10.1103/PhysRevD.80.104002}}.

\bibitem[{Hui} and {Nicolis}(2012)]{Hui_2012}
{Hui}, L.; {Nicolis}, A.
\newblock {Proposal for an Observational Test of the Vainshtein Mechanism}.
\newblock {\em \prl} {\bf 2012}, {\em 109},~051304.
\newblock {\url{https://doi.org/10.1103/PhysRevLett.109.051304}}.

\bibitem[{Hui} and {Nicolis}(2013)]{Hui_2013}
{Hui}, L.; {Nicolis}, A.
\newblock {No-Hair Theorem for the Galileon}.
\newblock {\em \prl} {\bf 2013}, {\em 110},~241104,
  \href{http://xxx.lanl.gov/abs/1202.1296}{{\normalfont
  [arXiv:hep-th/1202.1296]}}.
\newblock {\url{https://doi.org/10.1103/PhysRevLett.110.241104}}.

\bibitem[{Schmidt} \em{et~al.}(2010){Schmidt}, {Hu}, and {Lima}]{Schmidt_2010}
{Schmidt}, F.; {Hu}, W.; {Lima}, M.
\newblock {Spherical collapse and the halo model in braneworld gravity}.
\newblock {\em \prd} {\bf 2010}, {\em 81},~063005,
  \href{http://xxx.lanl.gov/abs/0911.5178}{{\normalfont
  [arXiv:astro-ph.CO/0911.5178]}}.
\newblock {\url{https://doi.org/10.1103/PhysRevD.81.063005}}.

\bibitem[{Sakstein} \em{et~al.}(2017){Sakstein}, {Jain}, {Heyl}, and
  {Hui}]{Sakstein_2017}
{Sakstein}, J.; {Jain}, B.; {Heyl}, J.S.; {Hui}, L.
\newblock {Tests of Gravity Theories Using Supermassive Black Holes}.
\newblock {\em \apjl} {\bf 2017}, {\em 844},~L14.
\newblock {\url{https://doi.org/10.3847/2041-8213/aa7e26}}.

\bibitem[Falck \em{et~al.}(2014)Falck, Koyama, Zhao, and Li]{Falck_2014}
Falck, B.; Koyama, K.; Zhao, G.b.; Li, B.
\newblock {The Vainshtein Mechanism in the Cosmic Web}.
\newblock {\em JCAP} {\bf 2014}, {\em 07},~058.
\newblock {\url{https://doi.org/10.1088/1475-7516/2014/07/058}}.

\bibitem[{Cardoso} \em{et~al.}(2008){Cardoso}, {Koyama}, {Seahra}, and
  {Silva}]{Cardoso_2008}
{Cardoso}, A.; {Koyama}, K.; {Seahra}, S.S.; {Silva}, F.P.
\newblock {Cosmological perturbations in the DGP braneworld: Numeric solution}.
\newblock {\em \prd} {\bf 2008}, {\em 77},~083512.
\newblock {\url{https://doi.org/10.1103/PhysRevD.77.083512}}.

\bibitem[{Schmidt}(2009)]{Schmidt_2009}
{Schmidt}, F.
\newblock {Self-consistent cosmological simulations of DGP braneworld gravity}.
\newblock {\em \prd} {\bf 2009}, {\em 80},~043001.
\newblock {\url{https://doi.org/10.1103/PhysRevD.80.043001}}.

\bibitem[{Khoury} and {Wyman}(2009)]{Khoury_2009}
{Khoury}, J.; {Wyman}, M.
\newblock {N-body simulations of DGP and degravitation theories}.
\newblock {\em \prd} {\bf 2009}, {\em 80},~064023.
\newblock {\url{https://doi.org/10.1103/PhysRevD.80.064023}}.

\bibitem[{Chan} and {Scoccimarro}(2009)]{Chan_2009}
{Chan}, K.C.; {Scoccimarro}, R.
\newblock {Large-scale structure in brane-induced gravity. II. Numerical
  simulations}.
\newblock {\em \prd} {\bf 2009}, {\em 80},~104005.
\newblock {\url{https://doi.org/10.1103/PhysRevD.80.104005}}.

\bibitem[Davis \em{et~al.}(2012)Davis, Li, Mota, and Winther]{Davis:2011pj}
Davis, A.C.; Li, B.; Mota, D.F.; Winther, H.A.
\newblock {Structure Formation in the Symmetron Model}.
\newblock {\em Astrophys. J.} {\bf 2012}, {\em 748},~61,
  \href{http://xxx.lanl.gov/abs/1108.3081}{{\normalfont
  [arXiv:astro-ph.CO/1108.3081]}}.
\newblock {\url{https://doi.org/10.1088/0004-637X/748/1/61}}.

\bibitem[Clampitt \em{et~al.}(2012)Clampitt, Jain, and Khoury]{Clampitt:2011mx}
Clampitt, J.; Jain, B.; Khoury, J.
\newblock {Halo Scale Predictions of Symmetron Modified Gravity}.
\newblock {\em JCAP} {\bf 2012}, {\em 1201},~030,
  \href{http://xxx.lanl.gov/abs/1110.2177}{{\normalfont
  [arXiv:astro-ph.CO/1110.2177]}}.
\newblock {\url{https://doi.org/10.1088/1475-7516/2012/01/030}}.

\bibitem[Brax \em{et~al.}(2012)Brax, Davis, Li, Winther, and Zhao]{Brax:2012nk}
Brax, P.; Davis, A.C.; Li, B.; Winther, H.A.; Zhao, G.B.
\newblock {Systematic Simulations of Modified Gravity: Symmetron and Dilaton
  Models}.
\newblock {\em JCAP} {\bf 2012}, {\em 1210},~002,
  \href{http://xxx.lanl.gov/abs/1206.3568}{{\normalfont
  [arXiv:astro-ph.CO/1206.3568]}}.
\newblock {\url{https://doi.org/10.1088/1475-7516/2012/10/002}}.

\bibitem[Llinares and Mota(2014)]{Llinares:2013jua}
Llinares, C.; Mota, D.F.
\newblock {Cosmological simulations of screened modified gravity out of the
  static approximation: effects on matter distribution}.
\newblock {\em Phys. Rev.} {\bf 2014}, {\em D89},~084023,
  \href{http://xxx.lanl.gov/abs/1312.6016}{{\normalfont
  [arXiv:astro-ph.CO/1312.6016]}}.
\newblock {\url{https://doi.org/10.1103/PhysRevD.89.084023}}.

\bibitem[Noller \em{et~al.}(2014)Noller, von Braun-Bates, and
  Ferreira]{Noller:2013wca}
Noller, J.; von Braun-Bates, F.; Ferreira, P.G.
\newblock {Relativistic scalar fields and the quasistatic approximation in
  theories of modified gravity}.
\newblock {\em Phys. Rev.} {\bf 2014}, {\em D89},~023521,
  \href{http://xxx.lanl.gov/abs/1310.3266}{{\normalfont
  [arXiv:astro-ph.CO/1310.3266]}}.
\newblock {\url{https://doi.org/10.1103/PhysRevD.89.023521}}.

\bibitem[Vardanyan(2019)]{Vardanyan:2019gjt}
Vardanyan, V.
\newblock {Aspects of cosmic acceleration}.
\newblock PhD thesis, Leiden U.,  2019.

\bibitem[Contigiani \em{et~al.}(2019)Contigiani, Vardanyan, and
  Silvestri]{Contigiani:2018hbn}
Contigiani, O.; Vardanyan, V.; Silvestri, A.
\newblock {Splashback radius in symmetron gravity}.
\newblock {\em Phys. Rev. D} {\bf 2019}, {\em 99},~064030,
  \href{http://xxx.lanl.gov/abs/1812.05568}{{\normalfont
  [arXiv:astro-ph.CO/1812.05568]}}.
\newblock {\url{https://doi.org/10.1103/PhysRevD.99.064030}}.

\bibitem[Upadhye(2013)]{Upadhye:2012rc}
Upadhye, A.
\newblock {Symmetron dark energy in laboratory experiments}.
\newblock {\em Phys. Rev. Lett.} {\bf 2013}, {\em 110},~031301,
  \href{http://xxx.lanl.gov/abs/1210.7804}{{\normalfont
  [arXiv:hep-ph/1210.7804]}}.
\newblock {\url{https://doi.org/10.1103/PhysRevLett.110.031301}}.

\bibitem[{Li} \em{et~al.}(2012){Li}, {Zhao}, {Teyssier}, and {Koyama}]{Li_2012}
{Li}, B.; {Zhao}, G.B.; {Teyssier}, R.; {Koyama}, K.
\newblock {ECOSMOG: an Efficient COde for Simulating MOdified Gravity}.
\newblock {\em \jcap} {\bf 2012}, {\em 2012},~051,
  \href{http://xxx.lanl.gov/abs/1110.1379}{{\normalfont
  [arXiv:astro-ph.CO/1110.1379]}}.
\newblock {\url{https://doi.org/10.1088/1475-7516/2012/01/051}}.

\bibitem[Llinares \em{et~al.}(2014)Llinares, Mota, and
  Winther]{Llinares:2013jza}
Llinares, C.; Mota, D.F.; Winther, H.A.
\newblock {ISIS: a new N-body cosmological code with scalar fields based on
  RAMSES. Code presentation and application to the shapes of clusters}.
\newblock {\em Astron. Astrophys.} {\bf 2014}, {\em 562},~A78,
  \href{http://xxx.lanl.gov/abs/1307.6748}{{\normalfont
  [arXiv:astro-ph.CO/1307.6748]}}.
\newblock {\url{https://doi.org/10.1051/0004-6361/201322412}}.

\bibitem[Puchwein \em{et~al.}(2013)Puchwein, Baldi, and
  Springel]{Puchwein:2013lza}
Puchwein, E.; Baldi, M.; Springel, V.
\newblock {Modified Gravity-GADGET: A new code for cosmological hydrodynamical
  simulations of modified gravity models}.
\newblock {\em Mon. Not. Roy. Astron. Soc.} {\bf 2013}, {\em 436},~348,
  \href{http://xxx.lanl.gov/abs/1305.2418}{{\normalfont
  [arXiv:astro-ph.CO/1305.2418]}}.
\newblock {\url{https://doi.org/10.1093/mnras/stt1575}}.

\bibitem[Winther \em{et~al.}(2015)Winther et~al.]{Winther:2015wla}
Winther, H.A.;  et~al.
\newblock {Modified Gravity N-body Code Comparison Project}.
\newblock {\em Mon. Not. Roy. Astron. Soc.} {\bf 2015}, {\em 454},~4208--4234,
  \href{http://xxx.lanl.gov/abs/1506.06384}{{\normalfont
  [arXiv:astro-ph.CO/1506.06384]}}.
\newblock {\url{https://doi.org/10.1093/mnras/stv2253}}.

\bibitem[{Briddon} \em{et~al.}(2021){Briddon}, {Burrage}, {Moss}, and
  {Tamosiunas}]{Briddon_2021}
{Briddon}, C.; {Burrage}, C.; {Moss}, A.; {Tamosiunas}, A.
\newblock {SELCIE: a tool for investigating the chameleon field of arbitrary
  sources}.
\newblock {\em \jcap} {\bf 2021}, {\em 2021},~043,
  \href{http://xxx.lanl.gov/abs/2110.11917}{{\normalfont
  [arXiv:gr-qc/2110.11917]}}.
\newblock {\url{https://doi.org/10.1088/1475-7516/2021/12/043}}.

\bibitem[Alnaes \em{et~al.}(2015)Alnaes, Blechta, Hake, A.~Johansson, Logg,
  Richardson, J.~Ring, and Wells]{AlnaesEtal2015}
Alnaes, M.S.; Blechta, J.; Hake, J.; A.~Johansson, B.K.; Logg, A.; Richardson,
  C.; J.~Ring, M.E.R.; Wells, G.N.
\newblock The {FEniCS} Project Version 1.5.
\newblock {\em Archive of Numerical Software} {\bf 2015}, {\em 3}.
\newblock {\url{https://doi.org/10.11588/ans.2015.100.20553}}.

\bibitem[A.~Logg(2012)]{LoggEtal2012}
A.~Logg, {K.-A.}~Mardal, G.N.W.e.a.
\newblock {\em Automated Solution of Differential Equations by the Finite
  Element Method}; Springer,  2012.
\newblock {\url{https://doi.org/10.1007/978-3-642-23099-8}}.

\bibitem[{Cimrman} \em{et~al.}(2018){Cimrman}, {Luke{\v{s}}}, and
  {Rohan}]{Cimrman2018}
{Cimrman}, R.; {Luke{\v{s}}}, V.; {Rohan}, E.
\newblock {Multiscale finite element calculations in Python using SfePy}.
\newblock {\em arXiv e-prints} {\bf 2018}, p. arXiv:1810.00674,
  \href{http://xxx.lanl.gov/abs/1810.00674}{{\normalfont
  [arXiv:cs.MS/1810.00674]}}.
\newblock {\url{https://doi.org/10.48550/arXiv.1810.00674}}.

\bibitem[{Braden} \em{et~al.}(2021){Braden}, {Burrage}, {Elder}, and
  {Saadeh}]{Braden_2021}
{Braden}, J.; {Burrage}, C.; {Elder}, B.; {Saadeh}, D.
\newblock {{\ensuremath{\varphi}}enics: Vainshtein screening with the finite
  element method}.
\newblock {\em \jcap} {\bf 2021}, {\em 2021},~010,
  \href{http://xxx.lanl.gov/abs/2011.07037}{{\normalfont
  [arXiv:gr-qc/2011.07037]}}.
\newblock {\url{https://doi.org/10.1088/1475-7516/2021/03/010}}.

\bibitem[{Tamosiunas} \em{et~al.}(2022{\natexlab{a}}){Tamosiunas}, {Briddon},
  {Burrage}, {Cutforth}, {Moss}, and {Vincent}]{Tamosiunas_2022_voids}
{Tamosiunas}, A.; {Briddon}, C.; {Burrage}, C.; {Cutforth}, A.; {Moss}, A.;
  {Vincent}, T.
\newblock {Chameleon screening in cosmic voids}.
\newblock {\em \jcap} {\bf 2022}, {\em 2022},~056,
  \href{http://xxx.lanl.gov/abs/2206.06480}{{\normalfont
  [arXiv:gr-qc/2206.06480]}}.
\newblock {\url{https://doi.org/10.1088/1475-7516/2022/11/056}}.

\bibitem[{Tamosiunas} \em{et~al.}(2022{\natexlab{b}}){Tamosiunas}, {Briddon},
  {Burrage}, {Cui}, and {Moss}]{Tamosiunas_2022_halos}
{Tamosiunas}, A.; {Briddon}, C.; {Burrage}, C.; {Cui}, W.; {Moss}, A.
\newblock {Chameleon screening depends on the shape and structure of NFW
  halos}.
\newblock {\em \jcap} {\bf 2022}, {\em 2022},~047,
  \href{http://xxx.lanl.gov/abs/2108.10364}{{\normalfont
  [arXiv:gr-qc/2108.10364]}}.
\newblock {\url{https://doi.org/10.1088/1475-7516/2022/04/047}}.

\bibitem[L\'evy \em{et~al.}(2022)L\'evy, Berg\'e, and Uzan]{Levy:2022xni}
L\'evy, H.; Berg\'e, J.; Uzan, J.P.
\newblock {Solving nonlinear Klein-Gordon equations on unbounded domains via
  the finite element method}.
\newblock {\em Phys. Rev. D} {\bf 2022}, {\em 106},~124021,
  \href{http://xxx.lanl.gov/abs/2209.07226}{{\normalfont
  [arXiv:gr-qc/2209.07226]}}.
\newblock {\url{https://doi.org/10.1103/PhysRevD.106.124021}}.

\bibitem[{Grosch} and {Orszag}(1977)]{Grosch1977}
{Grosch}, C.E.; {Orszag}, S.A.
\newblock {Numerical Solution of Problems in Unbounded Regions: Coordinate
  Transforms}.
\newblock {\em Journal of Computational Physics} {\bf 1977}, {\em
  25},~273--295.
\newblock {\url{https://doi.org/10.1016/0021-9991(77)90102-4}}.

\bibitem[Burrage \em{et~al.}(2018)Burrage, Copeland, Moss, and
  Stevenson]{Burrage:2017shh}
Burrage, C.; Copeland, E.J.; Moss, A.; Stevenson, J.A.
\newblock {The shape dependence of chameleon screening}.
\newblock {\em JCAP} {\bf 2018}, {\em 01},~056,
  \href{http://xxx.lanl.gov/abs/1711.02065}{{\normalfont
  [arXiv:astro-ph.CO/1711.02065]}}.
\newblock {\url{https://doi.org/10.1088/1475-7516/2018/01/056}}.

\bibitem[Burrage \em{et~al.}(2015)Burrage, Copeland, and
  Hinds]{Burrage:2014oza}
Burrage, C.; Copeland, E.J.; Hinds, E.A.
\newblock {Probing Dark Energy with Atom Interferometry}.
\newblock {\em JCAP} {\bf 2015}, {\em 03},~042,
  \href{http://xxx.lanl.gov/abs/1408.1409}{{\normalfont
  [arXiv:astro-ph.CO/1408.1409]}}.
\newblock {\url{https://doi.org/10.1088/1475-7516/2015/03/042}}.

\bibitem[Burrage and Copeland(2016)]{Burrage:2015lya}
Burrage, C.; Copeland, E.J.
\newblock {Using Atom Interferometry to Detect Dark Energy}.
\newblock {\em Contemp. Phys.} {\bf 2016}, {\em 57},~164--176,
  \href{http://xxx.lanl.gov/abs/1507.07493}{{\normalfont
  [arXiv:astro-ph.CO/1507.07493]}}.
\newblock {\url{https://doi.org/10.1080/00107514.2015.1060058}}.

\bibitem[Hamilton \em{et~al.}(2015)Hamilton, Jaffe, Haslinger, Simmons,
  M\"uller, and Khoury]{Hamilton:2015zga}
Hamilton, P.; Jaffe, M.; Haslinger, P.; Simmons, Q.; M\"uller, H.; Khoury, J.
\newblock {Atom-interferometry constraints on dark energy}.
\newblock {\em Science} {\bf 2015}, {\em 349},~849--851,
  \href{http://xxx.lanl.gov/abs/1502.03888}{{\normalfont
  [arXiv:physics.atom-ph/1502.03888]}}.
\newblock {\url{https://doi.org/10.1126/science.aaa8883}}.

\bibitem[Burrage \em{et~al.}(2016)Burrage, Kuribayashi-Coleman, Stevenson, and
  Thrussell]{Burrage:2016rkv}
Burrage, C.; Kuribayashi-Coleman, A.; Stevenson, J.; Thrussell, B.
\newblock {Constraining symmetron fields with atom interferometry}.
\newblock {\em JCAP} {\bf 2016}, {\em 1612},~041,
  \href{http://xxx.lanl.gov/abs/1609.09275}{{\normalfont
  [arXiv:astro-ph.CO/1609.09275]}}.
\newblock {\url{https://doi.org/10.1088/1475-7516/2016/12/041}}.

\bibitem[Jaffe \em{et~al.}(2017)Jaffe, Haslinger, Xu, Hamilton, Upadhye, Elder,
  Khoury, and M\"uller]{Jaffe:2016fsh}
Jaffe, M.; Haslinger, P.; Xu, V.; Hamilton, P.; Upadhye, A.; Elder, B.; Khoury,
  J.; M\"uller, H.
\newblock {Testing sub-gravitational forces on atoms from a miniature,
  in-vacuum source mass}.
\newblock {\em Nature Phys.} {\bf 2017}, {\em 13},~938,
  \href{http://xxx.lanl.gov/abs/1612.05171}{{\normalfont
  [arXiv:physics.atom-ph/1612.05171]}}.
\newblock {\url{https://doi.org/10.1038/nphys4189}}.

\bibitem[Sabulsky \em{et~al.}(2019)Sabulsky, Dutta, Hinds, Elder, Burrage, and
  Copeland]{Sabulsky:2018jma}
Sabulsky, D.O.; Dutta, I.; Hinds, E.A.; Elder, B.; Burrage, C.; Copeland, E.J.
\newblock {Experiment to detect dark energy forces using atom interferometry}.
\newblock {\em Phys. Rev. Lett.} {\bf 2019}, {\em 123},~061102,
  \href{http://xxx.lanl.gov/abs/1812.08244}{{\normalfont
  [arXiv:physics.atom-ph/1812.08244]}}.
\newblock {\url{https://doi.org/10.1103/PhysRevLett.123.061102}}.

\bibitem[Brax and Burrage(2011)]{Brax:2010gp}
Brax, P.; Burrage, C.
\newblock {Atomic Precision Tests and Light Scalar Couplings}.
\newblock {\em Phys. Rev. D} {\bf 2011}, {\em 83},~035020,
  \href{http://xxx.lanl.gov/abs/1010.5108}{{\normalfont
  [arXiv:hep-ph/1010.5108]}}.
\newblock {\url{https://doi.org/10.1103/PhysRevD.83.035020}}.

\bibitem[Brax \em{et~al.}(2023)Brax, Davis, and Elder]{Brax:2022olf}
Brax, P.; Davis, A.C.; Elder, B.
\newblock {Screened scalar fields in hydrogen and muonium}.
\newblock {\em Phys. Rev. D} {\bf 2023}, {\em 107},~044008,
  \href{http://xxx.lanl.gov/abs/2207.11633}{{\normalfont
  [arXiv:hep-ph/2207.11633]}}.
\newblock {\url{https://doi.org/10.1103/PhysRevD.107.044008}}.

\bibitem[Cronenberg \em{et~al.}(2018)Cronenberg, Brax, Filter, Geltenbort,
  Jenke, Pignol, Pitschmann, Thalhammer, and Abele]{Cronenberg:2018qxf}
Cronenberg, G.; Brax, P.; Filter, H.; Geltenbort, P.; Jenke, T.; Pignol, G.;
  Pitschmann, M.; Thalhammer, M.; Abele, H.
\newblock {Acoustic Rabi oscillations between gravitational quantum states and
  impact on symmetron dark energy}.
\newblock {\em Nature Phys.} {\bf 2018}, {\em 14},~1022--1026,
  \href{http://xxx.lanl.gov/abs/1902.08775}{{\normalfont
  [arXiv:hep-ph/1902.08775]}}.
\newblock {\url{https://doi.org/10.1038/s41567-018-0205-x}}.

\bibitem[Decca \em{et~al.}(2005)Decca, Lopez, Chan, Fischbach, Krause, and
  Jamell]{Decca:2005qz}
Decca, R.S.; Lopez, D.; Chan, H.B.; Fischbach, E.; Krause, D.E.; Jamell, C.R.
\newblock {Constraining new forces in the Casimir regime using the
  isoelectronic technique}.
\newblock {\em Phys. Rev. Lett.} {\bf 2005}, {\em 94},~240401,
  \href{http://xxx.lanl.gov/abs/hep-ph/0502025}{{\normalfont
  [arXiv:hep-ph/hep-ph/0502025]}}.
\newblock {\url{https://doi.org/10.1103/PhysRevLett.94.240401}}.

\bibitem[Decca \em{et~al.}(2003)Decca, Lopez, Fischbach, and
  Krause]{Decca:2003zz}
Decca, R.S.; Lopez, D.; Fischbach, E.; Krause, D.E.
\newblock {Measurement of the Casimir Force between Dissimilar Metals}.
\newblock {\em Phys. Rev. Lett.} {\bf 2003}, {\em 91},~050402,
  \href{http://xxx.lanl.gov/abs/quant-ph/0306136}{{\normalfont
  [arXiv:quant-ph/quant-ph/0306136]}}.
\newblock {\url{https://doi.org/10.1103/PhysRevLett.91.050402}}.

\bibitem[Almasi \em{et~al.}(2015)Almasi, Brax, Iannuzzi, and
  Sedmik]{Almasi:2015zpa}
Almasi, A.; Brax, P.; Iannuzzi, D.; Sedmik, R.I.P.
\newblock {Force sensor for chameleon and Casimir force experiments with
  parallel-plate configuration}.
\newblock {\em Phys. Rev.} {\bf 2015}, {\em D91},~102002,
  \href{http://xxx.lanl.gov/abs/1505.01763}{{\normalfont
  [arXiv:physics.ins-det/1505.01763]}}.
\newblock {\url{https://doi.org/10.1103/PhysRevD.91.102002}}.

\bibitem[Sedmik and Brax(2018)]{Sedmik:2018kqt}
Sedmik, R.; Brax, P.
\newblock {Status Report and first Light from Cannex: Casimir Force
  Measurements between flat parallel Plates}.
\newblock {\em {Proceedings, 10th International Conference on Precision Physics
  of Simple Atomic Systems (PSAS 2018): Vienna, Austria, May 14-18, 2018, J.
  Phys. Conf. Ser.}} {\bf 2018}, {\em 1138},~012014.
\newblock {\url{https://doi.org/10.1088/1742-6596/1138/1/012014}}.

\bibitem[Brax \em{et~al.}(2010)Brax, van~de Bruck, Davis, Shaw, and
  Iannuzzi]{Brax:2010xx}
Brax, P.; van~de Bruck, C.; Davis, A.C.; Shaw, D.J.; Iannuzzi, D.
\newblock {Tuning the Mass of Chameleon Fields in Casimir Force Experiments}.
\newblock {\em Phys. Rev. Lett.} {\bf 2010}, {\em 104},~241101,
  \href{http://xxx.lanl.gov/abs/1003.1605}{{\normalfont
  [arXiv:quant-ph/1003.1605]}}.
\newblock {\url{https://doi.org/10.1103/PhysRevLett.104.241101}}.

\bibitem[Adelberger(2002)]{Adelberger:2002ic}
Adelberger, E.G.
\newblock {Sub-millimeter tests of the gravitational inverse square law}.
\newblock In Proceedings of the {2nd Meeting on CPT and Lorentz Symmetry},
  2002, pp. 9--15,  \href{http://xxx.lanl.gov/abs/hep-ex/0202008}{{\normalfont
  [hep-ex/0202008]}}.
\newblock {\url{https://doi.org/10.1142/9789812778123_0002}}.

\bibitem[Brax \em{et~al.}(2022)Brax, Davis, and Elder]{Brax:2022uiv}
Brax, P.; Davis, A.C.; Elder, B.
\newblock {Casimir Tests of Scalar-Tensor Theories} {\bf 2022}.
\newblock  \href{http://xxx.lanl.gov/abs/2211.07840}{{\normalfont
  [arXiv:gr-qc/2211.07840]}}.

\bibitem[Sparnaay(1958)]{Sparnaay1958}
Sparnaay, M.J.
\newblock {Measurements of attractive forces between flat plates}.
\newblock {\em Physica} {\bf 1958}, {\em 24},~751--764.
\newblock {\url{https://doi.org/10.1016/S0031-8914(58)80090-7}}.

\bibitem[van Blokland and Overbeek(1978)]{Blokland1978}
van Blokland, P.H.G.M.; Overbeek, J.T.G.
\newblock van der Waals forces between objects covered with a chromium layer.
\newblock {\em J. Chem. Soc.{,} Faraday Trans. 1} {\bf 1978}, {\em
  74},~2637--2651.
\newblock {\url{https://doi.org/10.1039/F19787402637}}.

\bibitem[Lamoreaux(1997)]{Lamoreaux:1996wh}
Lamoreaux, S.K.
\newblock {Demonstration of the Casimir force in the 0.6 to 6 micrometers
  range}.
\newblock {\em Phys. Rev. Lett.} {\bf 1997}, {\em 78},~5--8.
\newblock [Erratum: Phys. Rev. Lett.81,5475(1998)],
  {\url{https://doi.org/10.1103/PhysRevLett.81.5475,
  10.1103/PhysRevLett.78.5}}.

\bibitem[Decca \em{et~al.}(2005)Decca, Lopez, Fischbach, Klimchitskaya, Krause,
  and Mostepanenko]{Decca:2005yk}
Decca, R.S.; Lopez, D.; Fischbach, E.; Klimchitskaya, G.L.; Krause, D.E.;
  Mostepanenko, V.M.
\newblock {Precise comparison of theory and new experiment for the Casimir
  force leads to stronger constraints on thermal quantum effects and long-range
  interactions}.
\newblock {\em Annals Phys.} {\bf 2005}, {\em 318},~37--80,
  \href{http://xxx.lanl.gov/abs/quant-ph/0503105}{{\normalfont
  [arXiv:quant-ph/quant-ph/0503105]}}.
\newblock {\url{https://doi.org/10.1016/j.aop.2005.03.007}}.

\bibitem[Mohideen and Roy(1998)]{Mohideen:1998iz}
Mohideen, U.; Roy, A.
\newblock {Precision measurement of the Casimir force from 0.1 to 0.9
  micrometers}.
\newblock {\em Phys. Rev. Lett.} {\bf 1998}, {\em 81},~4549--4552,
  \href{http://xxx.lanl.gov/abs/physics/9805038}{{\normalfont
  [physics/9805038]}}.
\newblock {\url{https://doi.org/10.1103/PhysRevLett.81.4549}}.

\bibitem[Harris \em{et~al.}(2000)Harris, Chen, and Mohideen]{Harris:2000zz}
Harris, B.W.; Chen, F.; Mohideen, U.
\newblock {Precision measurement of the Casimir force using gold surfaces}.
\newblock {\em Phys. Rev. A} {\bf 2000}, {\em 62},~052109,
  \href{http://xxx.lanl.gov/abs/quant-ph/0005088}{{\normalfont
  [quant-ph/0005088]}}.
\newblock {\url{https://doi.org/10.1103/PhysRevA.62.052109}}.

\bibitem[Chen \em{et~al.}(2016)Chen, Tham, Krause, Lopez, Fischbach, and
  Decca]{Chen:2014oda}
Chen, Y.J.; Tham, W.K.; Krause, D.E.; Lopez, D.; Fischbach, E.; Decca, R.S.
\newblock {Stronger Limits on Hypothetical Yukawa Interactions in the 30-8000
  nm Range}.
\newblock {\em Phys. Rev. Lett.} {\bf 2016}, {\em 116},~221102,
  \href{http://xxx.lanl.gov/abs/1410.7267}{{\normalfont
  [arXiv:hep-ex/1410.7267]}}.
\newblock {\url{https://doi.org/10.1103/PhysRevLett.116.221102}}.

\bibitem[Bressi \em{et~al.}(2002)Bressi, Carugno, Onofrio, and
  Ruoso]{Bressi:2002fr}
Bressi, G.; Carugno, G.; Onofrio, R.; Ruoso, G.
\newblock {Measurement of the Casimir force between parallel metallic
  surfaces}.
\newblock {\em Phys. Rev. Lett.} {\bf 2002}, {\em 88},~041804,
  \href{http://xxx.lanl.gov/abs/quant-ph/0203002}{{\normalfont
  [quant-ph/0203002]}}.
\newblock {\url{https://doi.org/10.1103/PhysRevLett.88.041804}}.

\bibitem[{Zou} \em{et~al.}(2013){Zou}, {Marcet}, {Rodriguez}, {Reid},
  {McCauley}, {Kravchenko}, {Lu}, {Bao}, {Johnson}, and {Chan}]{Zou2013}
{Zou}, J.; {Marcet}, Z.; {Rodriguez}, A.W.; {Reid}, M.T.H.; {McCauley}, A.P.;
  {Kravchenko}, I.I.; {Lu}, T.; {Bao}, Y.; {Johnson}, S.G.; {Chan}, H.B.
\newblock {Casimir forces on a silicon micromechanical chip}.
\newblock {\em Nature Communications} {\bf 2013}, {\em 4},~1845,
  \href{http://xxx.lanl.gov/abs/1207.6163}{{\normalfont
  [arXiv:cond-mat.mes-hall/1207.6163]}}.
\newblock {\url{https://doi.org/10.1038/ncomms2842}}.

\bibitem[{Neto} \em{et~al.}(2005){Neto}, {Lambrecht}, and {Reynaud}]{Neto_2005}
{Neto}, P.A.M.; {Lambrecht}, A.; {Reynaud}, S.
\newblock {Roughness correction to the Casimir force: Beyond the Proximity
  Force Approximation}.
\newblock {\em EPL (Europhysics Letters)} {\bf 2005}, {\em 69},~924--930,
  \href{http://xxx.lanl.gov/abs/quant-ph/0410101}{{\normalfont
  [arXiv:quant-ph/quant-ph/0410101]}}.
\newblock {\url{https://doi.org/10.1209/epl/i2004-10433-9}}.

\bibitem[Rodrigues \em{et~al.}(2006)Rodrigues, Neto, Lambrecht, and
  Reynaud]{Rodrigues_2006}
Rodrigues, R.B.; Neto, P.A.M.; Lambrecht, A.; Reynaud, S.
\newblock Lateral Casimir Force beyond the Proximity-Force Approximation.
\newblock {\em Phys. Rev. Lett.} {\bf 2006}, {\em 96},~100402.
\newblock {\url{https://doi.org/10.1103/PhysRevLett.96.100402}}.

\bibitem[{Lambrecht} and {Reynaud}(2000)]{Lambrecht_2000}
{Lambrecht}, A.; {Reynaud}, S.
\newblock {Casimir force between metallic mirrors}.
\newblock {\em European Physical Journal D} {\bf 2000}, {\em 8},~309--318,
  \href{http://xxx.lanl.gov/abs/quant-ph/9907105}{{\normalfont
  [arXiv:quant-ph/quant-ph/9907105]}}.
\newblock {\url{https://doi.org/10.1007/s100530050041}}.

\bibitem[Reynaud and Lambrecht(2017)]{Reynaud_2017}
Reynaud, S.; Lambrecht, A.
\newblock {407Casimir forces and vacuum energy}. In {\em {Quantum Optics and
  Nanophotonics}}; Oxford University Press,  2017;
  \href{http://xxx.lanl.gov/abs/https://academic.oup.com/book/0/chapter/203961737/chapter-pdf/45353061/oso-9780198768609-chapter-9.pdf}{{\normalfont
  [https://academic.oup.com/book/0/chapter/203961737/chapter-pdf/45353061/oso-9780198768609-chapter-9.pdf]}}.
\newblock {\url{https://doi.org/10.1093/oso/9780198768609.003.0009}}.

\bibitem[Dalvit and Onofrio(2009)]{Dalvit:2009gw}
Dalvit, D.A.R.; Onofrio, R.
\newblock {On the use of the proximity force approximation for deriving limits
  to short-range gravitational-like interactions from sphere-plane Casimir
  force experiments}.
\newblock {\em Phys. Rev. D} {\bf 2009}, {\em 80},~064025,
  \href{http://xxx.lanl.gov/abs/0909.3068}{{\normalfont
  [arXiv:quant-ph/0909.3068]}}.
\newblock {\url{https://doi.org/10.1103/PhysRevD.80.064025}}.

\bibitem[Krause \em{et~al.}(2007)Krause, Decca, Lopez, and
  Fischbach]{Krause:2007zz}
Krause, D.E.; Decca, R.S.; Lopez, D.; Fischbach, E.
\newblock {Experimental Investigation of the Casimir Force beyond the
  Proximity-Force Approximation}.
\newblock {\em Phys. Rev. Lett.} {\bf 2007}, {\em 98},~050403.
\newblock {\url{https://doi.org/10.1103/PhysRevLett.98.050403}}.

\bibitem[Brax \em{et~al.}(2007)Brax, van~de Bruck, Davis, Mota, and
  Shaw]{Brax:2007vm}
Brax, P.; van~de Bruck, C.; Davis, A.C.; Mota, D.F.; Shaw, D.J.
\newblock {Detecting chameleons through Casimir force measurements}.
\newblock {\em Phys. Rev.} {\bf 2007}, {\em D76},~124034,
  \href{http://xxx.lanl.gov/abs/0709.2075}{{\normalfont
  [arXiv:hep-ph/0709.2075]}}.
\newblock {\url{https://doi.org/10.1103/PhysRevD.76.124034}}.

\bibitem[Brax and Davis(2015)]{Brax:2014zta}
Brax, P.; Davis, A.C.
\newblock {Casimir, Gravitational and Neutron Tests of Dark Energy}.
\newblock {\em Phys. Rev. D} {\bf 2015}, {\em 91},~063503,
  \href{http://xxx.lanl.gov/abs/1412.2080}{{\normalfont
  [arXiv:hep-ph/1412.2080]}}.
\newblock {\url{https://doi.org/10.1103/PhysRevD.91.063503}}.

\bibitem[Upadhye \em{et~al.}(2006)Upadhye, Gubser, and Khoury]{Upadhye:2006vi}
Upadhye, A.; Gubser, S.S.; Khoury, J.
\newblock {Unveiling chameleons in tests of gravitational inverse-square law}.
\newblock {\em Phys. Rev. D} {\bf 2006}, {\em 74},~104024,
  \href{http://xxx.lanl.gov/abs/hep-ph/0608186}{{\normalfont
  [hep-ph/0608186]}}.
\newblock {\url{https://doi.org/10.1103/PhysRevD.74.104024}}.

\bibitem[Upadhye(2012)]{Upadhye:2012qu}
Upadhye, A.
\newblock {Dark energy fifth forces in torsion pendulum experiments}.
\newblock {\em Phys. Rev. D} {\bf 2012}, {\em 86},~102003,
  \href{http://xxx.lanl.gov/abs/1209.0211}{{\normalfont
  [arXiv:hep-ph/1209.0211]}}.
\newblock {\url{https://doi.org/10.1103/PhysRevD.86.102003}}.

\bibitem[Elder \em{et~al.}(2016)Elder, Khoury, Haslinger, Jaffe, M\"uller, and
  Hamilton]{Elder:2016yxm}
Elder, B.; Khoury, J.; Haslinger, P.; Jaffe, M.; M\"uller, H.; Hamilton, P.
\newblock {Chameleon Dark Energy and Atom Interferometry}.
\newblock {\em Phys. Rev. D} {\bf 2016}, {\em 94},~044051,
  \href{http://xxx.lanl.gov/abs/1603.06587}{{\normalfont
  [arXiv:astro-ph.CO/1603.06587]}}.
\newblock {\url{https://doi.org/10.1103/PhysRevD.94.044051}}.

\bibitem[{Koyama} and {Sakstein}(2015)]{Koyama_2015}
{Koyama}, K.; {Sakstein}, J.
\newblock {Astrophysical probes of the Vainshtein mechanism: Stars and
  galaxies}.
\newblock {\em \prd} {\bf 2015}, {\em 91},~124066,
  \href{http://xxx.lanl.gov/abs/1502.06872}{{\normalfont
  [arXiv:astro-ph.CO/1502.06872]}}.
\newblock {\url{https://doi.org/10.1103/PhysRevD.91.124066}}.

\bibitem[{Saito} \em{et~al.}(2015){Saito}, {Yamauchi}, {Mizuno}, {Gleyzes}, and
  {Langlois}]{Saito_2015}
{Saito}, R.; {Yamauchi}, D.; {Mizuno}, S.; {Gleyzes}, J.; {Langlois}, D.
\newblock {Modified gravity inside astrophysical bodies}.
\newblock {\em \jcap} {\bf 2015}, {\em 2015},~008--008,
  \href{http://xxx.lanl.gov/abs/1503.01448}{{\normalfont
  [arXiv:gr-qc/1503.01448]}}.
\newblock {\url{https://doi.org/10.1088/1475-7516/2015/06/008}}.

\bibitem[{Gleyzes} \em{et~al.}(2015){Gleyzes}, {Langlois}, {Piazza}, and
  {Vernizzi}]{Gleyze_2015}
{Gleyzes}, J.; {Langlois}, D.; {Piazza}, F.; {Vernizzi}, F.
\newblock {Exploring gravitational theories beyond Horndeski}.
\newblock {\em \jcap} {\bf 2015}, {\em 2015},~018--018,
  \href{http://xxx.lanl.gov/abs/1408.1952}{{\normalfont
  [arXiv:astro-ph.CO/1408.1952]}}.
\newblock {\url{https://doi.org/10.1088/1475-7516/2015/02/018}}.

\bibitem[{Sakstein}(2015{\natexlab{a}})]{Sakstein_2015}
{Sakstein}, J.
\newblock {Hydrogen Burning in Low Mass Stars Constrains Scalar-Tensor Theories
  of Gravity}.
\newblock {\em \prl} {\bf 2015}, {\em 115},~201101,
  \href{http://xxx.lanl.gov/abs/1510.05964}{{\normalfont
  [arXiv:astro-ph.CO/1510.05964]}}.
\newblock {\url{https://doi.org/10.1103/PhysRevLett.115.201101}}.

\bibitem[{Sakstein}(2015{\natexlab{b}})]{Sakstein_2015_dwarf}
{Sakstein}, J.
\newblock {Testing gravity using dwarf stars}.
\newblock {\em \prd} {\bf 2015}, {\em 92},~124045,
  \href{http://xxx.lanl.gov/abs/1511.01685}{{\normalfont
  [arXiv:astro-ph.CO/1511.01685]}}.
\newblock {\url{https://doi.org/10.1103/PhysRevD.92.124045}}.

\bibitem[{S{\'e}gransan} \em{et~al.}(2000){S{\'e}gransan}, {Delfosse},
  {Forveille}, {Beuzit}, {Udry}, {Perrier}, and {{may},or}]{Segransan_2000}
{S{\'e}gransan}, D.; {Delfosse}, X.; {Forveille}, T.; {Beuzit}, J.L.; {Udry},
  S.; {Perrier}, C.; {{may},or}, M.
\newblock {Accurate masses of very low mass stars. III. 16 new or improved
  masses}.
\newblock {\em \aap} {\bf 2000}, {\em 364},~665--673,
  \href{http://xxx.lanl.gov/abs/astro-ph/0010585}{{\normalfont
  [arXiv:astro-ph/astro-ph/0010585]}}.

\bibitem[{Paxton} \em{et~al.}(2011){Paxton}, {Bildsten}, {Dotter}, {Herwig},
  {Lesaffre}, and {Timmes}]{MESA_2011}
{Paxton}, B.; {Bildsten}, L.; {Dotter}, A.; {Herwig}, F.; {Lesaffre}, P.;
  {Timmes}, F.
\newblock {Modules for Experiments in Stellar Astrophysics (MESA)}.
\newblock {\em \apjs} {\bf 2011}, {\em 192},~3,
  \href{http://xxx.lanl.gov/abs/1009.1622}{{\normalfont
  [arXiv:astro-ph.SR/1009.1622]}}.
\newblock {\url{https://doi.org/10.1088/0067-0049/192/1/3}}.

\bibitem[{Paxton} \em{et~al.}(2013){Paxton}, {Cantiello}, {Arras}, {Bildsten},
  {Brown}, {Dotter}, {Mankovich}, {Montgomery}, {Stello}, {Timmes}, and
  {Townsend}]{MESA_2013}
{Paxton}, B.; {Cantiello}, M.; {Arras}, P.; {Bildsten}, L.; {Brown}, E.F.;
  {Dotter}, A.; {Mankovich}, C.; {Montgomery}, M.H.; {Stello}, D.; {Timmes},
  F.X.;  et~al.
\newblock {Modules for Experiments in Stellar Astrophysics (MESA): Planets,
  Oscillations, Rotation, and Massive Stars}.
\newblock {\em \apjs} {\bf 2013}, {\em 208},~4,
  \href{http://xxx.lanl.gov/abs/1301.0319}{{\normalfont
  [arXiv:astro-ph.SR/1301.0319]}}.
\newblock {\url{https://doi.org/10.1088/0067-0049/208/1/4}}.

\bibitem[{Paxton} \em{et~al.}(2015){Paxton}, {Marchant}, {Schwab}, {Bauer},
  {Bildsten}, {Cantiello}, {Dessart}, {Farmer}, {Hu}, {Langer}, {Townsend},
  {Townsley}, and {Timmes}]{MESA_2015}
{Paxton}, B.; {Marchant}, P.; {Schwab}, J.; {Bauer}, E.B.; {Bildsten}, L.;
  {Cantiello}, M.; {Dessart}, L.; {Farmer}, R.; {Hu}, H.; {Langer}, N.;  et~al.
\newblock {Modules for Experiments in Stellar Astrophysics (MESA): Binaries,
  Pulsations, and Explosions}.
\newblock {\em \apjs} {\bf 2015}, {\em 220},~15,
  \href{http://xxx.lanl.gov/abs/1506.03146}{{\normalfont
  [arXiv:astro-ph.SR/1506.03146]}}.
\newblock {\url{https://doi.org/10.1088/0067-0049/220/1/15}}.

\bibitem[{Paxton} \em{et~al.}(2018){Paxton}, {Schwab}, {Bauer}, {Bildsten},
  {Blinnikov}, {Duffell}, {Farmer}, {Goldberg}, {Marchant}, {Sorokina},
  {Thoul}, {Townsend}, and {Timmes}]{MESA_2018}
{Paxton}, B.; {Schwab}, J.; {Bauer}, E.B.; {Bildsten}, L.; {Blinnikov}, S.;
  {Duffell}, P.; {Farmer}, R.; {Goldberg}, J.A.; {Marchant}, P.; {Sorokina},
  E.;  et~al.
\newblock {Modules for Experiments in Stellar Astrophysics (MESA): Convective
  Boundaries, Element Diffusion, and Massive Star Explosions}.
\newblock {\em \apjs} {\bf 2018}, {\em 234},~34,
  \href{http://xxx.lanl.gov/abs/1710.08424}{{\normalfont
  [arXiv:astro-ph.SR/1710.08424]}}.
\newblock {\url{https://doi.org/10.3847/1538-4365/aaa5a8}}.

\bibitem[{Paxton} \em{et~al.}(2019){Paxton}, {Smolec}, {Schwab}, {Gautschy},
  {Bildsten}, {Cantiello}, {Dotter}, {Farmer}, {Goldberg}, {Jermyn}, {Kanbur},
  {Marchant}, {Thoul}, {Townsend}, {Wolf}, {Zhang}, and {Timmes}]{MESA_2019}
{Paxton}, B.; {Smolec}, R.; {Schwab}, J.; {Gautschy}, A.; {Bildsten}, L.;
  {Cantiello}, M.; {Dotter}, A.; {Farmer}, R.; {Goldberg}, J.A.; {Jermyn},
  A.S.;  et~al.
\newblock {Modules for Experiments in Stellar Astrophysics (MESA): Pulsating
  Variable Stars, Rotation, Convective Boundaries, and Energy Conservation}.
\newblock {\em \apjs} {\bf 2019}, {\em 243},~10,
  \href{http://xxx.lanl.gov/abs/1903.01426}{{\normalfont
  [arXiv:astro-ph.SR/1903.01426]}}.
\newblock {\url{https://doi.org/10.3847/1538-4365/ab2241}}.

\bibitem[{Jermyn} \em{et~al.}(2022){Jermyn}, {Bauer}, {Schwab}, {Farmer},
  {Ball}, {Bellinger}, {Dotter}, {Joyce}, {Marchant}, {Mombarg}, {Wolf},
  {Wong}, {Cinquegrana}, {Farrell}, {Smolec}, {Thoul}, {Cantiello}, {Herwig},
  {Toloza}, {Bildsten}, {Townsend}, and {Timmes}]{MESA_2022}
{Jermyn}, A.S.; {Bauer}, E.B.; {Schwab}, J.; {Farmer}, R.; {Ball}, W.H.;
  {Bellinger}, E.P.; {Dotter}, A.; {Joyce}, M.; {Marchant}, P.; {Mombarg},
  J.S.G.;  et~al.
\newblock {Modules for Experiments in Stellar Astrophysics (MESA):
  Time-Dependent Convection, Energy Conservation, Automatic Differentiation,
  and Infrastructure}.
\newblock {\em arXiv e-prints} {\bf 2022}, p. arXiv:2208.03651,
  \href{http://xxx.lanl.gov/abs/2208.03651}{{\normalfont
  [arXiv:astro-ph.SR/2208.03651]}}.
\newblock {\url{https://doi.org/10.48550/arXiv.2208.03651}}.

\bibitem[{Chang} and {Hui}(2011)]{Chang_2011}
{Chang}, P.; {Hui}, L.
\newblock {Stellar Structure and Tests of Modified Gravity}.
\newblock {\em \apj} {\bf 2011}, {\em 732},~25,
  \href{http://xxx.lanl.gov/abs/1011.4107}{{\normalfont
  [arXiv:astro-ph.CO/1011.4107]}}.
\newblock {\url{https://doi.org/10.1088/0004-637X/732/1/25}}.

\bibitem[Saltas and Lopes(2019)]{Saltas:2019ius}
Saltas, I.D.; Lopes, I.
\newblock {Obtaining Precision Constraints on Modified Gravity with
  Helioseismology}.
\newblock {\em Phys. Rev. Lett.} {\bf 2019}, {\em 123},~091103,
  \href{http://xxx.lanl.gov/abs/1909.02552}{{\normalfont
  [arXiv:astro-ph.CO/1909.02552]}}.
\newblock {\url{https://doi.org/10.1103/PhysRevLett.123.091103}}.

\bibitem[Saltas and Christensen-Dalsgaard(2022)]{Saltas:2022ybg}
Saltas, I.D.; Christensen-Dalsgaard, J.
\newblock {Searching for dark energy with the Sun}.
\newblock {\em Astron. Astrophys.} {\bf 2022}, {\em 667},~A115,
  \href{http://xxx.lanl.gov/abs/2205.14134}{{\normalfont
  [arXiv:astro-ph.SR/2205.14134]}}.
\newblock {\url{https://doi.org/10.1051/0004-6361/202244176}}.

\bibitem[{Jain} \em{et~al.}(2013){Jain}, {Vikram}, and {Sakstein}]{Jain_2013}
{Jain}, B.; {Vikram}, V.; {Sakstein}, J.
\newblock {Astrophysical Tests of Modified Gravity: Constraints from Distance
  Indicators in the Nearby Universe}.
\newblock {\em \apj} {\bf 2013}, {\em 779},~39,
  \href{http://xxx.lanl.gov/abs/1204.6044}{{\normalfont
  [arXiv:astro-ph.CO/1204.6044]}}.
\newblock {\url{https://doi.org/10.1088/0004-637X/779/1/39}}.

\bibitem[{Sakstein}(2013)]{Sakstein_2013}
{Sakstein}, J.
\newblock {Stellar oscillations in modified gravity}.
\newblock {\em \prd} {\bf 2013}, {\em 88},~124013,
  \href{http://xxx.lanl.gov/abs/1309.0495}{{\normalfont
  [arXiv:astro-ph.CO/1309.0495]}}.
\newblock {\url{https://doi.org/10.1103/PhysRevD.88.124013}}.

\bibitem[{Desmond} \em{et~al.}(2021){Desmond}, {Sakstein}, and
  {Jain}]{Desmond_2021}
{Desmond}, H.; {Sakstein}, J.; {Jain}, B.
\newblock {Five percent measurement of the gravitational constant in the Large
  Magellanic Cloud}.
\newblock {\em \prd} {\bf 2021}, {\em 103},~024028,
  \href{http://xxx.lanl.gov/abs/2012.05028}{{\normalfont
  [arXiv:astro-ph.CO/2012.05028]}}.
\newblock {\url{https://doi.org/10.1103/PhysRevD.103.024028}}.

\bibitem[{Desmond} \em{et~al.}(2019){Desmond}, {Jain}, and
  {Sakstein}]{Desmond_2019}
{Desmond}, H.; {Jain}, B.; {Sakstein}, J.
\newblock {Local resolution of the Hubble tension: The impact of screened fifth
  forces on the cosmic distance ladder}.
\newblock {\em \prd} {\bf 2019}, {\em 100},~043537,
  \href{http://xxx.lanl.gov/abs/1907.03778}{{\normalfont
  [arXiv:astro-ph.CO/1907.03778]}}.
\newblock {\url{https://doi.org/10.1103/PhysRevD.100.043537}}.

\bibitem[{Desmond} and {Sakstein}(2020)]{Desmond_2020_TRGB}
{Desmond}, H.; {Sakstein}, J.
\newblock {Screened fifth forces lower the TRGB-calibrated Hubble constant
  too}.
\newblock {\em \prd} {\bf 2020}, {\em 102},~023007,
  \href{http://xxx.lanl.gov/abs/2003.12876}{{\normalfont
  [arXiv:astro-ph.CO/2003.12876]}}.
\newblock {\url{https://doi.org/10.1103/PhysRevD.102.023007}}.

\bibitem[{Sakstein} \em{et~al.}(2019){Sakstein}, {Desmond}, and
  {Jain}]{Sakstein_2019}
{Sakstein}, J.; {Desmond}, H.; {Jain}, B.
\newblock {Screened fifth forces mediated by dark matter-baryon interactions:
  Theory and astrophysical probes}.
\newblock {\em \prd} {\bf 2019}, {\em 100},~104035,
  \href{http://xxx.lanl.gov/abs/1907.03775}{{\normalfont
  [arXiv:astro-ph.CO/1907.03775]}}.
\newblock {\url{https://doi.org/10.1103/PhysRevD.100.104035}}.

\bibitem[H\"og\r{a}s and M\"ortsell(2023)]{Hogas:2023vim}
H\"og\r{a}s, M.; M\"ortsell, E.
\newblock {Impact of symmetron screening on the Hubble tension: New constraints
  using cosmic distance ladder data}.
\newblock {\em Phys. Rev. D} {\bf 2023}, {\em 108},~024007,
  \href{http://xxx.lanl.gov/abs/2303.12827}{{\normalfont
  [arXiv:astro-ph.CO/2303.12827]}}.
\newblock {\url{https://doi.org/10.1103/PhysRevD.108.024007}}.

\bibitem[{Sakstein} \em{et~al.}(2017){Sakstein}, {Kenna-Allison}, and
  {Koyama}]{Sakstein_2017_osc}
{Sakstein}, J.; {Kenna-Allison}, M.; {Koyama}, K.
\newblock {Stellar pulsations in beyond Horndeski gravity theories}.
\newblock {\em \jcap} {\bf 2017}, {\em 2017},~007,
  \href{http://xxx.lanl.gov/abs/1611.01062}{{\normalfont
  [arXiv:gr-qc/1611.01062]}}.
\newblock {\url{https://doi.org/10.1088/1475-7516/2017/03/007}}.

\bibitem[{Zhao} \em{et~al.}(2011){Zhao}, {Li}, and {Koyama}]{Zhao_2011}
{Zhao}, G.B.; {Li}, B.; {Koyama}, K.
\newblock {Testing Gravity Using the Environmental Dependence of Dark Matter
  Halos}.
\newblock {\em \prl} {\bf 2011}, {\em 107},~071303,
  \href{http://xxx.lanl.gov/abs/1105.0922}{{\normalfont
  [arXiv:astro-ph.CO/1105.0922]}}.
\newblock {\url{https://doi.org/10.1103/PhysRevLett.107.071303}}.

\bibitem[{Schmidt}(2010)]{Schmidt_2010_dyn}
{Schmidt}, F.
\newblock {Dynamical masses in modified gravity}.
\newblock {\em \prd} {\bf 2010}, {\em 81},~103002,
  \href{http://xxx.lanl.gov/abs/1003.0409}{{\normalfont
  [arXiv:astro-ph.CO/1003.0409]}}.
\newblock {\url{https://doi.org/10.1103/PhysRevD.81.103002}}.

\bibitem[{Cabr{\'e}} \em{et~al.}(2012){Cabr{\'e}}, {Vikram}, {Zhao}, {Jain},
  and {Koyama}]{Cabre_2012}
{Cabr{\'e}}, A.; {Vikram}, V.; {Zhao}, G.B.; {Jain}, B.; {Koyama}, K.
\newblock {Astrophysical tests of gravity: a screening map of the nearby
  universe}.
\newblock {\em \jcap} {\bf 2012}, {\em 2012},~034,
  \href{http://xxx.lanl.gov/abs/1204.6046}{{\normalfont
  [arXiv:astro-ph.CO/1204.6046]}}.
\newblock {\url{https://doi.org/10.1088/1475-7516/2012/07/034}}.

\bibitem[{Yang} \em{et~al.}(2007){Yang}, {Mo}, {van den Bosch}, {Pasquali},
  {Li}, and {Barden}]{Yang_2007}
{Yang}, X.; {Mo}, H.J.; {van den Bosch}, F.C.; {Pasquali}, A.; {Li}, C.;
  {Barden}, M.
\newblock {Galaxy Groups in the SDSS DR4. I. The Catalog and Basic Properties}.
\newblock {\em \apj} {\bf 2007}, {\em 671},~153--170,
  \href{http://xxx.lanl.gov/abs/0707.4640}{{\normalfont
  [arXiv:astro-ph/0707.4640]}}.
\newblock {\url{https://doi.org/10.1086/522027}}.

\bibitem[{Abell} \em{et~al.}(1989){Abell}, {Corwin}, and {Olowin}]{Abell_1989}
{Abell}, G.O.; {Corwin}, Harold~G., J.; {Olowin}, R.P.
\newblock {A Catalog of Rich Clusters of Galaxies}.
\newblock {\em \apjs} {\bf 1989}, {\em 70},~1.
\newblock {\url{https://doi.org/10.1086/191333}}.

\bibitem[{Ebeling} \em{et~al.}(1996){Ebeling}, {Voges}, {Bohringer}, {Edge},
  {Huchra}, and {Briel}]{Ebeling_1996}
{Ebeling}, H.; {Voges}, W.; {Bohringer}, H.; {Edge}, A.C.; {Huchra}, J.P.;
  {Briel}, U.G.
\newblock {Properties of the X-ray-brightest Abell-type clusters of galaxies
  (XBACs) from ROSAT All-Sky Survey data - I. The sample}.
\newblock {\em \mnras} {\bf 1996}, {\em 281},~799--829,
  \href{http://xxx.lanl.gov/abs/astro-ph/9602080}{{\normalfont
  [arXiv:astro-ph/astro-ph/9602080]}}.
\newblock {\url{https://doi.org/10.1093/mnras/281.3.799}}.

\bibitem[{Karachentsev} \em{et~al.}(2004){Karachentsev}, {Karachentseva},
  {Huchtmeier}, and {Makarov}]{Karachentsev_2004}
{Karachentsev}, I.D.; {Karachentseva}, V.E.; {Huchtmeier}, W.K.; {Makarov},
  D.I.
\newblock {A Catalog of Neighboring Galaxies}.
\newblock {\em \aj} {\bf 2004}, {\em 127},~2031--2068.
\newblock {\url{https://doi.org/10.1086/382905}}.

\bibitem[{Lavaux} and {Hudson}(2011)]{Lavaux_2011}
{Lavaux}, G.; {Hudson}, M.J.
\newblock {The 2M++ galaxy redshift catalogue}.
\newblock {\em \mnras} {\bf 2011}, {\em 416},~2840--2856,
  \href{http://xxx.lanl.gov/abs/1105.6107}{{\normalfont
  [arXiv:astro-ph.CO/1105.6107]}}.
\newblock {\url{https://doi.org/10.1111/j.1365-2966.2011.19233.x}}.

\bibitem[{Zhao} \em{et~al.}(2011){Zhao}, {Li}, and {Koyama}]{Zhao_2011_nbody}
{Zhao}, G.B.; {Li}, B.; {Koyama}, K.
\newblock {N-body simulations for f(R) gravity using a self-adaptive
  particle-mesh code}.
\newblock {\em \prd} {\bf 2011}, {\em 83},~044007,
  \href{http://xxx.lanl.gov/abs/1011.1257}{{\normalfont
  [arXiv:astro-ph.CO/1011.1257]}}.
\newblock {\url{https://doi.org/10.1103/PhysRevD.83.044007}}.

\bibitem[{Desmond} \em{et~al.}(2018){Desmond}, {Ferreira}, {Lavaux}, and
  {Jasche}]{Desmond_2018_maps}
{Desmond}, H.; {Ferreira}, P.G.; {Lavaux}, G.; {Jasche}, J.
\newblock {Reconstructing the gravitational field of the local Universe}.
\newblock {\em \mnras} {\bf 2018}, {\em 474},~3152--3161,
  \href{http://xxx.lanl.gov/abs/1705.02420}{{\normalfont
  [arXiv:astro-ph.GA/1705.02420]}}.
\newblock {\url{https://doi.org/10.1093/mnras/stx3062}}.

\bibitem[{Behroozi} \em{et~al.}(2013){Behroozi}, {Wechsler}, and
  {Wu}]{Behroozi_2013}
{Behroozi}, P.S.; {Wechsler}, R.H.; {Wu}, H.Y.
\newblock {The ROCKSTAR Phase-space Temporal Halo Finder and the Velocity
  Offsets of Cluster Cores}.
\newblock {\em \apj} {\bf 2013}, {\em 762},~109,
  \href{http://xxx.lanl.gov/abs/1110.4372}{{\normalfont
  [arXiv:astro-ph.CO/1110.4372]}}.
\newblock {\url{https://doi.org/10.1088/0004-637X/762/2/109}}.

\bibitem[{Skillman} \em{et~al.}(2014){Skillman}, {Warren}, {Turk}, {Wechsler},
  {Holz}, and {Sutter}]{Skillman_2014}
{Skillman}, S.W.; {Warren}, M.S.; {Turk}, M.J.; {Wechsler}, R.H.; {Holz}, D.E.;
  {Sutter}, P.M.
\newblock {Dark Sky Simulations: Early Data Release}.
\newblock {\em arXiv e-prints} {\bf 2014}, p. arXiv:1407.2600,
  \href{http://xxx.lanl.gov/abs/1407.2600}{{\normalfont
  [arXiv:astro-ph.CO/1407.2600]}}.

\bibitem[{Lehmann} \em{et~al.}(2017){Lehmann}, {Mao}, {Becker}, {Skillman}, and
  {Wechsler}]{Lehman_2017}
{Lehmann}, B.V.; {Mao}, Y.Y.; {Becker}, M.R.; {Skillman}, S.W.; {Wechsler},
  R.H.
\newblock {The Concentration Dependence of the Galaxy-Halo Connection: Modeling
  Assembly Bias with Abundance Matching}.
\newblock {\em \apj} {\bf 2017}, {\em 834},~37,
  \href{http://xxx.lanl.gov/abs/1510.05651}{{\normalfont
  [arXiv:astro-ph.CO/1510.05651]}}.
\newblock {\url{https://doi.org/10.3847/1538-4357/834/1/37}}.

\bibitem[{Navarro} \em{et~al.}(1997){Navarro}, {Frenk}, and {White}]{NFW_1997}
{Navarro}, J.F.; {Frenk}, C.S.; {White}, S.D.M.
\newblock {A Universal Density Profile from Hierarchical Clustering}.
\newblock {\em \apj} {\bf 1997}, {\em 490},~493--508.
\newblock {\url{https://doi.org/10.1086/304888}}.

\bibitem[{Cole} and {Lacey}(1996)]{Cole_1996}
{Cole}, S.; {Lacey}, C.
\newblock {The structure of dark matter haloes in hierarchical clustering
  models}.
\newblock {\em \mnras} {\bf 1996}, {\em 281},~716,
  \href{http://xxx.lanl.gov/abs/astro-ph/9510147}{{\normalfont
  [arXiv:astro-ph/astro-ph/9510147]}}.
\newblock {\url{https://doi.org/10.1093/mnras/281.2.716}}.

\bibitem[{Jasche} and {Wandelt}(2012)]{BORG_1}
{Jasche}, J.; {Wandelt}, B.D.
\newblock {Bayesian inference from photometric redshift surveys}.
\newblock {\em \mnras} {\bf 2012}, {\em 425},~1042--1056,
  \href{http://xxx.lanl.gov/abs/1106.2757}{{\normalfont
  [arXiv:astro-ph.CO/1106.2757]}}.
\newblock {\url{https://doi.org/10.1111/j.1365-2966.2012.21423.x}}.

\bibitem[{Jasche} and {Wandelt}(2013)]{BORG_2}
{Jasche}, J.; {Wandelt}, B.D.
\newblock {Bayesian physical reconstruction of initial conditions from
  large-scale structure surveys}.
\newblock {\em \mnras} {\bf 2013}, {\em 432},~894--913,
  \href{http://xxx.lanl.gov/abs/1203.3639}{{\normalfont
  [arXiv:astro-ph.CO/1203.3639]}}.
\newblock {\url{https://doi.org/10.1093/mnras/stt449}}.

\bibitem[{Jasche} \em{et~al.}(2010){Jasche}, {Kitaura}, {Wandelt}, and
  {En{\ss}lin}]{BORG_3}
{Jasche}, J.; {Kitaura}, F.S.; {Wandelt}, B.D.; {En{\ss}lin}, T.A.
\newblock {Bayesian power-spectrum inference for large-scale structure data}.
\newblock {\em \mnras} {\bf 2010}, {\em 406},~60--85,
  \href{http://xxx.lanl.gov/abs/0911.2493}{{\normalfont
  [arXiv:astro-ph.CO/0911.2493]}}.
\newblock {\url{https://doi.org/10.1111/j.1365-2966.2010.16610.x}}.

\bibitem[{Jasche} \em{et~al.}(2015){Jasche}, {Leclercq}, and {Wandelt}]{BORG_4}
{Jasche}, J.; {Leclercq}, F.; {Wandelt}, B.D.
\newblock {Past and present cosmic structure in the SDSS DR7 main sample}.
\newblock {\em \jcap} {\bf 2015}, {\em 2015},~036,
  \href{http://xxx.lanl.gov/abs/1409.6308}{{\normalfont
  [arXiv:astro-ph.CO/1409.6308]}}.
\newblock {\url{https://doi.org/10.1088/1475-7516/2015/01/036}}.

\bibitem[{Lavaux} and {Jasche}(2016)]{Lavaux_2016}
{Lavaux}, G.; {Jasche}, J.
\newblock {Unmasking the masked Universe: the 2M++ catalogue through Bayesian
  eyes}.
\newblock {\em \mnras} {\bf 2016}, {\em 455},~3169--3179,
  \href{http://xxx.lanl.gov/abs/1509.05040}{{\normalfont
  [arXiv:astro-ph.CO/1509.05040]}}.
\newblock {\url{https://doi.org/10.1093/mnras/stv2499}}.

\bibitem[{Shao} \em{et~al.}(2019){Shao}, {Li}, {Cautun}, {Wang}, and
  {Wang}]{Shao_2019}
{Shao}, S.; {Li}, B.; {Cautun}, M.; {Wang}, H.; {Wang}, J.
\newblock {Screening maps of the local Universe I - Methodology}.
\newblock {\em \mnras} {\bf 2019}, {\em 489},~4912--4925,
  \href{http://xxx.lanl.gov/abs/1907.02081}{{\normalfont
  [arXiv:astro-ph.CO/1907.02081]}}.
\newblock {\url{https://doi.org/10.1093/mnras/stz2450}}.

\bibitem[{Li} \em{et~al.}(2013{\natexlab{a}}){Li}, {Zhao}, and
  {Koyama}]{Li_2013_VS}
{Li}, B.; {Zhao}, G.B.; {Koyama}, K.
\newblock {Exploring Vainshtein mechanism on adaptively refined meshes}.
\newblock {\em \jcap} {\bf 2013}, {\em 2013},~023,
  \href{http://xxx.lanl.gov/abs/1303.0008}{{\normalfont
  [arXiv:astro-ph.CO/1303.0008]}}.
\newblock {\url{https://doi.org/10.1088/1475-7516/2013/05/023}}.

\bibitem[{Li} \em{et~al.}(2013{\natexlab{b}}){Li}, {Barreira}, {Baugh},
  {Hellwing}, {Koyama}, {Pascoli}, and {Zhao}]{Li_2013_quartic}
{Li}, B.; {Barreira}, A.; {Baugh}, C.M.; {Hellwing}, W.A.; {Koyama}, K.;
  {Pascoli}, S.; {Zhao}, G.B.
\newblock {Simulating the quartic Galileon gravity model on adaptively refined
  meshes}.
\newblock {\em \jcap} {\bf 2013}, {\em 2013},~012,
  \href{http://xxx.lanl.gov/abs/1308.3491}{{\normalfont
  [arXiv:astro-ph.CO/1308.3491]}}.
\newblock {\url{https://doi.org/10.1088/1475-7516/2013/11/012}}.

\bibitem[{Wang} \em{et~al.}(2014){Wang}, {Mo}, {Yang}, {Jing}, and
  {Lin}]{Wang_2014}
{Wang}, H.; {Mo}, H.J.; {Yang}, X.; {Jing}, Y.P.; {Lin}, W.P.
\newblock {ELUCID{\textemdash}Exploring the Local Universe with the
  Reconstructed Initial Density Field. I. Hamiltonian Markov Chain Monte Carlo
  Method with Particle Mesh Dynamics}.
\newblock {\em \apj} {\bf 2014}, {\em 794},~94,
  \href{http://xxx.lanl.gov/abs/1407.3451}{{\normalfont
  [arXiv:astro-ph.CO/1407.3451]}}.
\newblock {\url{https://doi.org/10.1088/0004-637X/794/1/94}}.

\bibitem[{Wang} \em{et~al.}(2016){Wang}, {Mo}, {Yang}, {Zhang}, {Shi}, {Jing},
  {Liu}, {Li}, {Kang}, and {Gao}]{Wang_2016}
{Wang}, H.; {Mo}, H.J.; {Yang}, X.; {Zhang}, Y.; {Shi}, J.; {Jing}, Y.P.;
  {Liu}, C.; {Li}, S.; {Kang}, X.; {Gao}, Y.
\newblock {ELUCID - Exploring the Local Universe with ReConstructed Initial
  Density Field III: Constrained Simulation in the SDSS Volume}.
\newblock {\em \apj} {\bf 2016}, {\em 831},~164,
  \href{http://xxx.lanl.gov/abs/1608.01763}{{\normalfont
  [arXiv:astro-ph.CO/1608.01763]}}.
\newblock {\url{https://doi.org/10.3847/0004-637X/831/2/164}}.

\bibitem[{Naidoo} \em{et~al.}(2022){Naidoo}, {Hellwing}, {Bilicki},
  {Libeskind}, {Pfeifer}, and {Hoffman}]{Naidoo_2022}
{Naidoo}, K.; {Hellwing}, W.; {Bilicki}, M.; {Libeskind}, N.; {Pfeifer}, S.;
  {Hoffman}, Y.
\newblock {Constrained simulations of the local Universe with Modified
  Gravity}.
\newblock {\em arXiv e-prints} {\bf 2022}, p. arXiv:2209.14386,
  \href{http://xxx.lanl.gov/abs/2209.14386}{{\normalfont
  [arXiv:astro-ph.CO/2209.14386]}}.

\bibitem[{Kourkchi} \em{et~al.}(2020){Kourkchi}, {Tully}, {Eftekharzadeh},
  {Llop}, {Courtois}, {Guinet}, {Dupuy}, {Neill}, {Seibert}, {Andrews},
  {Chuang}, {Danesh}, {Gonzalez}, {Holthaus}, {Mokelke}, {Schoen}, and
  {Urasaki}]{Kourkchi_2020}
{Kourkchi}, E.; {Tully}, R.B.; {Eftekharzadeh}, S.; {Llop}, J.; {Courtois},
  H.M.; {Guinet}, D.; {Dupuy}, A.; {Neill}, J.D.; {Seibert}, M.; {Andrews}, M.;
   et~al.
\newblock {Cosmicflows-4: The Catalog of {\ensuremath{\sim}}10,000 Tully-Fisher
  Distances}.
\newblock {\em \apj} {\bf 2020}, {\em 902},~145,
  \href{http://xxx.lanl.gov/abs/2009.00733}{{\normalfont
  [arXiv:astro-ph.GA/2009.00733]}}.
\newblock {\url{https://doi.org/10.3847/1538-4357/abb66b}}.

\bibitem[{Doumler} \em{et~al.}(2013){Doumler}, {Gottl{\"o}ber}, {Hoffman}, and
  {Courtois}]{Doumler_2013_III}
{Doumler}, T.; {Gottl{\"o}ber}, S.; {Hoffman}, Y.; {Courtois}, H.
\newblock {Reconstructing cosmological initial conditions from galaxy peculiar
  velocities - III. Constrained simulations}.
\newblock {\em \mnras} {\bf 2013}, {\em 430},~912--923,
  \href{http://xxx.lanl.gov/abs/1212.2810}{{\normalfont
  [arXiv:astro-ph.CO/1212.2810]}}.
\newblock {\url{https://doi.org/10.1093/mnras/sts614}}.

\bibitem[{Zaroubi} \em{et~al.}(1995){Zaroubi}, {Hoffman}, {Fisher}, and
  {Lahav}]{Zaroubi_1995}
{Zaroubi}, S.; {Hoffman}, Y.; {Fisher}, K.B.; {Lahav}, O.
\newblock {Wiener Reconstruction of the Large-Scale Structure}.
\newblock {\em \apj} {\bf 1995}, {\em 449},~446,
  \href{http://xxx.lanl.gov/abs/astro-ph/9410080}{{\normalfont
  [arXiv:astro-ph/astro-ph/9410080]}}.
\newblock {\url{https://doi.org/10.1086/176070}}.

\bibitem[{Doumler} \em{et~al.}(2013){Doumler}, {Hoffman}, {Courtois}, and
  {Gottl{\"o}ber}]{Doumler_2013_I}
{Doumler}, T.; {Hoffman}, Y.; {Courtois}, H.; {Gottl{\"o}ber}, S.
\newblock {Reconstructing cosmological initial conditions from galaxy peculiar
  velocities - I. Reverse Zeldovich Approximation}.
\newblock {\em \mnras} {\bf 2013}, {\em 430},~888--901,
  \href{http://xxx.lanl.gov/abs/1212.2806}{{\normalfont
  [arXiv:astro-ph.CO/1212.2806]}}.
\newblock {\url{https://doi.org/10.1093/mnras/sts613}}.

\bibitem[{Winther} \em{et~al.}(2017){Winther}, {Koyama}, {Manera}, {Wright},
  and {Zhao}]{Winther_2017}
{Winther}, H.A.; {Koyama}, K.; {Manera}, M.; {Wright}, B.S.; {Zhao}, G.B.
\newblock {COLA with scale-dependent growth: applications to screened modified
  gravity models}.
\newblock {\em \jcap} {\bf 2017}, {\em 2017},~006,
  \href{http://xxx.lanl.gov/abs/1703.00879}{{\normalfont
  [arXiv:astro-ph.CO/1703.00879]}}.
\newblock {\url{https://doi.org/10.1088/1475-7516/2017/08/006}}.

\bibitem[{Tassev} \em{et~al.}(2013){Tassev}, {Zaldarriaga}, and
  {Eisenstein}]{Tassev_2013}
{Tassev}, S.; {Zaldarriaga}, M.; {Eisenstein}, D.J.
\newblock {Solving large scale structure in ten easy steps with COLA}.
\newblock {\em \jcap} {\bf 2013}, {\em 2013},~036,
  \href{http://xxx.lanl.gov/abs/1301.0322}{{\normalfont
  [arXiv:astro-ph.CO/1301.0322]}}.
\newblock {\url{https://doi.org/10.1088/1475-7516/2013/06/036}}.

\bibitem[Fasiello and Vlah(2017)]{Fasiello:2017bot}
Fasiello, M.; Vlah, Z.
\newblock {Screening in perturbative approaches to LSS}.
\newblock {\em Phys. Lett. B} {\bf 2017}, {\em 773},~236--241,
  \href{http://xxx.lanl.gov/abs/1704.07552}{{\normalfont
  [arXiv:astro-ph.CO/1704.07552]}}.
\newblock {\url{https://doi.org/10.1016/j.physletb.2017.08.032}}.

\bibitem[{Jain} and {VanderPlas}(2011)]{Jain_2011}
{Jain}, B.; {VanderPlas}, J.
\newblock {Tests of modified gravity with dwarf galaxies}.
\newblock {\em \jcap} {\bf 2011}, {\em 2011},~032,
  \href{http://xxx.lanl.gov/abs/1106.0065}{{\normalfont
  [arXiv:astro-ph.CO/1106.0065]}}.
\newblock {\url{https://doi.org/10.1088/1475-7516/2011/10/032}}.

\bibitem[{Vikram} \em{et~al.}(2013){Vikram}, {Cabr{\'e}}, {Jain}, and
  {VanderPlas}]{Vikram_2013}
{Vikram}, V.; {Cabr{\'e}}, A.; {Jain}, B.; {VanderPlas}, J.T.
\newblock {Astrophysical tests of modified gravity: the morphology and
  kinematics of dwarf galaxies}.
\newblock {\em \jcap} {\bf 2013}, {\em 2013},~020,
  \href{http://xxx.lanl.gov/abs/1303.0295}{{\normalfont
  [arXiv:astro-ph.CO/1303.0295]}}.
\newblock {\url{https://doi.org/10.1088/1475-7516/2013/08/020}}.

\bibitem[Giovanelli \em{et~al.}(2005)Giovanelli et~al.]{Giovanelli:2005ee}
Giovanelli, R.;  et~al.
\newblock {The Arecibo Legacy Fast ALFA Survey. 1. Science goals, survey design
  and strategy}.
\newblock {\em Astron. J.} {\bf 2005}, {\em 130},~2598--2612,
  \href{http://xxx.lanl.gov/abs/astro-ph/0508301}{{\normalfont
  [astro-ph/0508301]}}.
\newblock {\url{https://doi.org/10.1086/497431}}.

\bibitem[Abazajian \em{et~al.}(2009)Abazajian et~al.]{SDSS:2008tqn}
Abazajian, K.N.;  et~al.
\newblock {The Seventh Data Release of the Sloan Digital Sky Survey}.
\newblock {\em Astrophys. J. Suppl.} {\bf 2009}, {\em 182},~543--558,
  \href{http://xxx.lanl.gov/abs/0812.0649}{{\normalfont
  [arXiv:astro-ph/0812.0649]}}.
\newblock {\url{https://doi.org/10.1088/0067-0049/182/2/543}}.

\bibitem[{Desmond} \em{et~al.}(2018{\natexlab{a}}){Desmond}, {Ferreira},
  {Lavaux}, and {Jasche}]{Desmond_2018_sep}
{Desmond}, H.; {Ferreira}, P.G.; {Lavaux}, G.; {Jasche}, J.
\newblock {Fifth force constraints from the separation of galaxy mass
  components}.
\newblock {\em Phys.~Rev.~D} {\bf 2018}, {\em 98},~064015.
\newblock {\url{https://doi.org/10.1103/PhysRevD.98.064015}}.

\bibitem[{Desmond} \em{et~al.}(2018{\natexlab{b}}){Desmond}, {Ferreira},
  {Lavaux}, and {Jasche}]{Desmond_2018_warp}
{Desmond}, H.; {Ferreira}, P.G.; {Lavaux}, G.; {Jasche}, J.
\newblock {Fifth force constraints from galaxy warps}.
\newblock {\em Phys.~Rev.~D} {\bf 2018}, {\em 98},~083010.
\newblock {\url{https://doi.org/10.1103/PhysRevD.98.083010}}.

\bibitem[Desmond \em{et~al.}(2019)Desmond, Ferreira, Lavaux, and
  Jasche]{Desmond_2018_web}
Desmond, H.; Ferreira, P.G.; Lavaux, G.; Jasche, J.
\newblock {The Fifth Force in the Local Cosmic Web}.
\newblock {\em Mon. Not. Roy. Astron. Soc.} {\bf 2019}, {\em 483},~L64--L68.
\newblock {\url{https://doi.org/10.1093/mnrasl/sly221}}.

\bibitem[{Desmond} and {Ferreira}(2020)]{Desmond_2020}
{Desmond}, H.; {Ferreira}, P.G.
\newblock {Galaxy morphology rules out astrophysically relevant Hu-Sawicki f (R
  ) gravity}.
\newblock {\em \prd} {\bf 2020}, {\em 102},~104060,
  \href{http://xxx.lanl.gov/abs/2009.08743}{{\normalfont
  [arXiv:astro-ph.CO/2009.08743]}}.
\newblock {\url{https://doi.org/10.1103/PhysRevD.102.104060}}.

\bibitem[{Bartlett} \em{et~al.}(2021){Bartlett}, {Desmond}, and
  {Ferreira}]{Bartlett_2021_cal}
{Bartlett}, D.J.; {Desmond}, H.; {Ferreira}, P.G.
\newblock {Calibrating galaxy formation effects in galactic tests of
  fundamental physics}.
\newblock {\em \prd} {\bf 2021}, {\em 103},~123502,
  \href{http://xxx.lanl.gov/abs/2103.10356}{{\normalfont
  [arXiv:astro-ph.CO/2103.10356]}}.
\newblock {\url{https://doi.org/10.1103/PhysRevD.103.123502}}.

\bibitem[{Asvathaman} \em{et~al.}(2017){Asvathaman}, {Heyl}, and
  {Hui}]{Asvathaman_2017}
{Asvathaman}, A.; {Heyl}, J.S.; {Hui}, L.
\newblock {E{\"o}tv{\"o}s experiments with supermassive black holes}.
\newblock {\em \mnras} {\bf 2017}, {\em 465},~3261--3266.
\newblock {\url{https://doi.org/10.1093/mnras/stw2905}}.

\bibitem[{Bartlett} \em{et~al.}(2021{\natexlab{a}}){Bartlett}, {Desmond}, and
  {Ferreira}]{Bartlett_2021_VS}
{Bartlett}, D.J.; {Desmond}, H.; {Ferreira}, P.G.
\newblock {Constraints on Galileons from the positions of supermassive black
  holes}.
\newblock {\em \prd} {\bf 2021}, {\em 103},~023523,
  \href{http://xxx.lanl.gov/abs/2010.05811}{{\normalfont
  [arXiv:astro-ph.CO/2010.05811]}}.
\newblock {\url{https://doi.org/10.1103/PhysRevD.103.023523}}.

\bibitem[{Bartlett} \em{et~al.}(2021{\natexlab{b}}){Bartlett}, {Desmond},
  {Devriendt}, {Ferreira}, and {Slyz}]{Bartlett_2021_HAGN}
{Bartlett}, D.J.; {Desmond}, H.; {Devriendt}, J.; {Ferreira}, P.G.; {Slyz}, A.
\newblock {Spatially offset black holes in the Horizon-AGN simulation and
  comparison to observations}.
\newblock {\em \mnras} {\bf 2021}, {\em 500},~4639--4657,
  \href{http://xxx.lanl.gov/abs/2007.01353}{{\normalfont
  [arXiv:astro-ph.GA/2007.01353]}}.
\newblock {\url{https://doi.org/10.1093/mnras/staa3516}}.

\bibitem[{Khoury}(2013)]{Khoury_2013}
{Khoury}, J.
\newblock {Les Houches Lectures on Physics Beyond the Standard Model of
  Cosmology},  2013,  \href{http://xxx.lanl.gov/abs/1312.2006}{{\normalfont
  [1312.2006]}}.

\bibitem[{Murphy} \em{et~al.}(2012){Murphy}, {Adelberger}, {Battat}, {Hoyle},
  {Johnson}, {McMillan}, {Stubbs}, and {Swanson}]{Murphy_2012}
{Murphy}, T.~W., J.; {Adelberger}, E.G.; {Battat}, J.B.R.; {Hoyle}, C.D.;
  {Johnson}, N.H.; {McMillan}, R.J.; {Stubbs}, C.W.; {Swanson}, H.E.
\newblock {APOLLO: millimeter lunar laser ranging}.
\newblock {\em Classical and Quantum Gravity} {\bf 2012}, {\em 29},~184005.
\newblock {\url{https://doi.org/10.1088/0264-9381/29/18/184005}}.

\bibitem[Schmidt \em{et~al.}(2009)Schmidt, Lima, Oyaizu, and
  Hu]{Schmidt:2008tn}
Schmidt, F.; Lima, M.V.; Oyaizu, H.; Hu, W.
\newblock {Non-linear Evolution of f(R) Cosmologies III: Halo Statistics}.
\newblock {\em Phys. Rev. D} {\bf 2009}, {\em 79},~083518,
  \href{http://xxx.lanl.gov/abs/0812.0545}{{\normalfont
  [arXiv:astro-ph/0812.0545]}}.
\newblock {\url{https://doi.org/10.1103/PhysRevD.79.083518}}.

\bibitem[Li and Hu(2011)]{Li:2011uw}
Li, Y.; Hu, W.
\newblock {Chameleon Halo Modeling in f(R) Gravity}.
\newblock {\em Phys. Rev. D} {\bf 2011}, {\em 84},~084033,
  \href{http://xxx.lanl.gov/abs/1107.5120}{{\normalfont
  [arXiv:astro-ph.CO/1107.5120]}}.
\newblock {\url{https://doi.org/10.1103/PhysRevD.84.084033}}.

\bibitem[Pourhasan \em{et~al.}(2011)Pourhasan, Afshordi, Mann, and
  Davis]{Pourhasan:2011sm}
Pourhasan, R.; Afshordi, N.; Mann, R.B.; Davis, A.C.
\newblock {Chameleon Gravity, Electrostatics, and Kinematics in the Outer
  Galaxy}.
\newblock {\em JCAP} {\bf 2011}, {\em 12},~005,
  \href{http://xxx.lanl.gov/abs/1109.0538}{{\normalfont
  [arXiv:astro-ph.CO/1109.0538]}}.
\newblock {\url{https://doi.org/10.1088/1475-7516/2011/12/005}}.

\bibitem[Lombriser \em{et~al.}(2012)Lombriser, Koyama, Zhao, and
  Li]{Lombriser:2012nn}
Lombriser, L.; Koyama, K.; Zhao, G.B.; Li, B.
\newblock {Chameleon f(R) gravity in the virialized cluster}.
\newblock {\em Phys. Rev. D} {\bf 2012}, {\em 85},~124054,
  \href{http://xxx.lanl.gov/abs/1203.5125}{{\normalfont
  [arXiv:astro-ph.CO/1203.5125]}}.
\newblock {\url{https://doi.org/10.1103/PhysRevD.85.124054}}.

\bibitem[Lombriser \em{et~al.}(2013)Lombriser, Li, Koyama, and
  Zhao]{Lombriser:2013wta}
Lombriser, L.; Li, B.; Koyama, K.; Zhao, G.B.
\newblock {Modeling halo mass functions in chameleon f(R) gravity}.
\newblock {\em Phys. Rev. D} {\bf 2013}, {\em 87},~123511,
  \href{http://xxx.lanl.gov/abs/1304.6395}{{\normalfont
  [arXiv:astro-ph.CO/1304.6395]}}.
\newblock {\url{https://doi.org/10.1103/PhysRevD.87.123511}}.

\bibitem[Lombriser \em{et~al.}(2014)Lombriser, Koyama, and
  Li]{Lombriser:2013eza}
Lombriser, L.; Koyama, K.; Li, B.
\newblock {Halo modelling in chameleon theories}.
\newblock {\em JCAP} {\bf 2014}, {\em 03},~021,
  \href{http://xxx.lanl.gov/abs/1312.1292}{{\normalfont
  [arXiv:astro-ph.CO/1312.1292]}}.
\newblock {\url{https://doi.org/10.1088/1475-7516/2014/03/021}}.

\bibitem[Shi \em{et~al.}(2015)Shi, Li, Han, Gao, and Hellwing]{Shi:2015aya}
Shi, D.; Li, B.; Han, J.; Gao, L.; Hellwing, W.A.
\newblock {Exploring the liminality: properties of haloes and subhaloes in
  borderline $f(R)$ gravity}.
\newblock {\em Mon. Not. Roy. Astron. Soc.} {\bf 2015}, {\em 452},~3179--3191,
  \href{http://xxx.lanl.gov/abs/1503.01109}{{\normalfont
  [arXiv:astro-ph.CO/1503.01109]}}.
\newblock {\url{https://doi.org/10.1093/mnras/stv1549}}.

\bibitem[Mitchell \em{et~al.}(2019)Mitchell, Arnold, He, and
  Li]{Mitchell:2019qke}
Mitchell, M.A.; Arnold, C.; He, J.h.; Li, B.
\newblock {A general framework to test gravity using galaxy clusters II: A
  universal model for the halo concentration in $f(R)$ gravity}.
\newblock {\em Mon. Not. Roy. Astron. Soc.} {\bf 2019}, {\em 487},~1410--1425,
  \href{http://xxx.lanl.gov/abs/1901.06392}{{\normalfont
  [arXiv:astro-ph.CO/1901.06392]}}.
\newblock {\url{https://doi.org/10.1093/mnras/stz1389}}.

\bibitem[Barreira \em{et~al.}(2014)Barreira, Li, Hellwing, Lombriser, Baugh,
  and Pascoli]{Barreira:2014zza}
Barreira, A.; Li, B.; Hellwing, W.A.; Lombriser, L.; Baugh, C.M.; Pascoli, S.
\newblock {Halo model and halo properties in Galileon gravity cosmologies}.
\newblock {\em JCAP} {\bf 2014}, {\em 04},~029,
  \href{http://xxx.lanl.gov/abs/1401.1497}{{\normalfont
  [arXiv:astro-ph.CO/1401.1497]}}.
\newblock {\url{https://doi.org/10.1088/1475-7516/2014/04/029}}.

\bibitem[Mitchell \em{et~al.}(2021)Mitchell, Hern\'andez-Aguayo, Arnold, and
  Li]{Mitchell:2021aex}
Mitchell, M.A.; Hern\'andez-Aguayo, C.; Arnold, C.; Li, B.
\newblock {A general framework to test gravity using galaxy clusters IV:
  cluster and halo properties in DGP gravity}.
\newblock {\em Mon. Not. Roy. Astron. Soc.} {\bf 2021}, {\em 508},~4140--4156,
  \href{http://xxx.lanl.gov/abs/2106.13815}{{\normalfont
  [arXiv:astro-ph.CO/2106.13815]}}.
\newblock {\url{https://doi.org/10.1093/mnras/stab2817}}.

\bibitem[Pizzuti \em{et~al.}(2022)Pizzuti, Saltas, Umetsu, and
  Sartoris]{Pizzuti:2021brr}
Pizzuti, L.; Saltas, I.D.; Umetsu, K.; Sartoris, B.
\newblock {Probing vainsthein-screening gravity with galaxy clusters using
  internal kinematics and strong and weak lensing}.
\newblock {\em Mon. Not. Roy. Astron. Soc.} {\bf 2022}, {\em 512},~4280--4290,
  \href{http://xxx.lanl.gov/abs/2112.12139}{{\normalfont
  [arXiv:astro-ph.CO/2112.12139]}}.
\newblock {\url{https://doi.org/10.1093/mnras/stac746}}.

\bibitem[Falck \em{et~al.}(2015)Falck, Koyama, and Zhao]{Falck:2015rsa}
Falck, B.; Koyama, K.; Zhao, G.B.
\newblock {Cosmic Web and Environmental Dependence of Screening: Vainshtein vs.
  Chameleon}.
\newblock {\em JCAP} {\bf 2015}, {\em 07},~049,
  \href{http://xxx.lanl.gov/abs/1503.06673}{{\normalfont
  [arXiv:astro-ph.CO/1503.06673]}}.
\newblock {\url{https://doi.org/10.1088/1475-7516/2015/07/049}}.

\bibitem[Diemer and Kravtsov(2014)]{Diemer:2014xya}
Diemer, B.; Kravtsov, A.V.
\newblock {Dependence of the outer density profiles of halos on their mass
  accretion rate}.
\newblock {\em Astrophys. J.} {\bf 2014}, {\em 789},~1,
  \href{http://xxx.lanl.gov/abs/1401.1216}{{\normalfont
  [arXiv:astro-ph.CO/1401.1216]}}.
\newblock {\url{https://doi.org/10.1088/0004-637X/789/1/1}}.

\bibitem[More \em{et~al.}(2016)More et~al.]{More:2016vgs}
More, S.;  et~al.
\newblock {Detection of the Splashback Radius and Halo Assembly bias of Massive
  Galaxy Clusters}.
\newblock {\em Astrophys. J.} {\bf 2016}, {\em 825},~39,
  \href{http://xxx.lanl.gov/abs/1601.06063}{{\normalfont
  [arXiv:astro-ph.CO/1601.06063]}}.
\newblock {\url{https://doi.org/10.3847/0004-637X/825/1/39}}.

\bibitem[Baxter \em{et~al.}(2017)Baxter, Chang, Jain, Adhikari, Dalal,
  Kravtsov, More, Rozo, Rykoff, and Sheth]{Baxter:2017csy}
Baxter, E.; Chang, C.; Jain, B.; Adhikari, S.; Dalal, N.; Kravtsov, A.; More,
  S.; Rozo, E.; Rykoff, E.; Sheth, R.K.
\newblock {The Halo Boundary of Galaxy Clusters in the SDSS}.
\newblock {\em Astrophys. J.} {\bf 2017}, {\em 841},~18,
  \href{http://xxx.lanl.gov/abs/1702.01722}{{\normalfont
  [arXiv:astro-ph.CO/1702.01722]}}.
\newblock {\url{https://doi.org/10.3847/1538-4357/aa6ff0}}.

\bibitem[Chang \em{et~al.}(2018)Chang et~al.]{Chang:2017hjt}
Chang, C.;  et~al.
\newblock {The Splashback Feature around DES Galaxy Clusters: Galaxy Density
  and Weak Lensing Profiles}.
\newblock {\em Astrophys. J.} {\bf 2018}, {\em 864},~83,
  \href{http://xxx.lanl.gov/abs/1710.06808}{{\normalfont
  [arXiv:astro-ph.CO/1710.06808]}}.
\newblock {\url{https://doi.org/10.3847/1538-4357/aad5e7}}.

\bibitem[Shin \em{et~al.}(2018)Shin et~al.]{Shin:2018pic}
Shin, T.;  et~al.
\newblock {Measurement of the Splashback Feature around SZ-selected Galaxy
  Clusters with DES, SPT and ACT} {\bf 2018}.
\newblock  \href{http://xxx.lanl.gov/abs/1811.06081}{{\normalfont
  [arXiv:astro-ph.CO/1811.06081]}}.

\bibitem[Umetsu and Diemer(2017)]{Umetsu:2016cun}
Umetsu, K.; Diemer, B.
\newblock {Lensing Constraints on the Mass Profile Shape and the Splashback
  Radius of Galaxy Clusters}.
\newblock {\em Astrophys. J.} {\bf 2017}, {\em 836},~231,
  \href{http://xxx.lanl.gov/abs/1611.09366}{{\normalfont
  [arXiv:astro-ph.CO/1611.09366]}}.
\newblock {\url{https://doi.org/10.3847/1538-4357/aa5c90}}.

\bibitem[Contigiani \em{et~al.}(2018)Contigiani, Hoekstra, and
  Bahé]{Contigiani:2018qxn}
Contigiani, O.; Hoekstra, H.; Bahé, Y.M.
\newblock {Weak lensing constraints on splashback around massive clusters} {\bf
  2018}.
\newblock  \href{http://xxx.lanl.gov/abs/1809.10045}{{\normalfont
  [arXiv:astro-ph.CO/1809.10045]}}.

\bibitem[Adhikari \em{et~al.}(2018)Adhikari, Sakstein, Jain, Dalal, and
  Li]{Adhikari:2018izo}
Adhikari, S.; Sakstein, J.; Jain, B.; Dalal, N.; Li, B.
\newblock {Splashback in galaxy clusters as a probe of cosmic expansion and
  gravity}.
\newblock {\em JCAP} {\bf 2018}, {\em 11},~033,
  \href{http://xxx.lanl.gov/abs/1806.04302}{{\normalfont
  [arXiv:astro-ph.CO/1806.04302]}}.
\newblock {\url{https://doi.org/10.1088/1475-7516/2018/11/033}}.

\bibitem[{Fillmore} and {Goldreich}(1984)]{Fillmore:1984}
{Fillmore}, J.A.; {Goldreich}, P.
\newblock {Self-similar gravitational collapse in an expanding universe}.
\newblock {\em \apj} {\bf 1984}, {\em 281},~1--8.
\newblock {\url{https://doi.org/10.1086/162070}}.

\bibitem[{Bertschinger}(1985)]{Bertschinger:1985pd}
{Bertschinger}, E.
\newblock {Self-similar secondary infall and accretion in an Einstein-de Sitter
  universe}.
\newblock {\em \apjs} {\bf 1985}, {\em 58},~39--65.
\newblock {\url{https://doi.org/10.1086/191028}}.

\bibitem[Lithwick and Dalal(2011)]{Lithwick:2010ej}
Lithwick, Y.; Dalal, N.
\newblock {Self-Similar Solutions of Triaxial Dark Matter Halos}.
\newblock {\em Astrophys. J.} {\bf 2011}, {\em 734},~100,
  \href{http://xxx.lanl.gov/abs/1010.3723}{{\normalfont
  [arXiv:astro-ph.CO/1010.3723]}}.
\newblock {\url{https://doi.org/10.1088/0004-637X/734/2/100}}.

\bibitem[Shi(2016)]{Shi:2016lwp}
Shi, X.
\newblock {The outer profile of dark matter haloes: an analytical approach}.
\newblock {\em Mon. Not. Roy. Astron. Soc.} {\bf 2016}, {\em 459},~3711--3720,
  \href{http://xxx.lanl.gov/abs/1603.01742}{{\normalfont
  [arXiv:astro-ph.CO/1603.01742]}}.
\newblock {\url{https://doi.org/10.1093/mnras/stw925}}.

\bibitem[Amendola \em{et~al.}(2018)Amendola et~al.]{Amendola:2016saw}
Amendola, L.;  et~al.
\newblock {Cosmology and fundamental physics with the Euclid satellite}.
\newblock {\em Living Rev. Rel.} {\bf 2018}, {\em 21},~2,
  \href{http://xxx.lanl.gov/abs/1606.00180}{{\normalfont
  [arXiv:astro-ph.CO/1606.00180]}}.
\newblock {\url{https://doi.org/10.1007/s41114-017-0010-3}}.

\bibitem[Gubitosi \em{et~al.}(2013)Gubitosi, Piazza, and
  Vernizzi]{Gubitosi:2012hu}
Gubitosi, G.; Piazza, F.; Vernizzi, F.
\newblock {The Effective Field Theory of Dark Energy}.
\newblock {\em JCAP} {\bf 2013}, {\em 02},~032,
  \href{http://xxx.lanl.gov/abs/1210.0201}{{\normalfont
  [arXiv:hep-th/1210.0201]}}.
\newblock {\url{https://doi.org/10.1088/1475-7516/2013/02/032}}.

\bibitem[Gleyzes \em{et~al.}(2013)Gleyzes, Langlois, Piazza, and
  Vernizzi]{Gleyzes:2013ooa}
Gleyzes, J.; Langlois, D.; Piazza, F.; Vernizzi, F.
\newblock {Essential Building Blocks of Dark Energy}.
\newblock {\em JCAP} {\bf 2013}, {\em 08},~025,
  \href{http://xxx.lanl.gov/abs/1304.4840}{{\normalfont
  [arXiv:hep-th/1304.4840]}}.
\newblock {\url{https://doi.org/10.1088/1475-7516/2013/08/025}}.

\bibitem[Creminelli \em{et~al.}(2014)Creminelli, Gleyzes, Hui, Simonovi\'c, and
  Vernizzi]{Creminelli:2013nua}
Creminelli, P.; Gleyzes, J.; Hui, L.; Simonovi\'c, M.; Vernizzi, F.
\newblock {Single-Field Consistency Relations of Large Scale Structure. Part
  III: Test of the Equivalence Principle}.
\newblock {\em JCAP} {\bf 2014}, {\em 06},~009,
  \href{http://xxx.lanl.gov/abs/1312.6074}{{\normalfont
  [arXiv:astro-ph.CO/1312.6074]}}.
\newblock {\url{https://doi.org/10.1088/1475-7516/2014/06/009}}.

\bibitem[Bellini and Sawicki(2014)]{Bellini:2014fua}
Bellini, E.; Sawicki, I.
\newblock {Maximal freedom at minimum cost: linear large-scale structure in
  general modifications of gravity}.
\newblock {\em JCAP} {\bf 2014}, {\em 07},~050,
  \href{http://xxx.lanl.gov/abs/1404.3713}{{\normalfont
  [arXiv:astro-ph.CO/1404.3713]}}.
\newblock {\url{https://doi.org/10.1088/1475-7516/2014/07/050}}.

\bibitem[Hu \em{et~al.}(2014)Hu, Raveri, Frusciante, and Silvestri]{Hu:2013twa}
Hu, B.; Raveri, M.; Frusciante, N.; Silvestri, A.
\newblock {Effective Field Theory of Cosmic Acceleration: an implementation in
  CAMB}.
\newblock {\em Phys. Rev. D} {\bf 2014}, {\em 89},~103530,
  \href{http://xxx.lanl.gov/abs/1312.5742}{{\normalfont
  [arXiv:astro-ph.CO/1312.5742]}}.
\newblock {\url{https://doi.org/10.1103/PhysRevD.89.103530}}.

\bibitem[Raveri \em{et~al.}(2014)Raveri, Hu, Frusciante, and
  Silvestri]{Raveri:2014cka}
Raveri, M.; Hu, B.; Frusciante, N.; Silvestri, A.
\newblock {Effective Field Theory of Cosmic Acceleration: constraining dark
  energy with CMB data}.
\newblock {\em Phys. Rev. D} {\bf 2014}, {\em 90},~043513,
  \href{http://xxx.lanl.gov/abs/1405.1022}{{\normalfont
  [arXiv:astro-ph.CO/1405.1022]}}.
\newblock {\url{https://doi.org/10.1103/PhysRevD.90.043513}}.

\bibitem[Zumalac\'arregui \em{et~al.}(2017)Zumalac\'arregui, Bellini, Sawicki,
  Lesgourgues, and Ferreira]{Zumalacarregui:2016pph}
Zumalac\'arregui, M.; Bellini, E.; Sawicki, I.; Lesgourgues, J.; Ferreira, P.G.
\newblock {hi\_class: Horndeski in the Cosmic Linear Anisotropy Solving
  System}.
\newblock {\em JCAP} {\bf 2017}, {\em 08},~019,
  \href{http://xxx.lanl.gov/abs/1605.06102}{{\normalfont
  [arXiv:astro-ph.CO/1605.06102]}}.
\newblock {\url{https://doi.org/10.1088/1475-7516/2017/08/019}}.

\bibitem[Bellini \em{et~al.}(2020)Bellini, Sawicki, and
  Zumalac\'arregui]{Bellini:2019syt}
Bellini, E.; Sawicki, I.; Zumalac\'arregui, M.
\newblock {hi\_class: Background Evolution, Initial Conditions and
  Approximation Schemes}.
\newblock {\em JCAP} {\bf 2020}, {\em 02},~008,
  \href{http://xxx.lanl.gov/abs/1909.01828}{{\normalfont
  [arXiv:astro-ph.CO/1909.01828]}}.
\newblock {\url{https://doi.org/10.1088/1475-7516/2020/02/008}}.

\end{thebibliography}

\end{document}